\documentclass[aps,prd,preprint,superscriptaddress,showpacs]{revtex4}

\usepackage{amsmath}
\usepackage{amssymb}
\usepackage{graphicx,color}
\usepackage{array}

\makeatletter
  
  \@addtoreset{equation}{section}
  \makeatother

\newcommand{\Slash}[1]{{\ooalign{\hfil#1\hfil\crcr\raise.167ex\hbox{/}}}}

\def\thline{\noalign{\hrule height 1pt}}
\def\tvline{\vrule width 1pt} 
\newcommand {\beq}{\begin{eqnarray}}
\newcommand {\eeq}{\end{eqnarray}}

\newcommand{\NF}{N_{\rm F}}
\newcommand{\NC}{N_{\rm C}}

\newcommand{\hs}[1]{\hspace{#1 mm}}
\newcommand{\bpm}{\begin{pmatrix}}
\newcommand{\epm}{\end{pmatrix}}

\begin{document}

\title{
Classifying bions in Grassmann sigma models\\ 
and non-Abelian gauge theories by D-branes
}

\author{Tatsuhiro Misumi}
\email{misumi(at)phys-h.keio.ac.jp}

\author{Muneto Nitta}
\email{nitta(at)phys-h.keio.ac.jp}

\author{Norisuke Sakai}
\email{norisuke.sakai(at)gmail.com}

\affiliation{Department of Physics, and Research and Education Center for Natural Sciences, 
Keio University, Hiyoshi 4-1-1, Yokohama, Kanagawa 223-8521, Japan}

\begin{abstract} 
We classify bions in the Grassmann $Gr_{N_{\rm F},N_{\rm C}}$ 
sigma model (including the ${\mathbb C}P^{N_{\rm F}-1}$ model) 
on ${\mathbb R}^{1}\times S^{1}$ with twisted boundary conditions.
We formulate these models as $U(\NC)$ gauge theories 
with $\NF$ flavors in the fundamental representations. 
These theories can be promoted to supersymmetric gauge theories 
and further can be embedded into D-brane configurations in 
type II superstring theories. 
We focus on specific configurations composed of multiple 
fractional instantons, termed neutral bions and charged bions, 
which are identified as perturbative infrared renormalons 
by \"{U}nsal and his collaborators.
We show that D-brane configurations as well as the moduli matrix 
offer a very useful tool to classify all possible bion 
configurations in these models.  
Contrary to the ${\mathbb C}P^{N_{\rm F}-1}$ model, 
there exist Bogomol'nyi-Prasad-Sommerfield (BPS) fractional 
instantons with topological charge greater than unity 
(of order $N_{\rm C}$) that cannot be reduced to a composite 
of an instanton and fractional instantons.  
As a consequence, we find that the Grassmann sigma model admits 
neutral bions made of BPS and anti-BPS 
fractional instantons each of which has topological charge 
greater (less) than one (minus one), that are not decomposable 
into instanton anti-instanton and the rests. 
The ${\mathbb C}P^{N_{\rm F}-1}$ model is found to have no 
charged bions. 
In contrast, we find that the Grassmann sigma model admits 
charged bions, for which we construct exact non-BPS solutions 
of the field equations. 
\end{abstract}

\maketitle

\newpage



\section{Introduction}
\label{sec:Intro}

The extensive studies on QCD 
with adjoint fermions (adj.) on compactified spacetime 
have recently revealed great significance of fractional 
multi-instanton configurations 
with zero instanton charge, called ``bions"
\cite{Unsal:2007vu, Unsal:2007jx, Shifman:2008ja, 
Poppitz:2009uq, Poppitz:2012sw,Argyres:2012vv, Argyres:2012ka, 
Dunne:2012ae, Dunne:2012zk, Dabrowski:2013kba, Dunne:2013ada, 
Cherman:2013yfa, Basar:2013eka, Dunne:2014bca, Cherman:2014ofa,
Bolognesi:2013tya, Misumi:2014jua,  Misumi:2014raa}.
It is known that there exist two types of such bion 
configurations, including ``magnetic (charged) bions" and 
``neutral bions".
While magnetic bions bring about the semiclassical confinement 
in QCD(adj.) on ${\mathbb R}^{3} \times S^{1}$ 
\cite{Hosotani:1983xw, Hosotani:1988bm, Myers:2007vc, 
Myers:2009df, Cossu:2009sq, Meisinger:2009ne, Nishimura:2009me, 
Anber:2011de, Ogilvie:2012is, Kashiwa:2013rmg, Cossu:2013ora}, 
neutral bions (zero topological charge and zero magnetic charge), 
which are identified as the infrared renormalons in the theory 
\cite{Argyres:2012vv, Argyres:2012ka, Dunne:2012ae, Dunne:2012zk, 
Dabrowski:2013kba, Dunne:2013ada, Cherman:2013yfa, Basar:2013eka, 
Dunne:2014bca, Cherman:2014ofa, 'tHooft:1977am, Fateev:1994ai, Fateev:1994dp},
induce a center stabilizing potential for Wilson line holonomy, 
and play an essential role in unambiguous and self-consistent 
semiclassical definition of quantum field theories through the 
process known as ``resurgence": 
It has been shown that imaginary ambiguities arising in neutral 
bion's amplitude and those arising in non-Borel-summable 
perturbative series (renormalon ambiguities) cancel against 
each other in the small compactification-scale regime of 
QCD(adj.) on ${\mathbb R}^{3} \times S^{1}$.
It implies that the full semi-classical expansion including 
perturbative and non-perturbative sectors, ``resurgent" 
expansion \cite{Ec1}, leads to unambiguous and self-consistent 
definition of field theories 
in the same manner as the Bogomol'nyi-Zinn-Justin (BZJ) 
prescription in quantum mechanics 
\cite{Bogomolny:1980ur, ZinnJustin:1981dx, ZinnJustin:2004ib}.

Bions and the resurgence in the low-dimensional models have 
been extensively investigated for the ${\mathbb C}P^{N-1}$ model
\cite{Dunne:2012ae, Dunne:2012zk, Dabrowski:2013kba,
Bolognesi:2013tya, Misumi:2014jua},  
principal chiral models \cite{Cherman:2013yfa,Cherman:2014ofa}, 
and quantum mechanics \cite{Dunne:2013ada, Basar:2013eka, Dunne:2014bca}. 
In Refs.~\cite{Dunne:2012ae, Dunne:2012zk}, generic arguments 
on bion configurations were given in the ${\mathbb C}P^{N-1}$ 
model on ${\mathbb R}^1 \times S^1$ with ${\mathbb Z}_{N}$ 
twisted boundary conditions, which is a corresponding situation 
to $U(1)^{N-1}$ center-symmetric phase in QCD(adj.), 
based on the independent instanton description 
taking account of interactions between far-separated fractional 
instantons and anti-instantons. 
According to the study, the imaginary ambiguity in the 
amplitude of neutral bions has the same magnitude with an 
opposite sign as the leading ambiguity ($\sim \mp i\pi e^{-2S_{I}/N}$) 
arising from the non-Borel-summable series expanded around 
the perturbative vacuum.
The ambiguities at higher orders are cancelled by amplitudes 
of bion molecules (2-bion, 3-bion,...), 
and the full trans-series expansion around 
the perturbative and non-perturbative vacua results in 
unambiguous semiclassical definition of field theories.

Among other things, the two dimensional ${\mathbb C}P^{N-1}$ model 
enjoys common features with four-dimensional Yang-Mills theory
\cite{Polyakov} 
such as asymptotic freedom, dynamical mass generation, 
and the presence of instantons \cite{Polyakov:1975yp,Din:1980jg}.
Fractional instantons in the ${\mathbb C}P^{N-1}$ model 
on ${\mathbb R}^1 \times S^1$
with twisted boundary conditions 
were found in Ref.~\cite{Eto:2004rz} 
(see also Refs.~\cite{Bruckmann:2007zh}).
Fractional instantons in the Grasssmann sigma model 
were also found in Ref.~\cite{Eto:2006mz}.
Explicit solutions or ansatze corresponding to bion 
configurations in the ${\mathbb C}P^{N-1}$ model 
have been investigated recently 
\cite{Dabrowski:2013kba, Bolognesi:2013tya, Misumi:2014jua}. 
Although fractional instantons are Bogomol'nyi-Prasad-Sommerfield 
(BPS) solutions \cite{Bogomolny:1975de,Prasad:1975kr}, bions 
are non-BPS as composite of fractional instantons and anti-instantons. 
In Ref.~\cite{Dabrowski:2013kba}, such non-BPS solutions were found 
out in the ${\mathbb C}P^{N-1}$ model on ${\mathbb R}^1 \times S^1$ 
with the ${\mathbb Z}_{N}$ twisted boundary condition 
by using the method of Ref.~\cite{Din:1980jg}, which 
can be saddle 
points for the trans-series expansion. 
In our previous work \cite{Misumi:2014jua} we have studied an 
ansatz corresponding to neutral bions in the ${\mathbb C}P^{N-1}$ 
model beyond exact solutions, 
and have shown that our ansatz is consistent with the result 
from the far-separated instanton gas calculus 
\cite{Dunne:2012ae, Dunne:2012zk} even from short to large 
separations.

The purpose of our present work is to classify ansatze corresponding to 
all possible bion configurations in the ${\mathbb C}P^{N-1}$ and Grassmann sigma models
on ${\mathbb R}^{1}\times S^{1}$ with twisted boundary conditions.
We study mainly the ${\mathbb Z}_{N}$ twisted boundary condition for simplicity, 
although we can easily extend our study to more general twisted boundary conditions.
In our study, we introduce a new viewpoint based on D-brane configurations
to study bion configurations:
The ${\mathbb C}P^{N-1}$ and 
Grassmann sigma models are formulated as 
a $U(\NC)$ gauge theories with $\NF$ flavors in the fundamental 
representations \cite{D'Adda:1978kp,Aoyama:1979zj,Higashijima:1999ki,
Lindstrom:1983rt,Hitchin:1986ea,Antoniadis:1996ra,Arai:2003tc}, 
which can be embedded into supersymmetric gauge theories by 
adding fermions (and scalar fields) appropriately.
Sigma model instantons (lumps) in the Grassmann sigma model 
are promoted to non-Abelian vortices \cite{Hanany:2003hp, 
Auzzi:2003fs, Shifman:2004dr, Hanany:2004ea, Eto:2005yh} 
(see Refs.~\cite{Tong:2005un,Eto:2006pg,Shifman:2007ce} as a review)
in gauge theories, especially of semi-local type 
\cite{Shifman:2006kd,Eto:2007yv}. 
By doing so, the moduli space of BPS vortices (lumps) can be 
clarified completely in terms of the moduli matrix \cite{Eto:2006pg}.
These theories can be further embedded into Hanany-Witten type 
D-brane configurations in type II string theories \cite{Hanany:1996ie,Giveon:1998sr}, 
where vortices can be identified with certain D-branes \cite{Hanany:2003hp}.
The T-duality transformation along $S^1$ maps vortices to domain 
walls \cite{Isozumi:2004jc}, which can be described by kinky 
D-branes \cite{Lambert:1999ix,Eto:2004vy}. 
These D-brane configurations were used to study moduli space 
of non-Abelian vortices before \cite{Eto:2006mz}. 

In this paper, we show that these D-brane configurations as 
well as the moduli matrix offer very useful tools to classify 
all possible bion configurations in the Grassmann sigma model  
including the ${\mathbb C}P^{N-1}$ model, and the corresponding 
non-Abelian gauge theories. 
We unexpectedly find that the Grassmann sigma model admits 
neutral bions made of BPS and anti-BPS fractional instantons 
each of which has a topological charge greater (less) than 
one (minus one), but it cannot be decomposed into instanton 
anti-instanton and the rests.   
We find that the Grassmann sigma model admits charged bions, 
while the ${\mathbb C}P^{N-1}$ model does not. 
There are many different species of fractional instantons in 
the Grassmann sigma model. 
Among them, we can choose species of BPS fractional instantons 
and anti-BPS fractional instantons that are noninteracting and 
can coexist stably. In such cases, we obtain exact non-BPS 
solutions representing charged bions. 
We also calculate the energy density and topological charge 
density of the bion configurations in these models numerically 
to obtain their interaction energies, which give valuable 
informations on the interactions between constituent fractional 
instantons, such as the sign and magnitude of the strength, 
and the dependence on the separations between constituent 
fractional instantons.

In Sec.~\ref{mdl}, we formulate the Grassmann sigma model 
$Gr_{N_{\rm F},N_{\rm C}}$ including the 
${\mathbb C}P^{N_{\rm F}-1}$ model as $U(N_{\rm C})$ gauge 
theory with $N_{\rm F}$ Higgs scalar fields in the fundamental 
representation. 
We also present BPS equations for BPS vortices or lumps in 
these theories and the moduli matrix which exhaust moduli 
parameters of BPS solutions. 
In Sec.~\ref{sec:D-brane}, we introduce D-brane configurations 
in type II string theories, that realize our theory on certain 
D-brane world-volumes. 
We then study fractional instantons in Grassmann sigma model 
$Gr_{N_{\rm F},N_{\rm C}}$ including the 
${\mathbb C}P^{N_{\rm F}-1}$ model in terms of D-brane 
configurations and the moduli matrix. 
In Sec.~\ref{sec:CPN}, we classify neutral bions in 
the ${\mathbb C}P^{N_{\rm F}-1}$ model. 
In Sec.~\ref{sec:Gr}, we classify neutral and charged bions in 
the Grassmann sigma model $Gr_{N_{\rm F},N_{\rm C}}$. 
In Sec.~\ref{sec:int}, we discuss interaction energy for 
bions with changing the distance between 
fractional instanton constituents. 
Sec.~\ref{sec:SD} is devoted to summary and discussion.
In Appendix \ref{app:constraint}, we discuss solutions of 
constraint of Grassmann sigma model.

\section{The $U(N_{\rm C})$ gauge theory 
and Grassmann sigma model}
\label{mdl}

\subsection{Gauge theory and moduli matrix 
}
\label{sc:mdl:model}

Target spaces of supersymmetric 
nonlinear sigma models must be K\"ahler 
for four supercharges \cite{Zumino:1979et} 
and hyper-K\"ahler 
for eight supercharges \cite{Alvarez-Gaume:1981hm}. 
The ${\mathbb C}P^{N_{\rm F}-1}$ 
and Grassmann sigma models can be obtained from supersymmetric gauge theories with 
four supercharges \cite{Aoyama:1979zj,D'Adda:1978kp,Higashijima:1999ki} 
and eight supercharges 
\cite{Lindstrom:1983rt,Hitchin:1986ea,Antoniadis:1996ra,Arai:2003tc}. 
In this subsection we consider two-dimensional euclidean 
gauge field theories in the flat $x^1$-$x^2$ plane with 
$U(N_{\rm C})$ gauge group and $N_{\rm F}$ flavors of scalar 
fields in the fundamental representation denoted as an 
$N_{\rm C} \times N_{\rm F}$ matrix $H$. The Lagrangian is 
given as 
\begin{eqnarray}
{\cal L}_{\rm gauge} = 
{\rm Tr}\left[ {1\over 2g^2}F_{\mu\nu}F_{\mu\nu} 
+{\cal D}_\mu H \left({\cal D}_\mu H\right)^\dagger \right]
+{\rm Tr}
\left[
\frac{g^2}{4}
\left(v^2\mathbf{1}_{N_{\rm C}}
-H  H^{\dagger} 
\right)^2 
\right] . 
\label{eq:mdl:total_lagrangian}
\end{eqnarray}
where $g$ is the gauge coupling, $v$ is a real positive parameter 
(Fayet-Iliopoulos parameter in the context of supersymmetry) 
\cite{Eto:2006pg}. 
The covariant derivative ${\cal D}_\mu$ with the gauge field $W_\mu$ 
and field strength $F_{\mu\nu}$ are defined as 
${\cal D}_\mu H=(\partial_\mu + iW_\mu)H, \qquad 
F_{\mu\nu}=-i[{\cal D}_\mu,\,{\cal D}_\nu]$. 
We use a matrix notation such as 
$W_\mu=W_\mu^I T_I$, 
where $T_I \ (I = 0, 1,2,\cdots, N_{\rm C}^2-1)$ 
are matrix generators of the gauge 
group $G$ in the fundamental representation 
satisfying 
${\rm Tr}(T_I T_J) = \frac12\delta_{IJ}$, 
$[ T_I , T_J]=i f_{IJ}{}^{K} T_K$ with $T^0$ as the 
$U(1)$ generator. 
The gauge couplings for $U(1)$ and $SU(N_{\rm C})$ 
are independent, but we have chosen identical values for 
them to discuss classical field configurations in simple terms. 

Since the scalar fields $H$ are massless, the Lagrangian has 
a global symmetry $SU(N_{\rm F})$. 
It can be embedded into a supersymmetric theory with eight 
supercharges \cite{Eto:2006pg}. 
Consequently it admits BPS solitons\cite{Bogomolny:1975de,Prasad:1975kr} 
which preserve a part of supercharges\cite{Witten:1978mh}. 
Vacuum in this model is characterized by the vanishing vacuum 
energy 
\begin{eqnarray} 
H  H^{\dagger}  =v^2\mathbf{1}_{N_{\rm C}}. 
\label{eq:mdl:D-term-cond}
\end{eqnarray}
This condition necessitates some of scalar fields $H$ 
to be non-vanishing (rank$H=N_{\rm C}$), implying that the 
gauge symmetry is completely broken (Higgs phase). 
This vacuum is called the color-flavor locked vacuum 
where $N_{\rm C}$ out of $N_{\rm F}$ flavors should be chosen 
to be non-vanishing and leaves only a diagonal $SU(N_{\rm C})$ 
of color $SU(N_{\rm C})$ and $SU(N_{\rm G})$ subgroup of flavor 
$SU(N_{\rm F})$ group beside the remaining 
$SU(N_{\rm F}-N_{\rm C})\times U(1)$ as the global symmetry. 

Since we consider two euclidean dimensions, 
instantons are the usual vortices with codimension two. 
The Bogomol'nyi completion\cite{Bogomolny:1975de} can be applied 
to the Lagrangian ${\cal L}$ to give a bound 
\begin{eqnarray}
{\cal L}&=&{\rm Tr}\left[\frac1{g^2}
\left(B_3 + {g^2 \over 2} (v^2 {\bf 1}_{N_{\rm C}} - H H^\dagger)\right)^2
+\left({\cal D}_1 H + i {\cal D}_2 H\right)
\left({\cal D}_1 H + i {\cal D}_2 H\right)^\dagger\right]
\nonumber \\
&&
+{\rm Tr}\left[-v^2\,B_3
+2i\partial _{[1}H{\cal D}_{2]}H^\dagger \right]
\ge 
{\rm Tr}\left[-v^2\,B_3
+2i\partial _{[1}H{\cal D}_{2]}H^\dagger \right]
\label{eq:vtx:energy}
\end{eqnarray}
with a magnetic field $B_3 \equiv F_{12}$. 
The bound is saturated if the following BPS vortex 
equations hold \cite{Hanany:2003hp, Auzzi:2003fs} 
\begin{eqnarray}
 0 &=& {\cal D}_1 H + i {\cal D}_2 H, 
   \label{eq:vtx:BPSeq1}
\\ 
 0 &=& B_3 + {g^2 \over 2} (v^2 {\bf 1}_{N_{\rm C}} - H H^\dagger) .
   \label{eq:vtx:BPSeq2}
\end{eqnarray}
When these BPS equations are satisfied, the total energy $T$ 
is given by 
\begin{eqnarray}
 T \equiv \int d^2x {\cal L} 
= 2\pi v^2 \; Q, 
  \label{eq:vtx:tension}
\end{eqnarray}
with the topological charge $Q$ (instanton number) 
defined by 
\begin{eqnarray}
Q
=-\frac{1}{2\pi}\int d^2 x\ {\rm Tr} B_3  
=-\frac{1}{2\pi}\int d^2 x\ {\rm Tr} 
\left(\frac{1}{2}\epsilon_{\mu\nu}F_{\mu\nu}\right) . 
  \label{eq:vtx:topologicalCharge}
\end{eqnarray}
measuring the winding number of the $U(1)$ part of the broken 
$U(N_{\rm C})$ gauge symmetry \footnote{
Our normalization convention of $Q$ corresponds to that in 
our previous work \cite{Misumi:2014jua} divided by $2\pi$, 
and gives integer values for instantons. 
}. 

Let us define $S=S(z,\bar z) \in GL(N_{\rm C},{\mathbb C})$ 
using a complex coordinate $z \equiv x^1+ix^2$ 
\begin{eqnarray}
 W_1+iW_2 =-i2S^{-1}{\bar \partial}_z S . 
  \label{eq:vtx:defS}
\end{eqnarray}
We can solve\cite{Eto:2005yh} the first of the BPS equations 
(\ref{eq:vtx:BPSeq1}) in terms of $S$ 
\begin{eqnarray}
 H =S^{-1} H_0(z) ,  
  \label{eq:vtx:solution}
\end{eqnarray}
where $H_0(z)$ is an arbitrary $N_{\rm C}$ by $N_{\rm F}$ 
matrix whose components are holomorphic with respect to $z$, 
which is called the {\it moduli matrix} of BPS solitons. 
By defining a gauge invariant quantity 
\begin{eqnarray}  
 \Omega (z,\bar z) \equiv S(z,\bar z) S^\dagger (z,\bar z) , 
  \label{eq:vtx:omega} 
\end{eqnarray} 
the second BPS 
equations (\ref{eq:vtx:BPSeq2}) can be rewritten as 
\begin{eqnarray}
 \partial_z (\Omega^{-1} \bar \partial_z \Omega ) 
 = {g^2 \over 4} (v^2{\bf 1}_{N_{\rm C}} - \Omega^{-1} H_0 H_0^\dagger) .
 \label{eq:vtx:master}
\end{eqnarray}
We call this the {\it master equation} for BPS solitons 
\footnote{The master equation reduces to the so-called Taubes 
equation \cite{Taubes:1979tm} in the case of ANO vortices 
($N_{\rm C} = N_{\rm F} = 1$) by rewriting 
$v^2\Omega(z,\bar z)=|H_0|^2 e^{-\xi (z,\bar z)}$ with 
$H_0 = \prod_i(z-z_i)$. Note that $\log \Omega$ is regular everywhere 
while $\xi$ is singular at vortex positions. Non-integrability of 
the master equation has been shown in \cite{Inami:2006wr}. }. 
This equation is expected to give no additional moduli parameters. 
It was proved for $N_{\rm F} = N_{\rm C}=1$ 
(the ANO vortices)\cite{Taubes:1979tm} and is consistent 
with the index theorem\cite{Hanany:2003hp} in general 
$N_{\rm C}$ and $N_{\rm F}$. 
Moreover, this fact can be easily proved in the strong 
coupling limit where the gauge theories reduce to the 
Grassmann sigma model, as we show in the next 
subsection. 
Thus we assume that the moduli matrix $H_0$ describes the 
moduli space completely. 

However we note that there exists a redundancy in the solution 
(\ref{eq:vtx:solution}): physical quantities $H$ and $W_{1,2}$ 
are invariant under the following $V$-transformations 
\begin{eqnarray}
\hspace{-1cm}
 H_0 (z)\to H_0' (z)= V(z) H_0(z), 
\quad S(z,\bar z) \to S'(z,\bar z)= V(z) S(z,\bar z) , 
  \label{eq:vtx:V-trans}
\end{eqnarray}
with $V(z) \in GL(N_{\rm C},{\mathbb C})$ for 
${}^\forall z\in {\mathbb C}$, 
whose elements are holomorphic with respect to $z$.  
Let us note that $\Omega$ 
is invariant under $U(N_{\rm C})$ gauge transformations, 
but is covariant under the $V$-transformations 
\begin{eqnarray}
 \Omega\to V \Omega V^\dagger 
  \label{eq:vtx:V-transOmega}
\end{eqnarray}
Incorporating all possible boundary conditions, 
we find that the total moduli space of BPS solitons 
${\cal M}^{\rm total}_{N_{\rm C},N_{\rm F}}$ is given by 
\begin{eqnarray} 
 {\cal M}^{\rm total}_{N_{\rm C},N_{\rm F}}&=&
 {\left\{H_0(z)|H_0(z)\in M_{N_{\rm C},N_{\rm F}}\right\}\over 
 \left\{V(z)|V(z)\in M_{N_{\rm C},N_{\rm C}}, 
 {\rm det}V(z)\not=0 \right\}}
  \label{eq:vtx:totalquotient}
\end{eqnarray}
where $M_{N,N'}$ denotes a set of holomorphic $N\times N'$ 
matrices \cite{Eto:2005yh,Eto:2006pg}.

The Lagrangian (\ref{eq:vtx:energy}) evaluated for the BPS 
solutions in Eq.(\ref{eq:vtx:solution}) can be rewritten in 
terms of the gauge invariant matrix $\Omega$ in 
Eq.(\ref{eq:vtx:omega}) as   
\begin{eqnarray}
{\cal L}|_{\rm BPS}&=&{\rm Tr}\left[-v^2\,B_3
+2i\partial _{[1}H{\cal D}_{2]}H^\dagger \right] 
\Big|_{\rm BPS}\nonumber\\
&=&2v^2\,{\bar \partial _z}{\partial _z}
\left(1-{4\over g^2c}{\bar \partial _z}{\partial _z}
\right)\log \det \Omega .
  \label{eq:vtx:energy2}
\end{eqnarray}
The last four-derivative term above does not contribute to the 
total energy if the configuration approaches to a vacuum on 
the boundary.
Eq.(\ref{eq:vtx:master}) implies the asymptotic behavior 
of the gauge invariant quantity $\Omega \to \frac{1}{v^2} H_0 H_0^\dagger$ 
at the boundary as $z \to \infty$.   
Therefore the topological charge $Q$ and the total energy $T$ 
of the BPS solitons are given in terms of the modul matrix 
$H_0$ as 
\begin{eqnarray}
 T|_{\rm BPS}=2\pi v^2\,Q 
=-{v^2\over 2}i\oint dz {\partial _z}\log\det(H_0H_0^\dagger )+{\rm c.c.} 
\label{eq:vtx:tension2}
\end{eqnarray}
It is important to recognize that the simple formulas 
in Eqs.(\ref{eq:vtx:energy2}) and (\ref{eq:vtx:tension2}) 
are valid only for BPS or anti-BPS solitons. 
We need to use the original definition of energy density 
(\ref{eq:mdl:total_lagrangian}) to obtain the energy of 
bions, since bions are non-BPS configurations as composites 
of BPS and anti-BPS solitons.

\subsection{Grassmann sigma model as a strong coupling limit 
}
\label{sc:mdl:grassmannModel}

Gauge theories reduce to nonlinear sigma models 
with target spaces as Grassmann manifolds in the strong 
gauge coupling limit $g^2 \to \infty$. 
When they are embedded into supersymmetric gauge theories with 
four (eight) supercharges, they become (hyper-)K\"ahler (HK) nonlinear sigma models~\cite{Zumino:1979et,Alvarez-Gaume:1981hm}
on the Higgs branch~\cite{Argyres:1996eh,Antoniadis:1996ra} 
of gauge theories as their target spaces. 
This construction of (hyper-)K\"ahler manifold is called 
a  (hyper-)K\"ahler quotient~\cite{Lindstrom:1983rt,Hitchin:1986ea}. 
In order to have finite energy configuration, it is necessary 
to be at the minimum of the potential leading to a constraint 
\begin{eqnarray}
H H^\dagger =v^2{\bf 1}_{N_{\rm C}}. 
 \label{eq:mdl:constraintHH}
\end{eqnarray}
Since the gauge kinetic terms for $W_\mu$ 
disappear in the limit of infinite 
coupling, gauge fields $W_\mu$ become auxiliary fields which can 
be expressed in terms of scalar fields $H$ through their 
field equations 
\begin{eqnarray}
W_\mu
= \frac{i}{2v^2} (
{\partial}_\mu H H^{\dagger}
-
H {\partial}_\mu H^{\dagger} )  
= \frac{i}{v^2} 
{\partial}_\mu H H^{\dagger}
.  
\label{eq:mdl:constr_gauge}
\end{eqnarray}
After eliminating $W_\mu$, the Lagrangian (\ref{eq:mdl:reduced-L}) 
with the constraints (\ref{eq:mdl:D-term-cond}) becomes 
a nonlinear sigma model, 
\begin{eqnarray}
\mathcal{L}_{\rm grassmann}
&=&{\rm Tr}\left[(\partial_\mu+iW_\mu) H 
((\partial_\mu+iW_\mu) H)^\dagger \right]
 \nonumber 
\\
\!\!\!\!\!\!&\!=\!&
{\rm Tr}\left[\left(\partial_\mu H
-\frac{1}{v^2}{\partial}_\mu H H^{\dagger}H \right) 
\left(\partial_\mu H^\dagger -\frac{1}{v^2}H^\dagger
H {\partial}_\mu H^{\dagger}\right) 
\right] , 
\label{eq:mdl:reduced-L}
\end{eqnarray}
with the complex 
Grassmann manifold $Gr_{N_{\rm F},N_{\rm C}}$ as a target space 
\begin{eqnarray}
Gr_{N_{\rm F},N_{\rm C}}
\simeq 
{SU(N_{\rm F}) \over 
 SU(N_{\rm C}) \times SU(N_{\rm F}-N_{\rm C}) \times U(1) 
}.
\label{eq:mdl:Gr}
\end{eqnarray}
This is the Grassmann sigma model which is the main focus of our study. 
Now one can see that the parameter $1/v$ serves as the coupling 
constant of the Grassmann sigma model. 
Let us also note that the ${\mathbb C}P^{N_{\rm F}-1}$ 
sigma 
model is obtained as a special case of the Grassmann sigma model: 
$Gr_{N_{\rm F},N_{\rm C}=1}
={\mathbb C}P^{N_{\rm F}-1}$. 
The topological charge $Q$ in Eq.(\ref{eq:vtx:topologicalCharge}) 
can also be expressed in terms of scalar fields $H$ as 
\begin{eqnarray}
Q
=\frac{i}{2\pi v^2}\int d^2 x {\rm Tr} 
\left(\epsilon_{\mu\nu}{\cal D}_\mu H 
({\cal D}_{\nu}H)^\dagger\right) 
=\frac{i}{2\pi v^2}\int d^2 x \epsilon_{\mu\nu}{\partial}_\mu{\rm Tr} 
\left( H 
{\partial}_{\nu}H^\dagger\right) . 
  \label{eq:vtx:topologicalChargeNlsm}
\end{eqnarray}
One should note the procedure leading to the Grassmann sigma 
model is unrelated to supersymmetry. 
Therefore constraints in Eqs.(\ref{eq:mdl:constraintHH}) and 
(\ref{eq:mdl:constr_gauge}) have to be obeyed irrespective of 
BPS or non-BPS field configurations.

Let us first consider (anti-)BPS solutions. 
For finite gauge coupling, analytic solutions of BPS equations 
are not possible. 
For Grassmann sigma model corresponding to the 
infinite gauge coupling, one of the BPS equations, the master 
equation (\ref{eq:vtx:master}) becomes identical to the constraint 
of the Grassmann sigma model in Eq.~(\ref{eq:mdl:constraintHH}) 
and can be easily solved algebraically 
\begin{eqnarray}
\Omega = v^{-2} H_0 H_0^\dagger .
\label{eq:mdl:nlsmConstr}
\end{eqnarray}
As described in Appendix \ref{app:constraint}, we can express 
the nonnegative hermitian matrix $\Omega$ in terms of a 
unitary matrix $U$ and a nonnegative diagonal matrix 
$\Omega_{\rm d}$ as 
\begin{equation}
\Omega = U \Omega_{\rm d} U^{\dagger}, 
\quad UU^\dagger ={\bf 1}_{N_{\rm C}}, 
 \label{eq:diagonal_omega}
\end{equation}
We can define the inverse square root 
$\Omega^{-1/2}=v(H_0H_0^\dagger)^{-1/2}=U\Omega_{\rm d}^{-1/2}U^{\dagger}$ 
and use it as a possible $U(N_{\rm C})$-gauge choice of $S^{-1}$ 
to obtain the physical scalar field $H$ as 
\begin{equation}
H
=  U \Omega_{\rm d}^{-1/2} U^{\dagger} H_0, 
 \label{eq:H_formula}
\end{equation}
which satisfies the constraint of Grassmann sigma model 
 $HH^\dagger=v^2{\bf 1}_{N_{\rm C}}$ in Eq.~(\ref{eq:mdl:constraintHH}), 
as shown in Appendix \ref{app:constraint}. 

Now let us consider general (non-BPS) field configurations of 
Grassmann sigma model aiming at bion configurations. 
Any field configurations $H$ in Grassmann sigma model should 
satisfy the constraint in Eq.(\ref{eq:mdl:constraintHH}). 
From experiences in ${\mathbb C}P^{N_{\rm F}-1}$ model 
\cite{Dabrowski:2013kba, Misumi:2014jua}, 
we know that bion configurations need not be a solution of 
field equations. 
On the other hand, we wish to consider bion field configurations 
to become solutions of field equations asymptotically when 
constituent fractional instantons are far apart. 
Therefore our strategy is the following: 
we consider non-holomorphic (functions of both $z, \bar z$) 
moduli matrices $H_0$ corresponding to composites of fractional 
instantons and anti-instantons, which become non-BPS exact solutions 
asymptotically as separation goes to infinity. 
The only additional condition to satisfy is the constraint of 
Grassmann sigma model in Eq.(\ref{eq:mdl:constraintHH}). 
This is achieved by the formula in Eq.~(\ref{eq:H_formula}). 
It is important to realize that this formula can be regarded 
merely as a solution of the constraint 
$HH^\dagger=v^2{\bf 1}_{N_{\rm C}}$, without 
any reference to the BPS condition.

To obtain the most general $S$ for any gauge choice of $U(N_{\rm C})$ 
gauge invariance, we have a freedom of using another unitary 
matrix $\tilde U\in U(N_{\rm C})$ as 
\begin{equation}
S = \tilde U \Omega^{1/2}, 
\quad \tilde U \tilde U^\dagger={\bf 1}_{N_{\rm C}}. 
 \label{eq:general_S}
\end{equation}
This unitary matrix $\tilde U$ is precisely the freedom of 
the $U(N_{\rm C})$ gauge transformations of the underlying 
gauge theory.

From the target manifold (\ref{eq:mdl:Gr}) one can easily 
see that there exists a Seiberg-like duality 
between theories with the same number of flavors and 
with two different gauge groups 
in the case of the infinite gauge 
coupling~\cite{Argyres:1996eh,Antoniadis:1996ra}: 
\begin{eqnarray}
 U(N_{\rm C}) \leftrightarrow 
 U(N_{\rm F}-N_{\rm C}) .
\label{eq:mdl:duality}
\end{eqnarray}
This duality is exact in the strong coupling limit of the 
gauge theory, namely it holds for the entire Lagrangian of the 
Grassmann sigma models. 
One should note that 
for each BPS solution of the $U(N_{\rm C})$ gauge 
theory with $N_{\rm F}$ flavors of scalar fields $H$ in the 
fundamental representations, there exists a corresponding 
anti-BPS solution of the $U(N_{\rm F}-N_{\rm C})$ gauge 
theory with the same number of flavors of scalar fields 
$\tilde H$ in the fundamental representations and vice versa. 
Besides the Grassmann sigma model constraint for $H$ 
\begin{eqnarray} 
 HH{}^\dagger =v^2{\bf 1}_{N_{\rm C}},
\end{eqnarray} 
and for $\tilde H$ at strong coupling 
\begin{eqnarray}
 \tilde H\tilde H{}^\dagger =v^2{\bf 1}_{N_{\rm F}-N_{\rm C}},
\end{eqnarray} 
these solutions must satisfy the following orthogonality constraints 
\begin{eqnarray}
 H \tilde H{}^\dagger =0.
\label{orthogonal}
\end{eqnarray}
The boundary conditions for the BPS solution $H$ and the 
anti-BPS solution $\tilde H$ should be chosen to be associated 
with complementary vacua~\cite{Isozumi:2004jc,Ohta:2013tma}.

Duality between $U(N_{\rm C})$ gauge theories and $U(N_{\rm F}-N_{\rm C})$ 
gauge theories can be formulated in terms of the corresponding moduli 
matrices $H_0$ and $\tilde H_0$ as 
\begin{eqnarray}
 H_0 \tilde H_0^\dagger = 0 
 \label{eq:wll:duality}
\end{eqnarray}
Together with the complementary boundary conditions, 
this relation determines $\tilde H_0$ uniquely 
from $H_0$ up to the $V$-equivalence (\ref{eq:vtx:V-trans}).
Although this duality is not exact for finite coupling 
there still exists a one-to-one dual map 
by the relation 
among the moduli matrix $H_0$ in the original gauge theory 
and the $(N_{\rm F} - N_{\rm C}) \times N_{\rm F}$ 
moduli matrix $\tilde H_0$ of the dual gauge theory.


\subsection{
${\mathbb Z}_{N_{\rm F}}$ twisted boundary conditions 
and fractional instantons 
}
\label{sec:ZN}

In the  present subsection, we introduce a 
${\mathbb Z}_{N_{\rm F}}$ twisted boundary condition 
in the $U(N_{\rm C})$ gauge theory with $N_{\rm F}$ flavors or 
the Grassmann sigma model as its strong coupling limit 
on ${\mathbb R}^1 \times S^1$. 
The ${\mathbb Z}_{N_{\rm F}}$ twisted boundary conditions in 
a compactified direction is expressed in terms of a twisting 
matrix $B$ as
\cite{Dunne:2012zk, Dabrowski:2013kba}
\begin{equation}
H(x^{1}, x^{2}+L) = B
\,H(x^{1},x^{2})\,,
\,\,\,\,\,\,\,\,\,\,\,\,\,
B
={\rm diag.}\left[1, e^{2\pi i/N_{\rm F}}, e^{4\pi i/N_{\rm F}},
\cdot\cdot\cdot, 
e^{2(N_{\rm F}-1)\pi i/N_{\rm F}}  \right]\,.
\label{ZNC}
\end{equation}
The ${\mathbb Z}_{N_{\rm F}}$ twisted boundary condition 
breaks the global $SU(N_{\rm F})$ symmetry down to 
${\mathbb Z}_{N_{\rm F}}$. 
Fractional instantons (kink instantons) carry integer 
multiple of the minimum topological charge $1/N_{\rm F}$ 
in the Grassmann sigma models on ${\mathbb R}^1 \times S^1$ with a 
${\mathbb Z}_{N_{\rm F}}$ twisted boundary condition 
\cite{Eto:2004rz,Eto:2006mz}.

When $x^2$ is compactified with the period $L$, the lowest mass of 
Kaluza-Klein modes is $2\pi/L$. 
In the case of ${\mathbb Z}_{N_{\rm F}}$ twisted boundary condition, 
the lowest mass is also fractionalized to give $2\pi/(N_{\rm F}L)$. 
The fractional instanton is in one-to-one correspondence with 
the kink \cite{Eto:2004rz,Eto:2006mz} as a function of $x^1$. 
The study of BPS equations for kinks in the strong coupling limit 
reveals that the size of the fractional instanton is 
given by the inverse of the mass difference associated to the 
adjacent vacua \cite{Shifman:2002jm,Eto:2006pg}. 
Therefore the size of elementary fractional instanton with the 
instanton charge $1/N_{\rm F}$ is given by $N_{\rm F}L/(2\pi)$. 
When two fractional instantons are compressed together, they 
can form a compressed fractional instanton, whose size should 
be a half of the individual fractional instantons. 
By the same token, $n$ fractional instanton can be compressed 
together to form a compressed $n$-fractional-instantons whose 
size should be $N_{\rm F}L/(2\pi n)$.

From next subsection we make all the dimensionful quantities 
and parameters dimensionless by using the compact scale $L$ 
($L\to 1$) unless we have a special reason to recover it.


\section{D-brane configurations for fractional 
instantons}\label{sec:D-brane}

\subsection{D-brane configurations
\label{subsec:dbrane}}
\begin{figure}[h]
 \begin{center}
  \includegraphics[width= 60mm]{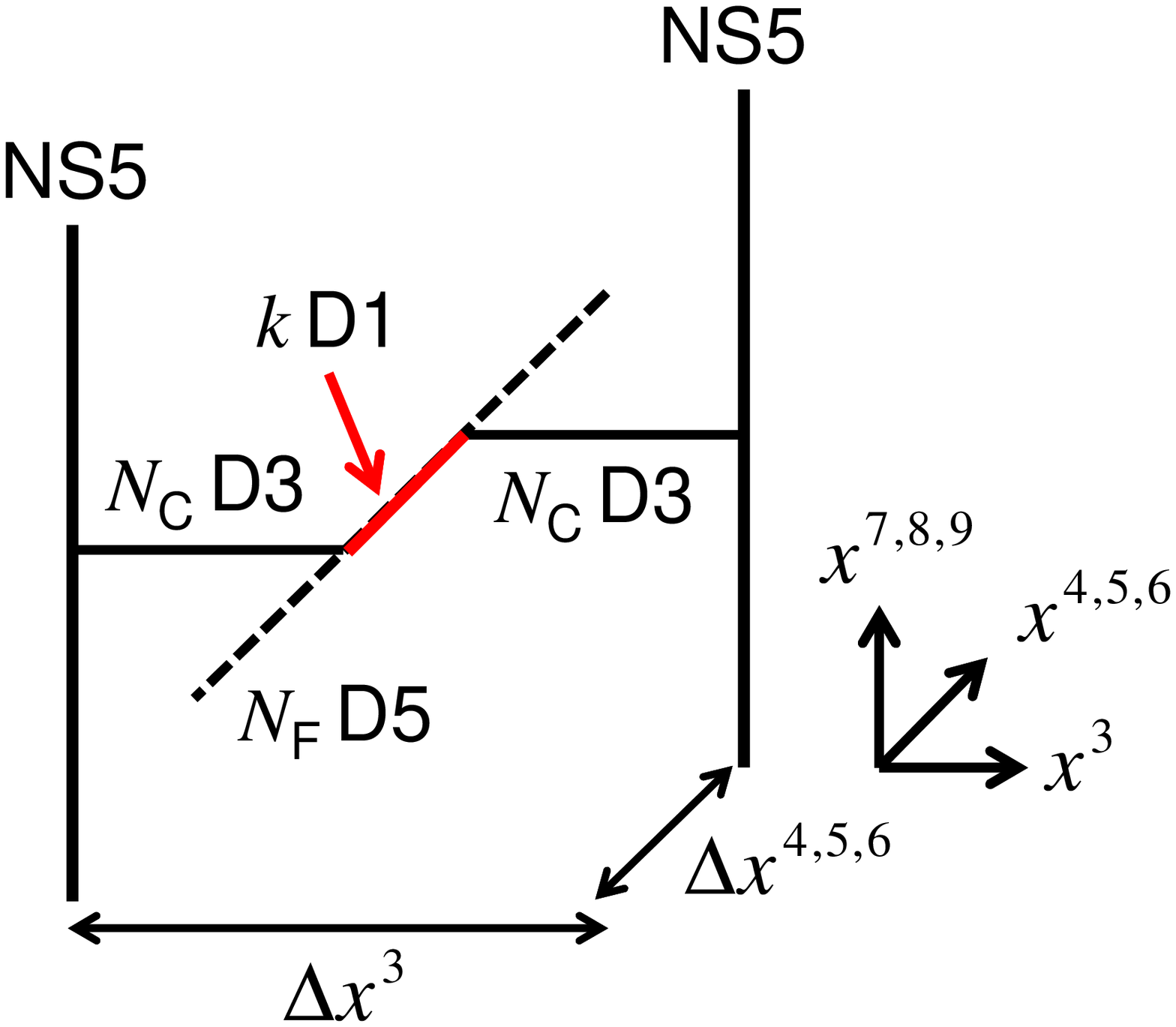}
  \includegraphics[width= 70mm]{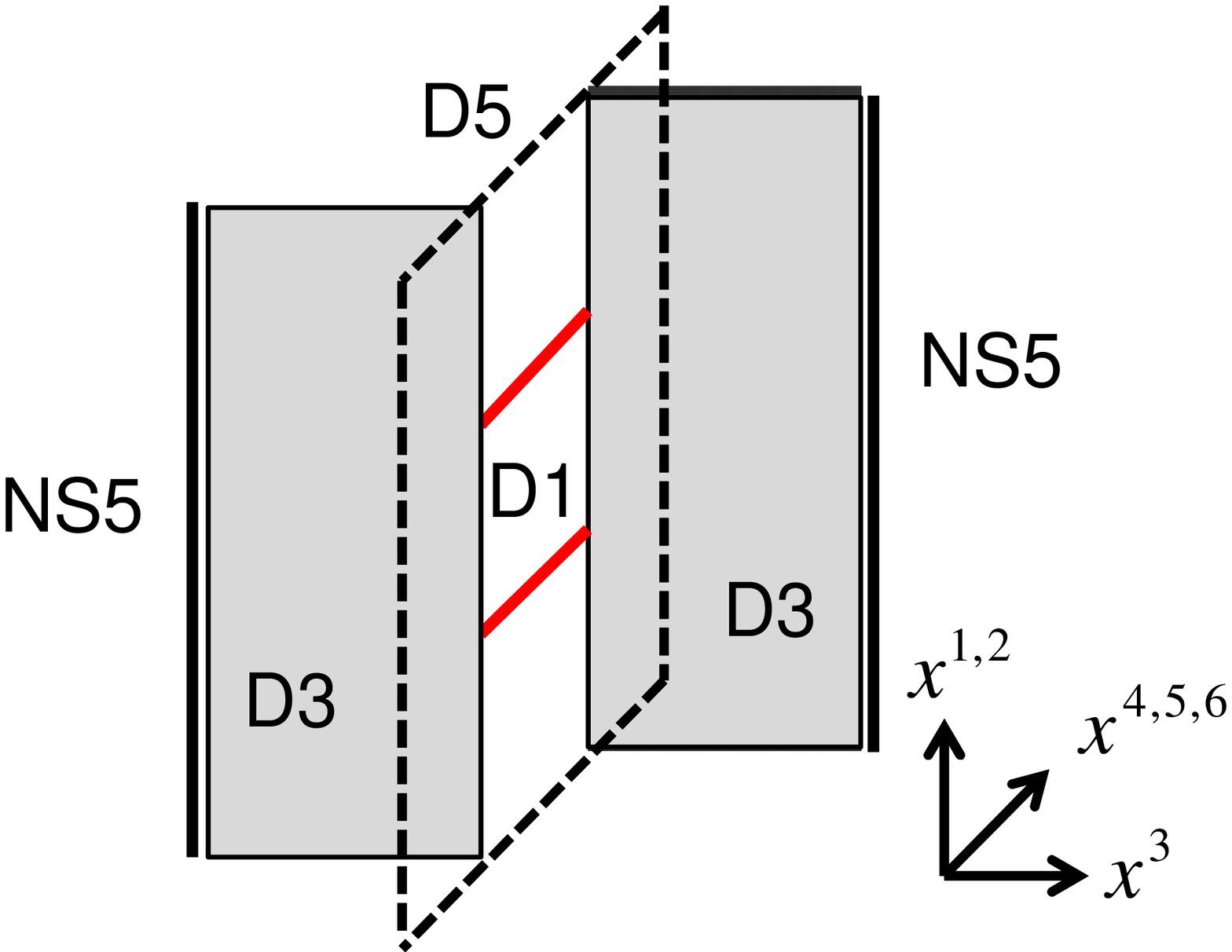}
 \end{center}
 \caption{ Brane configuration for $k$ vortices. 
As for separation of the two NS5-branes, 
$\Delta x^3$ corresponds to $1/g^2$ 
and $(\Delta x^4,\Delta x^5,\Delta x^6)$ 
correspond to the triplet of the FI parameters.}
 \label{fig:vortices}
\end{figure}

The gauge theory introduced in the last section 
can be made ${\cal N}=2$ supersymmetric (with eight supercharges)  
by doubling the Higgs scalar fields $H$ 
and adding fermionic superpartners (Higgsino and gaugino) and adjoint 
scalars (dimensionally reduced gauge fields) \cite{Eto:2006pg}.
Then, the theory can be realized by a D-brane configuration 
\cite{Eto:2004vy}.
We first consider the Hanany-Witten brane configuration 
\cite{Hanany:1996ie,Giveon:1998sr}.
We are interested in euclidean space ${\mathbb R} \times S^1$, 
but we consider a brane configuration in 2+1 dimensions 
by adding ``time" direction. 
In Table \ref{tab:vortices}, we summarize 
the directions in which the D-branes extend. 
In Fig.~\ref{fig:vortices} the brane configuration is 
schematically drawn. 
\begin{table}[h]
\begin{center}
\begin{tabular}{!{\tvline}c!{\tvline}cccccccccc!{\tvline}}
\thline
&$x^0$&$x^1$&$x^2$&$x^3$&$x^4$&$x^5$&$x^6$&$x^7$&$x^8$&$x^9$\\
\thline
$\NC$ $\rm D3$&$\circ$&$\circ$&$\circ$&$\circ$&$-$&$-$&$-$&$-$&$-$&$-$ \\
\hline
$\NF$ $\rm D5$&$\circ$&$\circ$&$\circ$&$-$&$\circ$&$\circ$&$\circ$&$-$&$-$&$-$ \\
\hline 
$2\hs{2}{\rm NS}5$&$\circ$&$\circ$&$\circ$&$-$&$-$&$-$&$-$&$\circ$&$\circ$&$\circ$ \\
\hline
$k\hs{2}\rm D1$&$\circ$&$\times$&$\times$&$-$&$\circ$&$-$&$-$&$-$&$-$&$-$ \\
\thline
\end{tabular}
\end{center}
\caption{ Brane configuration for $k$ vortices: Branes are extended along directions denoted by $\circ$, and are not extended
along directions denoted by $-$. The symbol $\times$ denotes the
codimensions of the $k$ D1-branes on the worldvolume of the D3-branes
excluding the $x^3$ which is a finite line segment.}
\label{tab:vortices}
\end{table}
The $U(\NC)$ gauge theory is realized on the $\NC$ 
coincident D3-brane world-volume 
which are 
stretched between two ${\rm NS}5$ branes. 
The $\rm D3$ brane world-volume 
have the finite length $\Delta x^3$ between 
two ${\rm NS}5$ branes,  and therefore 
the $\rm D3$ brane world-volume theory is
$(2+1)$-dimensional 
$U(\NC)$ gauge theory with a gauge coupling 
$\frac{1}{g^2} = |\Delta x^3| \tau_3 l_s^4 
= \frac{|\Delta x^3|}{g^{(B)}_s}$, with  
the string coupling constant $g^{(B)}_s$ in type IIB string theory
and the D3-brane tension  $\tau_3 = 1/(g^{(B)}_s l_s^4)$.
The positions of the $\NF$ $\rm D5$ branes 
in the $x^7$-, $x^8$- and $x^9$-directions coincide 
with those of the $\rm D3$ branes. 
Strings which connect between D3 and D5 branes give rise to 
the $\NF$ hypermultiplets (the Higgs fields $H$ and Higgsinos) 
in the D3 brane worldvolume theory.  
The two ${\rm NS}5$ branes are separated into the 
$x^4$-, $x^5$- and $x^6$-directions, 
which give the triplet of the FI parameters $c^a$ 
\cite{Hanany:2003hp,Hanany:2004ea}.  
We choose it as $c^a = (0,0,v^2=\Delta x^4/(g^{(B)}_s l_s^2) >0)$, 
with the string length $l_s$. 

Now, we consider BPS non-Abelian vortices in this setup~\cite{Hanany:2003hp,Hanany:2004ea}. 
$k$ vortices are represented by $k$ D1-branes 
stretched between D3-branes from the following reasons.
(1) The D1-branes preserve half supersymmetry 
of  the D3-brane world-volume theory.  
(2) The endpoints of the D1-branes are of codimension two
in D3-brane world-volume, 
denoted by the symbol $\times$ in Table \ref{tab:vortices},
since the $x^3$ direction of the $\rm D3$ brane worldvolume 
is finite between the two NS5-branes. 
(3) The energy 
$\tau_1 \Delta x^4=\frac{\Delta x^4}{g^{(B)}_s l_s^2}=v^2$ 
of each $\rm D1$ brane 
coincides with that of a vortex. 
Therefore, one concludes that 
the $k \hs{2} \rm D1$ branes correspond to $k$ 
vortices in the $\rm D3$ brane world-volume theory.  
When $\NF=\NC$, the vortices are called local vortices, 
while for$\NF>\NC$ the vortices are called semi-local vortices. 

Next, we compactify the $x^2$-direction on $S^1$ with the 
period $L$ for our purpose. 
First, we turn on a constant background gauge field  
\beq 
  A_2 = {\rm diag} (m_1,\cdots,m_{\NF})  \label{eq:Wilson}
\eeq
on the $\rm D5$ brane worldvolume 
as a non-trivial Wilson loop around $S^1$ on the $\rm D5$ branes. 
This precisely gives a twisted boundary condition in our context. 
In this paper, we consider the ${\mathbb Z}_N$ symmetric 
twisted boundary condition corresponding to $m_n=2\pi n/(LN_{\rm F}), 
n=1,\cdots,N_{\rm F}$. 
We then take T-duality 
along the $x^2$ direction. 
Table \ref{tab:walls} shows 
the directions in which the branes extend
after T-duality transformation. 
\begin{table}[h]
\begin{center}
\begin{tabular}{!{\tvline}c!{\tvline}cccccccccc!{\tvline}}
\thline
&$x^0$&$x^1$&$x^2$&$x^3$&$x^4$&$x^5$&$x^6$&$x^7$&$x^8$&$x^9$\\
\thline
$\NC$ $\rm D2$&$\circ$&$\circ$&$-$&$\circ$&$-$&$-$&$-$&$-$&$-$&$-$ \\
\hline
$\NF$ $\rm D4$&$\circ$&$\circ$&$-$&$-$&$\circ$&$\circ$&$\circ$&$-$&$-$&$-$ \\
\hline 
$2\hs{2}{\rm NS}5$&$\circ$&$\circ$&$\circ$&$-$&$-$&$-$
&$-$&$\circ$&$\circ$&$\circ$ \\
\hline
$k\hs{2}\rm D2'$&$\circ$&$\times$&$\circ$&$-$&$\circ$&$-$&$-$&$-$&$-$&$-$ \\
\hline
$k\hs{2}\rm D2^{\ast}$&$\circ$&$\times$&$\circ$&$\circ$
&$-$&$-$&$-$&$-$&$-$&$-$ \\
\thline
\end{tabular}
\end{center}
\caption{T-dualized configuration: Branes are extended along 
directions denoted by $\circ$, and are not extended
along directions denoted by $-$. The symbol $\times$ denotes the
codimensions of the k D2'-branes on the worldvolume of the D2-branes,
excluding the $x^3$ which is a finite line segment.
\label{tab:walls}
}
\end{table}
With this Wilson loop in Eq.~(\ref{eq:Wilson}), 
the positions of the $\rm D4$ branes 
are split into the $x^2$-direction by an amount 
\beq
 X_{A}=2 \pi l_s^2 m_A \hs{2} (A = 1,\cdots,\NF).
\eeq
These separations give hypermultiplet masses in the 
$\rm D2$ brane worldvolume theory. 
The worldvolume theory of the $\rm D2$ branes is 
$(1+1)$-dimensional gauge theory with a gauge coupling 
$\frac{1}{\hat g^2} = \frac{\Delta x^3 l_s}{g^{(A)}_s} = 
\frac{\Delta x^3 R}{g^{(B)}_s}$
and the FI parameter
$\hat c = \frac{\Delta x^4}{g^{(A)}_s l_s} = \frac{\Delta x^4 R}{g^{(B)}_s l^2_s} $, where $g^{(A)}_s$ is the string coupling constant in type IIA string theory.

\begin{figure}[h]
\begin{center}
\includegraphics[width= 60mm]{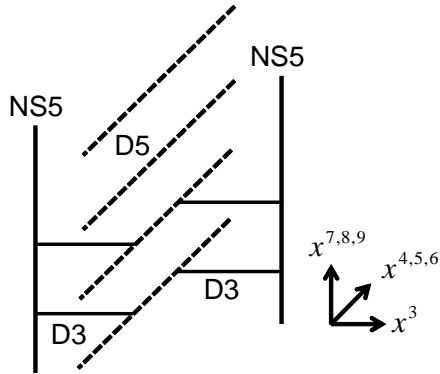}
\end{center}
\caption{vacuum configuration $(k=0)$}
\label{fig:k=0}
\end{figure}
First, let us consider vacua without vortices ($k=0$). 
As shown in Fig.~\ref{fig:k=0}, each $\rm D2$ 
brane ends on one of the $\rm D4$ branes, 
on each of which 
at most one $\rm D2$ brane can end,  
which is known as the s-rule~\cite{Hanany:1996ie}. 
There are ${}_{\NF}C_{\NC}=\NF !/\NC !(\NF-\NC) !$ vacua in 
the Grassmann sigma model 
\cite{Arai:2003tc}.

\begin{figure}[h]
\begin{center}
\includegraphics[width=140mm]{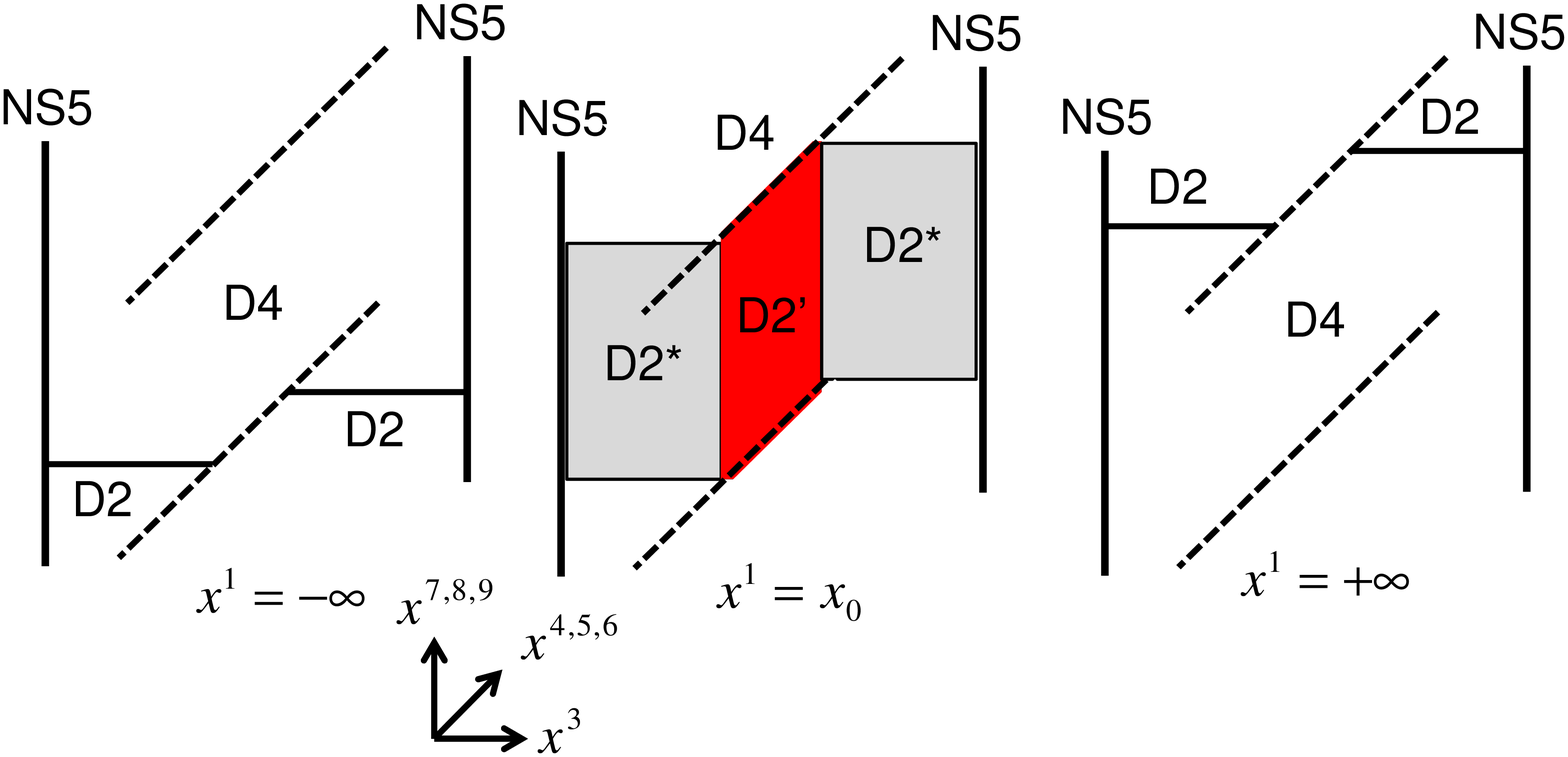}
\end{center}
\caption{ T-dualized configurations for $k=1$: 
the $\rm D2'$ brane is assumed to be located at $x^1=x_{0}$.
\label{fig:k=1}
}
 \begin{center}
  \includegraphics[width= 70mm]{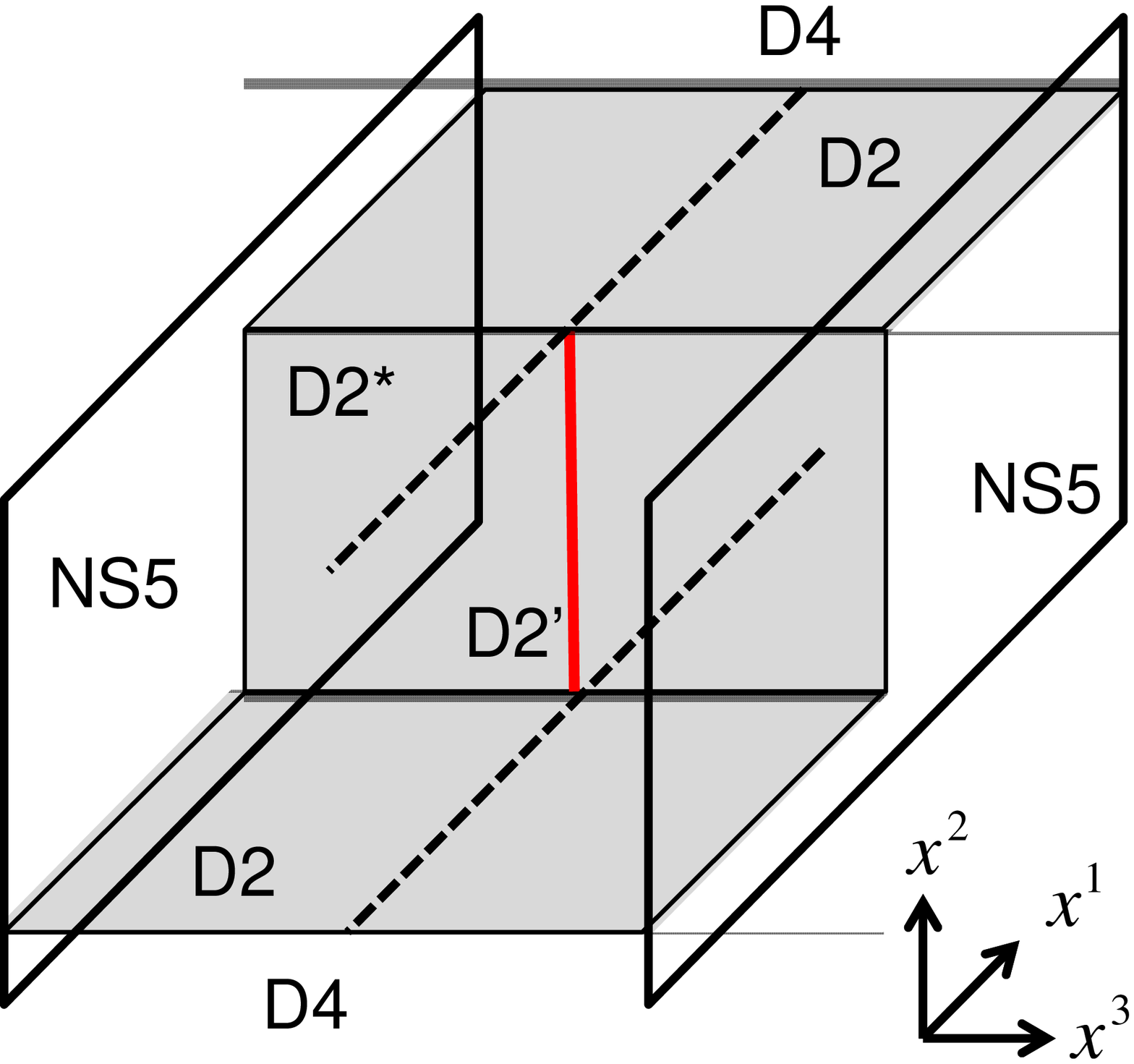}
 \end{center}
 \caption{Brane configuration for a wall. 
\label{fig:wall} }
\end{figure}

Let us consider vortices ($k \not = 0$). 
The $\rm D1$ branes representing vortices 
are mapped by the T-duality to 
$\rm D2$ branes, which we denote as $\rm D2'$, 
 stretched between the $\rm D4$ branes, 
as shown in the middle figure in Fig.~\ref{fig:k=1}, 
where the position of the D2$'$ brane in 
the $x^1$ coordinate is denoted as $x_{0}$.
The $\rm D2$ branes are attached to different 
$\rm D4$ branes at $x^1=-\infty$ and $x^1=+\infty$, 
and there 
must exist $\rm D2$ branes which connect the $\rm D2$ 
branes ending on different $\rm D4$ branes 
at some point in the $x^1$-coordinate. 
These $\rm D2$ branes correspond to $\rm D2'$ in 
Fig.~\ref{fig:k=1}. 
Since the D2$'$ branes do not end in the $x^1$-direction,
they must be bent to the $x^3$-direction to end on 
the NS5-branes.
We denote these D2-branes by $\rm D2^\ast$ 
in Fig.~\ref{fig:k=1}. 
In Fig.~\ref{fig:wall}, the brane configuration 
in the $x^1$, $x^2$, $x^3$-coordinates is shown. 
This is nothing but 
a brane configuration \cite{Eto:2004vy}
of a BPS kink (domain wall) \cite{Isozumi:2004jc} 
in the $\rm D2$ brane theory.
The energy of the kinks  can be calculated 
from this brane configuration in the strong coupling limit. 
Since the gauge coupling $\frac{1}{\hat g^2}$ is proportional to $\Delta x^3$, 
the $\rm D2^{\ast}$ branes disappear in the limit 
$\hat g \rightarrow \infty$. 
The $\rm D2'$ branes have the energy $\tau_2 \hs{1} \Delta x^4  
\hs{1} l_s^2 \Delta m = \frac{\Delta x^4 \Delta m}{g^{(A)}_s l_s}
=\hat c \hs{1} \Delta m$ coinciding with the energy of a kink. 
A set of D2$'$+D2$^{\ast}$-branes between two D4-branes  
 is a kink as a fractional vortex (or a lump), 
which corresponds to a fractional instanton in euclidean 
${\mathbb R} \times S^1$ space 
in our context.

\begin{figure}[h]
\begin{center}
\includegraphics[width=80mm]{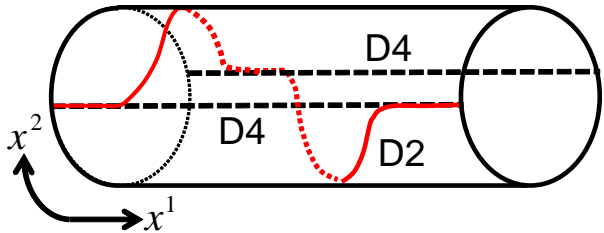} 
%
\end{center}
\caption{ T-dual picture for $\NF=2, \NC=1, k=1$.
 \label{fig:tk=1}}
\end{figure}

The unit vortex corresponds to the \rm D2 brane winding 
around the $S^1$ of the cylinder 
with exhibiting a kink  
as in Fig.~\ref{fig:tk=1} (a). 
The size of the kink in the $x^1$-direction 
is that $1/g\sqrt c$ of  
an Abrikosov-Nielsen-Olesen (ANO) vortex.  
Note that the scalar field $\hat \Sigma(x^1)$ has period $1/R$.
This vortex can be decomposed into two walls 
by changing the size of the vortex.
In this configuration, the \rm D2 brane is attached to the same \rm D4 brane at $x^1 \rightarrow \pm \infty$ 
with exhibiting kinks twice as in Fig.~\ref{fig:tk=1} (b).
The relative distance between the two kinks 
can be interpreted as 
the size moduli of the single semi-local vortex (or lumps). 
The small size limit 
of the configuration reduces to the ANO vortex with 
the ANO size $1/(gv)$. 
In other words, the small lump singularity in the strong 
coupling limit $g \to \infty$ 
is resolved by the size of the ANO vortex for finite $g$. 
By using the brane picture presented here, 
the moduli space of multiple non-Abelian vortices 
was classified in Ref.~\cite{Eto:2006mz}.
In this paper, we use this kinky D-brane picture to 
classify all possible bion configurations. 
We can visualize the kink exhibited in Fig.~\ref{fig:tk=1} 
by using the Wilson loop around the $S^1$ along $x^2$ 
\cite{Eto:2006mz}  
\begin{equation}
\Sigma (x^1) = -\frac{1}{L}\log\Bigl[ \mathbf{P} \ 
{\rm exp}\int_0^{L}dx^2 W_2 (x_1, x_2) \Bigl],
\label{eq:defsigma}
\end{equation} 
where $\mathbf{P}$ is the path-ordering and the gauge field $W_2$ 
is given in Eq.~(\ref{eq:mdl:constr_gauge}) in the Grassmann 
sigma model. Since this is a matrix, the intuitive meaning of the 
brane picture can be best visualized when the matrix $H_0H_0^\dagger$ 
and $\Sigma$ are nearly diagonal.

Before doing that, we make a comment on a brane picture of 
the Seiberg-like duality, which exchanges the gauge group as 
$U(N_{\rm C}) \leftrightarrow U(N_{\rm F}-N_{\rm C})$ 
in Eq.~(\ref{eq:mdl:duality}).
In the Hanany-Witten setup, 
this is achieved by the exchange of the positions in 
$x^3$ of the two NS5-branes. 
When an NS5-brane passes through a D5-brane,  
a D3-branes is created (annihilated) 
if it is (not) stretched between 
the NS5-brane and the D5-brane before the crossing, 
due to the Hanany-Witten effect 
\cite{Hanany:1996ie,Giveon:1998sr}.
This changes the number of the D3-branes 
from $N_{\rm C}$ to $N_{\rm F}-N_{\rm C}$.
This exchange flips the sign of the FI parameters 
$c^a \leftrightarrow - c^a$. 
The Seiberg dual in the presence of vortices 
was studied in Ref.~\cite{Eto:2007yv}.
\begin{figure}[htbp]
\begin{center}
 \includegraphics[width=0.8\textwidth]{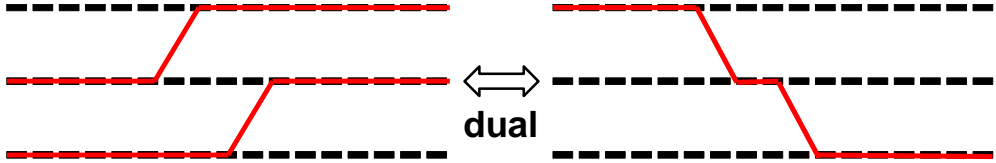}
\end{center}
\caption{Brane picture of the Seiberg dual for 
kinks, 
where the presence and absence of color branes
is exchanged. 
  Here we give an example of the duality between 
$Gr_{3,2}$ and 
$Gr_{3,1} \simeq {\mathbb C}P^2$.
The horizontal and vertical directions are 
$x_1$ and $x_2$, respectively. 
}
\label{fig:Seiberg-dual}
\end{figure}
In the T-dual configuration, the presence and absence 
of D2-branes on D4-branes are exchanged 
for vacuum configurations.
This exchange must holds in the presence 
of kinky D2-branes 
since the exchange occurs at every $x^1$ 
\cite{Eto:2004vy}. 
This is illustrated in Fig.~\ref{fig:Seiberg-dual}. 
In this dual transformation, 
the positions and number of the kinks are unchanged. 
While the shapes of the kinks 
would be different for finite gauge couplings, 
they coincide at strong gauge couplings,  
where,  in the original picture, the positions of the NS5-branes 
in $x^3$ coincide with those of the D5-branes.

Now let us summerize how to construct field configurations 
with the ${\mathbb Z}_{N_{\rm F}}$ twisted boundary condition 
in Grssmann sigma model. 
\begin{enumerate}
\item
Because of the color-flavor locking and ${\mathbb Z}_{N_{\rm F}}$ 
twisted boundary condition, we need to consider moduli matrix 
$H_0$ with each row representing each color line 
exactly similar to the case of 
the ${\mathbb C}P^{N_{\rm F}-1}$ 
model. 
\item
The $V$-equivalence allows us to multiply any complex 
number for each row of $H_0$. 
This complex number must be a (anti-)holomorphic function if 
we wish to construct a (anti-)BPS solution. 
\item
The (anti-)BPS solution should have $H_0$ with 
monotonically increasing (decreasing) color lines. 
\item
Because of s-rule~\cite{Hanany:1996ie}, no color line 
occupies the same flavor line in any region of $x^1$. 
Consequently no crossing is allowed for color lines  except 
at isolated points (see reconnection phenomena in later sections). 
\end{enumerate}

\subsection{Fractional instantons in 
the ${\mathbb C}P^{N_{\rm F}-1}$ model ($N_{\rm C}=1$)}

In terms of the complex coordinate $z = x_{1}+ix_{2}$ on 
${\mathbb R}^{1}\times S^{1}$ with $0\le x_2 < L=1$, the 
fractional instantons for the ${\mathbb C}P^1$ model 
satisfying the ${\mathbb Z}_2$ twisted boundary condition 
can be parameterized by the following moduli matrices with 
real moduli parameters $\lambda_L, \lambda_R > 0$ and 
$\theta_L, \theta_R$ 
\begin{eqnarray}
&& H_{0L}
 = \left(\lambda_L e^{i\theta_L}e^{-\pi z}, 
1 \right), \quad
 H_{0R}
 = \left(\lambda_R e^{i\theta_R}e^{\pi z}, 
1 \right), \nonumber \\
&& 
H_{0L}^*
 = \left(\lambda_L e^{-i\theta_L}e^{-\pi \bar z}, 
1 \right), \quad
 H_{0R}^*
 = \left(\lambda_R e^{-i\theta_R}e^{\pi \bar z}, 
1 \right). 
 \label{eq:fractional}
\end{eqnarray}
One should note that the twisted boundary condition automatically 
introduces nontrivial $x^1$ dependence in $H_0$. 
We call these as the elementary fractional instantons.

\begin{figure}[htbp]
\begin{center}
\begin{tabular}{cc}
 \includegraphics[width=0.4\textwidth]{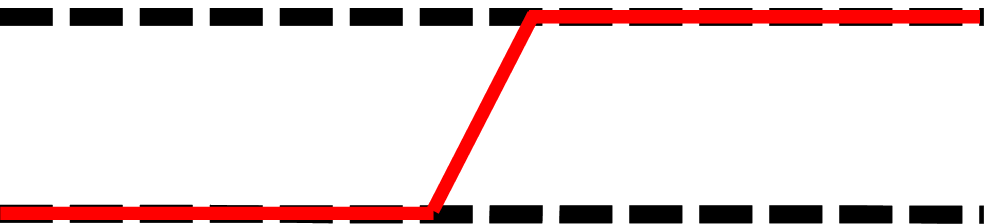}
& \includegraphics[width=0.4\textwidth]{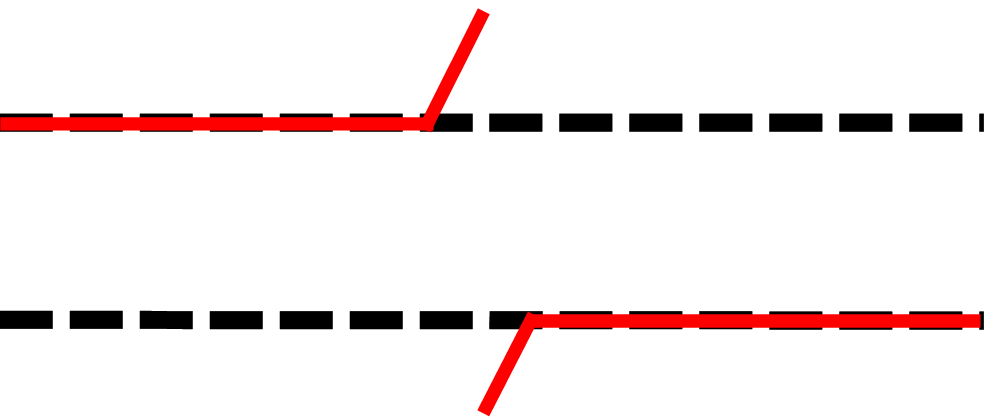}\\
(a) $H_{0L}$ & (b) $H_{0R}$\\
 \includegraphics[width=0.4\textwidth]{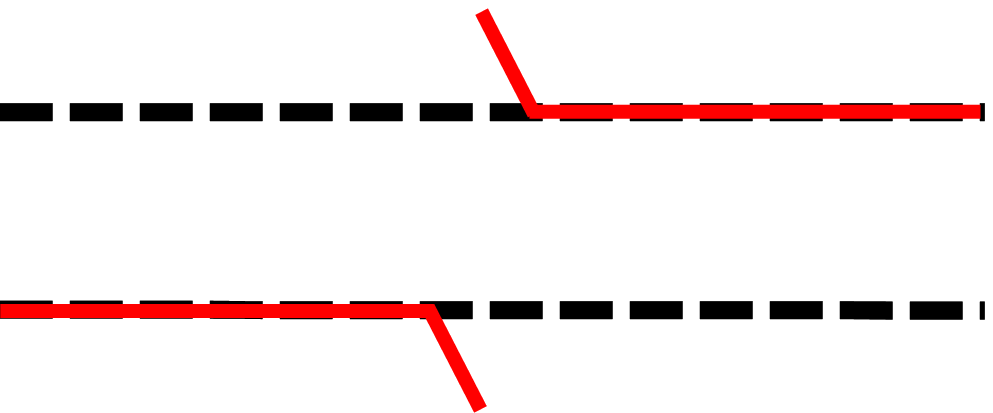}
& \includegraphics[width=0.4\textwidth]{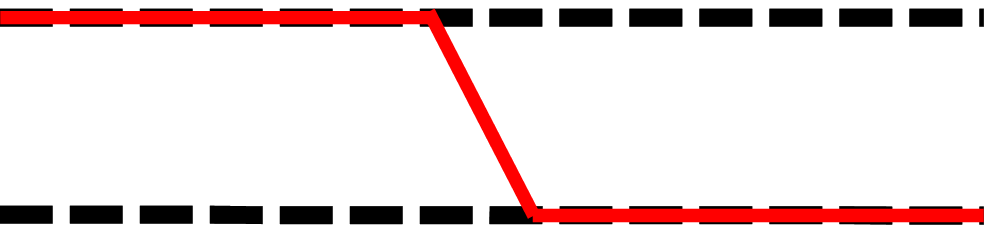}\\
(c) $H_{0L}^*$ & (d) $H_{0R}^*$
\end{tabular}
\end{center}
\caption{
Brane configurations of fractional instantons in 
the ${\mathbb C}P^1$ model with the ${\mathbb Z}_2$ twisted 
boundary condition, 
corresponding to (a) $H_{0L}$, (b) $H_{0R}$, 
(c) $H_{0L}^*$, and (d) $H_{0R}^*$. 
The horizontal and vertical directions are 
$x_1$ and $x_2$, respectively. 
The instanton charges $Q$ are  (a) $+1/2$, (b) $+1/2$, 
(c) $-1/2$, (d) $-1/2$, respectively.
}
\label{fig:fractional-config}
\end{figure}

For sufficiently far left in $x^1\to-\infty$, $H_{0L}$ 
(consequently the physical field $H_{L}$ also) is dominated 
by the first component, namely the configuration is in the 
first vacuum $H_{L}\sim(ve^{i\theta_1},0)$. 
In contrast, $H_{0L}$ is dominated by the second component at 
far right, and $H_{L}$ is in the second vacuum 
$H_{L}\sim (0,ve^{i\theta_2})$. 
The moduli parameter $\theta_{L}$ represents the relative 
phase $\theta_{L}=\theta_1-\theta_2$ between adjacent vacua. 
In terms of the kinky brane picture, the location of the kink 
can be defined as the point in $x^1$ where two components have 
the same magnitude: 
\begin{eqnarray}
\lambda e^{-\pi x^1}= 1 \quad \to \quad 
x^1= \frac{1}{\pi}\log\frac{1}{\lambda}. 
 \label{eq:position_instanton}
\end{eqnarray}
From the T-duality, this is precisely the location of the BPS 
fractional instanton. 
Therefore,  
the fractional (anti-)instanton $H_{0L}$ $(H_{0L}^*)$ is 
situated at $x_1 = \frac{1}{\pi}\log\lambda_L$, and the 
fractional (anti-)instanton 
$H_{0R}$ $(H_{0R}^*)$ is at $x_1 = \frac{1}{\pi}\log\frac{1}{\lambda_R}$. 
Since the moduli matrices $H_{0L}$ and $H_{0R}$ are holomorphic 
(depend on $z$ only), they give BPS solutions with instanton 
charge $Q=+1/2$. 
These two BPS solutions are distinguished only by the label of 
flavor brane on which the color brane is residing, and are of 
the same type. 
On the other hand, $H_{0L}^*$ and $H_{0R}^*$ are anti-holomorphic 
(depend on $\bar z$ only) and give anti-BPS solutions with instanton 
charge $Q=-1/2$. 
All of them are (anti-)BPS ${\mathbb C}P^1$ kinks 
\cite{Abraham:1992vb,Arai:2002xa}. 
The brane configurations of these solutions are shown in 
Fig.~\ref{fig:fractional-config}.

In the case of the ${\mathbb C}P^{N_{\rm F}-1}$ model, we have 
several kinds of elementary fractoinal instantons, connecting 
different adjacent flavors as ${\mathbb C}P^{N-1}$ kinks 
\cite{Gauntlett:2000ib}. 
There can exist elementary and composite fractional instantons.
There are $N_{\rm F}$ different species of elementary 
fractional instantons: one connecting $n$-th flavor and 
$(n+1)$-th flavor with $1 \le n \le N_{\rm F}$ 
($N_{\rm F}+1$ identified with $1$) is parameterized as 
\begin{eqnarray}
&& H_{0n}
 = \left(0, \cdots, 0, 
\lambda_n e^{i\theta_n}e^{-2\pi z/N_{\rm F}} , 
1, 
0,\cdots,0 \right)\,, 
\nonumber\\
&& 
H_{0n}^*
 = \left(0, \cdots, 0,
\lambda_n e^{-i\theta_n}e^{-2\pi \bar z/N_{\rm F}} , 
1,
0, \cdots, 0 \right)\,, 
\label{eq:elementary_frac_inst_cpN}
\end{eqnarray}
where the value $1$ corresponds to the $(n+1)$-th flavor 
\footnote{
In the case of ${\mathbb C}P^1$, the moduli matrix $H_{0R}$ is 
contained in (\ref{eq:elementary_frac_inst_cpN}) as $n=2$ case, 
as one can see by making a $V$-transformation 
$(\lambda_R e^{i\theta_R}e^{\pi z}, 1 )\sim 
(1, \lambda_R^{-1} e^{-i\theta_R}e^{-\pi z})$.} 
One should note that the twisted boundary condition automatically 
introduces nontrivial $x^1$ dependence in $H_0$. 
The BPS solution given by $H_{0n}$ carries an 
instanton charge $1/N_{\rm F}$, and its conjugate $H_{0n}^*$ 
carries an instanton charge $-1/N_{\rm F}$. 
We call these BPS solutions as the elementary fractional instantons. 
The fractional instantons for $H_{0n}$ and $H_{0n}^*$ 
are located at $\frac{N_{\rm F}}{2\pi}\log\lambda_n$. 
The other moduli $\theta_n$ represents the relative phase of 
the $n$-th and $(n+1)$-th vacua. 
All these elementary fractional instantons are physically distinct, 
and are needed to form an instanton with unit charge 
as a composite of fractional instantons, as shown in Fig.~\ref{fig:tk=1}.
In the particular case of the $Z_{N_{\rm F}}$ twisted boundary 
condition, they are distinct only by the vacuum label $n$ 
($n=1,\cdots,N_{\rm F}$) and have identical properties. 
In that sense, we will exhibit only one of them 
as the representative in the following.

For each topological charge $n/N_{\rm F}, n=1, \cdots, N_{\rm F}-1$, 
there are $N_{\rm F}$ distinct BPS composite fractional instantons, 
but we will exhibit only one of them as a kink connecting 
the first flavor to the $n+1$-th flavor, since all other solutions 
starting from other vacua have identical properties. 
When the topological charge reaches unity, it becomes a 
genuine 
instanton. Those BPS solitons containing at least one instanton 
are not counted as fractional instantons here. 
The BPS composite fractional instanton with the maximal 
topological charge $(N_{\rm F}-1)/N_{\rm F}$ is given by 
the moduli matrix 
\begin{eqnarray}
&& H_{0}
 = \left( 
\lambda_1 e^{i\theta_1}e^{-{(N_{\rm F}-1)\over N_{\rm F}}2\pi z} \,,
\lambda_2 e^{i\theta_2}e^{-{(N_{\rm F}-2)\over N_{\rm F}}2\pi z} ,\cdots ,
\lambda_{N_{\rm F}-1} e^{i\theta_{N_{\rm F}-1}}e^{-{2\pi\over N_{\rm F}} z} 
, 1 \right)\,
\end{eqnarray}
with $2N_{\rm F}-2$ moduli parameters. 
Constituent fractional instantons are located at 
$\frac{N_{\rm F}}{2\pi}\log\frac{\lambda_1}{\lambda_2}, 
\frac{N_{\rm F}}{2\pi}\log\frac{\lambda_2}{\lambda_3}, \cdots, 
\frac{N_{\rm F}}{2\pi}
\log\frac{\lambda_{N_{{\rm F}-2}}}{\lambda_{N_{{\rm F}-1}}}, 
\frac{N_{\rm F}}{2\pi}\log\lambda_{N_{{\rm F}-1}}$, provided 
they are ordered as 
$\frac{N_{\rm F}}{2\pi}\log\frac{\lambda_1}{\lambda_2} <  
\frac{N_{\rm F}}{2\pi}\log\frac{\lambda_2}{\lambda_3} < \cdots 
< \frac{N_{\rm F}}{2\pi}
\log\frac{\lambda_{N_{{\rm F}-2}}}{\lambda_{N_{{\rm F}-1}}} 
<\frac{N_{\rm F}}{2\pi}\log\lambda_{N_{{\rm F}-1}}$. 
If any one of these inequalities are not satisfied, for instance, 
$\frac{N_{\rm F}}{2\pi}\log\frac{\lambda_{n}}{\lambda_{n+1}} 
\ge 
\frac{N_{\rm F}}{2\pi}\log\frac{\lambda_{n+1}}{\lambda_{n+2}} 
$, 
two fractional instantons are merged into one. 
In the limit of negative infinite relative separation, 
$\lambda_{n+1}\to 0$, the solution becomes a compressed 
fractional instantons located at the common center 
$\frac{N_{\rm F}}{4\pi}\log\frac{\lambda_{n}}{\lambda_{n+2}}$. 
In the limit, the size of the compressed fractional instantons 
becomes half of that of the individual fractional instanton. 

\begin{figure}[htbp]
\begin{center}
\begin{tabular}{ccc}
 \includegraphics[width=0.33\textwidth]{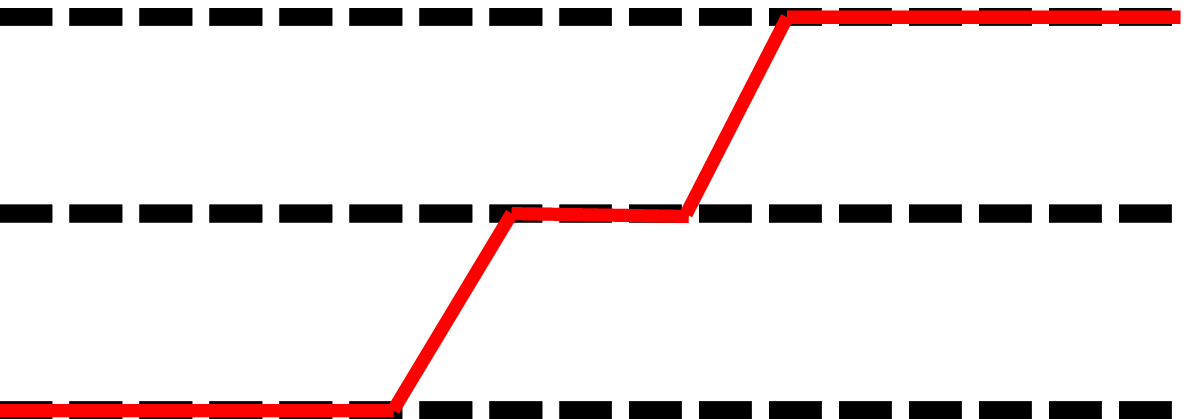}
& \includegraphics[width=0.33\textwidth]{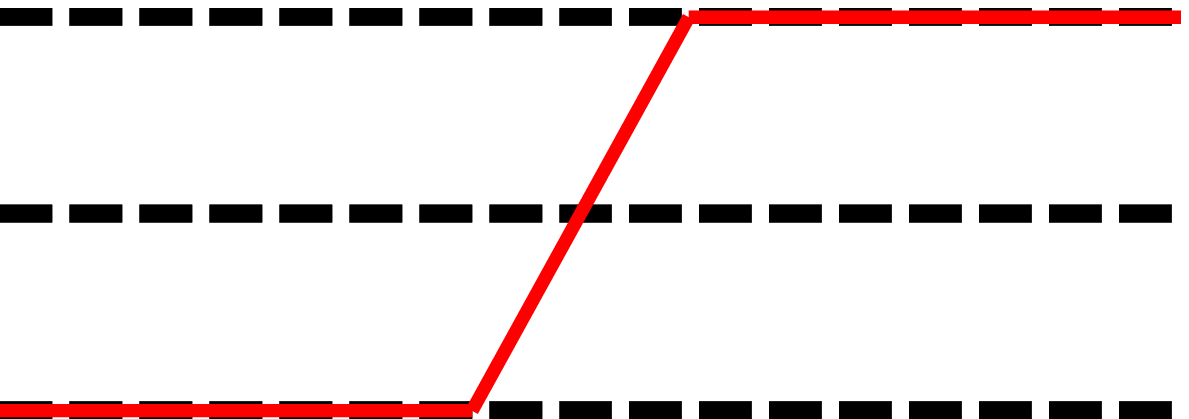}
& \includegraphics[width=0.33\textwidth]{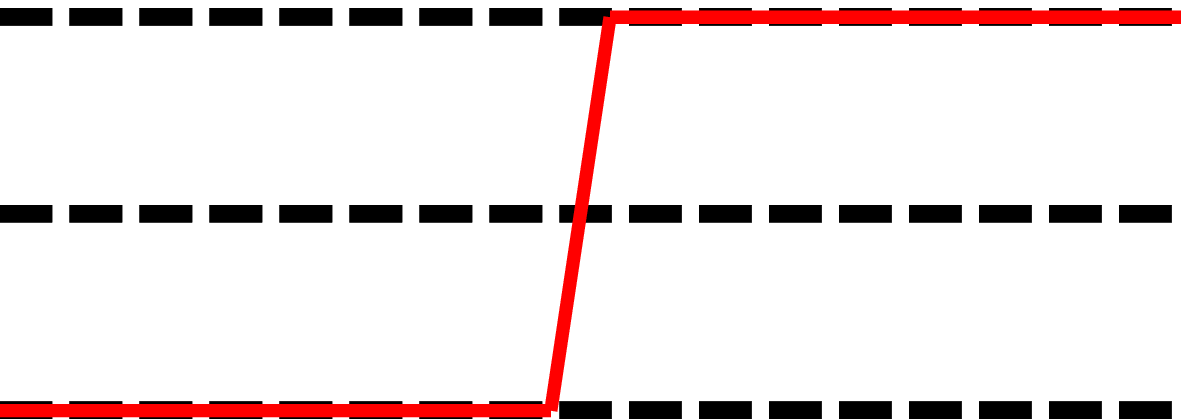}\\
(a)  & (b) & (c) 
\end{tabular}
\end{center}
\caption{
Composite fractional instantons in the 
${\mathbb C}P^2$ model. 
Keeping the cnter of mass position 
$\frac{3}{4\pi}\log\lambda_1$ fixed, 
(a) Generic separated fractional 
instanton solution ($\lambda_1 \approx 1, \lambda_2 \gg 1$) 
(b) Two fractional instantons touching and beginning to merge 
together ($\lambda_1\approx 1, \lambda_2\approx 1$). 
(c) Compressed fractional instanton solution 
($\lambda_1\approx 1, \lambda_2\to 0$). 
}
\label{fig:CPN-wall}
\end{figure}
Fig.~\ref{fig:CPN-wall} shows an example in the case of the 
${\mathbb C}P^2$ model : Two fractional 
instantons represented by the moduli matrix 
\begin{eqnarray} 
H_{0} = \left( \lambda_1 e^{i\theta_1}e^{-4\pi z/3},
\lambda_2 e^{i\theta_2}e^{-2\pi z/3}, 1 \right). 
\end{eqnarray}
Fractional instantons are located at 
$\frac{3}{2\pi}\log\frac{\lambda_1}{\lambda_2}$ and  
$\frac{3}{2\pi}\log{\lambda_2}$ in $x^1$, when 
$\lambda_2 \gg 1$, as shown in Fig.~\ref{fig:CPN-wall}(a). 
Keeping the center of mass position $\frac{3}{4\pi}\log\lambda_1$
fixed, we can decrease the relative separation 
$\frac{3}{2\pi}\log\frac{\lambda_1}{\lambda_2^2}$. 
When $\lambda_2 \approx 1$, two fractional instantons 
are touching and begin to merge as shown in Fig.~\ref{fig:CPN-wall}(b). 
When $\lambda_2 \to 0$, moduli matrix becomes 
$H_{0} = \left(\lambda_1 e^{i\theta_1}e^{-4\pi z/3}, 0, 1 \right)$ 
and two fractional instantons are compressed completely 
to become a single compressed fractional instanton with a width of 
the half of individual fractional instanton.

\begin{figure}[htbp]
\begin{center}
\begin{tabular}{cc}
 \includegraphics[width=0.4\textwidth]{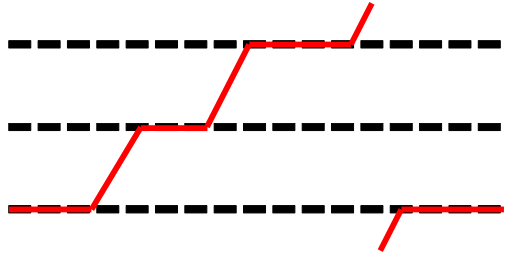} & \\
\end{tabular}
\end{center}
\caption{
A ${\mathbb C}P^{2}$ instanton with the unit instanton charge.
}
\label{fig:CPN-instanton}
\end{figure}

The ${\mathbb C}P^{N_{\rm F}-1}$ instanton with 
the unit instanton charge 
can be obtained with the moduli matrix 
\begin{eqnarray} 
\!\!\!\!H_{0}
\! = \!\!
\left(\lambda_1 e^{i\theta_1}e^{-\frac{N_{\rm F}-1}{N_{\rm F}}2\pi z}
+ \lambda_{N_{\rm F}} e^{i\theta_{N_{\rm F}}} e^{\frac{2\pi z}{N_{\rm F}}}, 
\lambda_2 e^{i\theta_2}e^{-\frac{N_{\rm F}-2}{N_{\rm F}}2\pi z} ,
\cdots, \lambda_{N_{\rm F}-1} 
e^{i\theta_{N_{\rm F}-1}}e^{-\frac{2\pi z}{N_{\rm F}}}, 1 
\right).
\end{eqnarray}
Fig.~\ref{fig:CPN-instanton} shows the BPS instanton in the 
case of the ${\mathbb C}P^2$ model. 

\subsection{Fractional instantons in the Grassmann sigma model
\label{subsec:frac_inst_grassmann}}

Now let us move to the Grassmann sigma model, 
admiting the Grassmann kinks \cite{Isozumi:2004jc}, 
which are interpreted as fractional instantons \cite{Eto:2006mz}. 
All BPS fractional instanton solutions can be obtained by 
holomorphic moduli matrices. 
As described at the end of Sec.~\ref{subsec:dbrane}, 
each of the $N_{\rm C}$ rows corresponding to each of the 
$N_{\rm C}$ color lines of $U(N_{\rm C})$ 
gauge group can be constructed following precisely the same 
procedure as $N_{\rm C}=1$ case (${\mathbb C}P^{N_{\rm F}-1}$ 
model). 
It is relatively straightforward to construct a moduli matrix 
$H_0$ corresponding to the BPS fractional instanton solutions, 
although computing the inverse square root of the gauge 
invariant matrix $(H_0 H_0^\dagger)^{-1/2}$ becomes nontrivial 
as $N_{\rm C}$ increases.

For the $a$-th color ($a=1,\cdots,N_{\rm C}$) line in a single 
row, the increment of flavor 
label from the left infinity to the right infinity is denoted as 
$k_a$, which corresponds to the numbers of elementary 
fractional instantons (the instanton charge of $k_a/N_{\rm F}$). 
Then the (anti-)BPS solutions are characterized by a set of 
nonnegative (nonpositive) integers 
\begin{eqnarray} 
(k_1, k_2, \cdots, k_{N_{\rm C}}),  
\label{eq:frac_inst_number_vector}
\end{eqnarray}
 corresponding to 
the number of fractional instantons for each color. 
This vector of fractional instanton number is used to specify 
BPS fractional instantons.

To write down the moduli matrix, we need to specify the vacuum 
on which the soliton is constructed. 
To classify BPS solutions using the moduli matrix, we introduce 
a vector of flavor labels $f_a$ occupied by the $a$-th color 
line in the left vacuum as 
\begin{eqnarray} 
\left<f_1, f_2, \cdots, f_{N_{\rm C}}\right>. 
\label{eq:flavor_label_vector}
\end{eqnarray}
This vacuum label for the left vacuum and the vector of fractional 
instanton number in Eq.~(\ref{eq:frac_inst_number_vector}) give 
a complete characterization of distinct fractional instantons. 
This characterization is valid even in the case of the 
non-${\mathbb Z}_{N_{\rm F}}$-symmetric boundary condition, unless some of flavors are degenerated.

Although the flavor label $f_1$ occupied by the first color 
is physically distinct from other labels, we only exhibit 
the case of $f_1=1$ here, since cyclic rotations of 
entire flavor labels give the configurations that have the 
same properties. 
For instance, we exhibit only one elementary BPS fractional 
instanton, although there are $N_{\rm F}$ distinct elementary 
BPS fractional instantons with identical properties starting 
from different vacua. 
Hence we choose 
\begin{eqnarray} 
1=f_1 < f_2 < \cdots < f_{N_{\rm C}} \le N_{\rm F}. 
\label{eq:ordering_flavor_label}
\end{eqnarray}
The s-rule implies that there should be no crossing of color lines 
and 
\begin{eqnarray} 
1+k_1=f_1+k_1 < f_{2}+k_{2}< \cdots < 
f_{N_{\rm C}}+k_{N_{\rm C}} < f_{1}+k_{1}+N_{\rm F}. 
\label{eq:s-rule_ordering}
\end{eqnarray}
To enumerate genuine fractional instantons, we identify 
BPS instanton with unit instanton charge as having the 
identical sets of flavors occupied by color lines for left and 
right vacua  
\begin{eqnarray} 
1+k_1=
f_2 , 
\cdots,  
f_{N_{\rm C}-1}+k_{N_{\rm C}-1}=f_{N_{\rm C}}, \; 
f_{N_{\rm C}}+k_{N_{\rm C}} =
N_{\rm F}+1, 
\label{eq:instanton_label}
\end{eqnarray}
because of $f_1=1$.

Similarly to the ${\mathbb C}P^{N_{\rm F}-1}$ model, many BPS 
fractional instantons starting from different vacua have 
identical properties. 
In the case of the ${\mathbb Z}_{N_{\rm F}}$ twisted boundary 
condition, the global ${\mathbb Z}_{N_{\rm F}}$ symmetry survives. 
By acting the ${\mathbb Z}_{N_{\rm F}}$ global transformations, 
the vacua of Grssmann sigma model are classified into several 
equivalence classes. 
Those BPS solitons starting from the left vacua in the same 
equivalence class have identical properties, although they are 
distinct physical solitons. 
Therefore we exhibit only the representative for each 
equivalence class. 
One should note that all vacua of the 
${\mathbb C}P^{N_{\rm F}-1}$ model ($N_{\rm C}=1$) are 
connected by the ${\mathbb Z}_{N_{\rm F}}$ action and are in one 
equivalence class.


We need $N_{\rm F} \ge N_{\rm C}$ in order to have a ground 
state with zero energy (supersymmetric vacuum when embedded 
into a supersymmetric theory). 
Taking the limit of strong coupling $g^2 \to \infty$ to obtain 
Grassmann sigma model requires $N_{\rm F} > N_{\rm C}$. 
Since the $N_{\rm C}=1$ case reduces to the previously studied 
${\mathbb C}P^{N_{\rm F}-1}$ models, we consider $N_{\rm C} \ge 2$. 
The exact duality between $N_{\rm C}$ and $N_{\rm F}-N_{\rm C}$ 
in the case of Grassmann sigma model implies that 
$Gr_{N_{\rm F},N_{\rm C}=N_{\rm F}-1}=Gr_{N_{\rm F},1}
={\mathbb C}P^{N_{\rm F}-1}$. 
Therefore the simplest Grassman model beyond the ${\mathbb C}P^{N_{\rm F}-1}$ 
is the case of $N_{\rm C}=2, N_{\rm F}=4$. 
The vacua of Grassmann sigma model $Gr_{4,2}$ falls into two 
equivalence class connected by ${\mathbb Z}_{N_{\rm F}}$ 
transformations: two colors occupy two 
adjacent flavors, or two non-adjacent flavors.

\begin{figure}[htbp]
\begin{center}
\begin{tabular}{ccc}
   \includegraphics[width=0.33\textwidth]{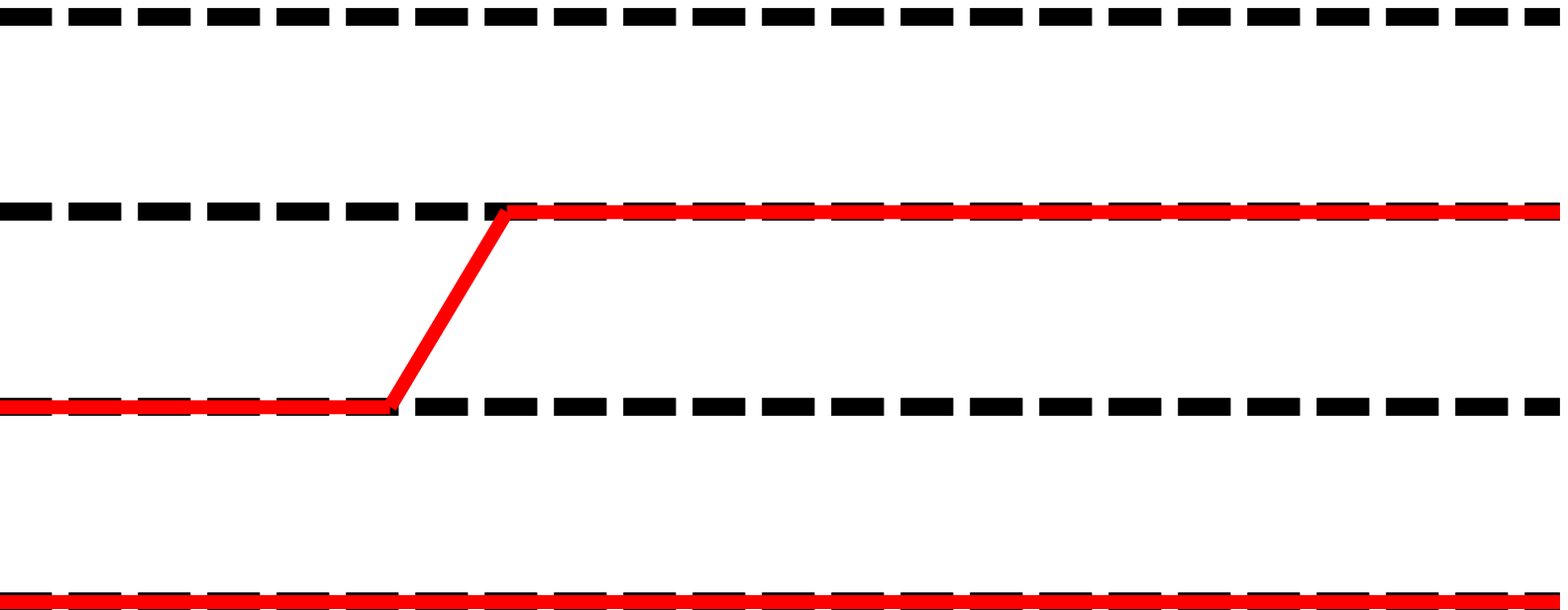} 
& \includegraphics[width=0.33\textwidth]{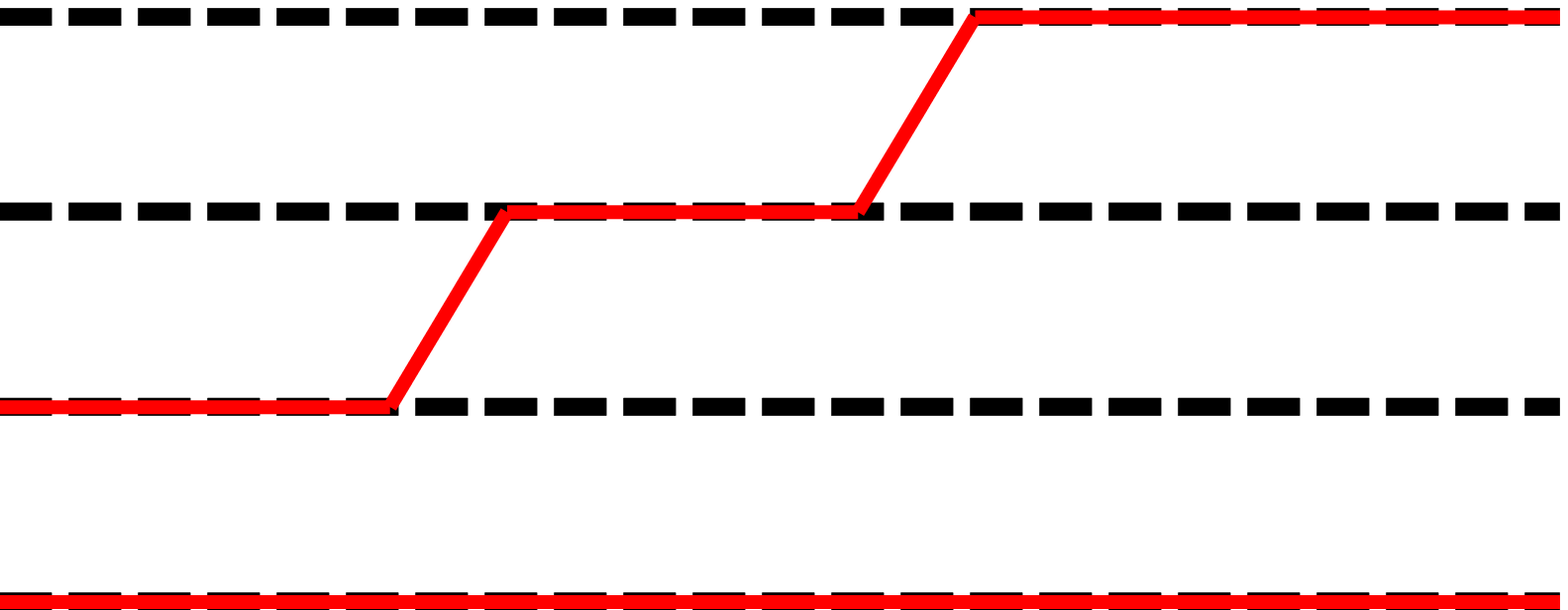}
& \includegraphics[width=0.33\textwidth]{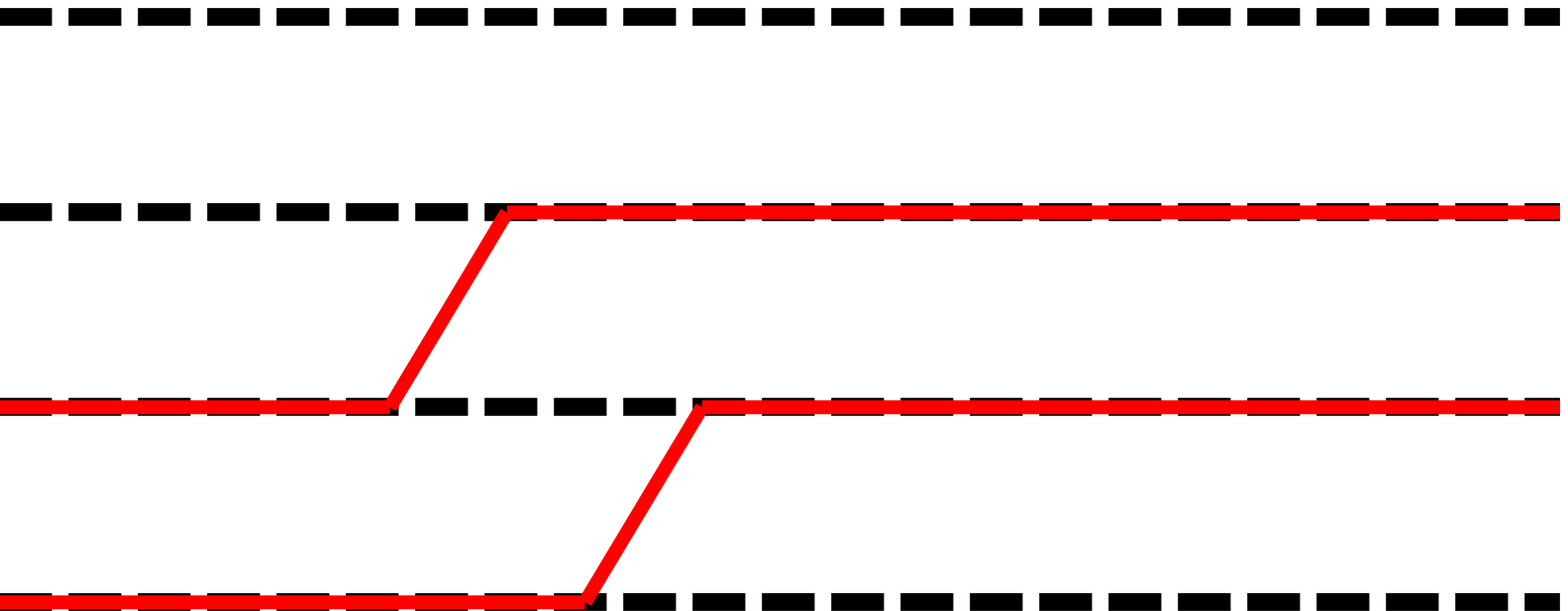}
\\
(a) $(0,1)$ &(b) $(0,2)$ &(c) $(1,1)$\\
   \includegraphics[width=0.33\textwidth]{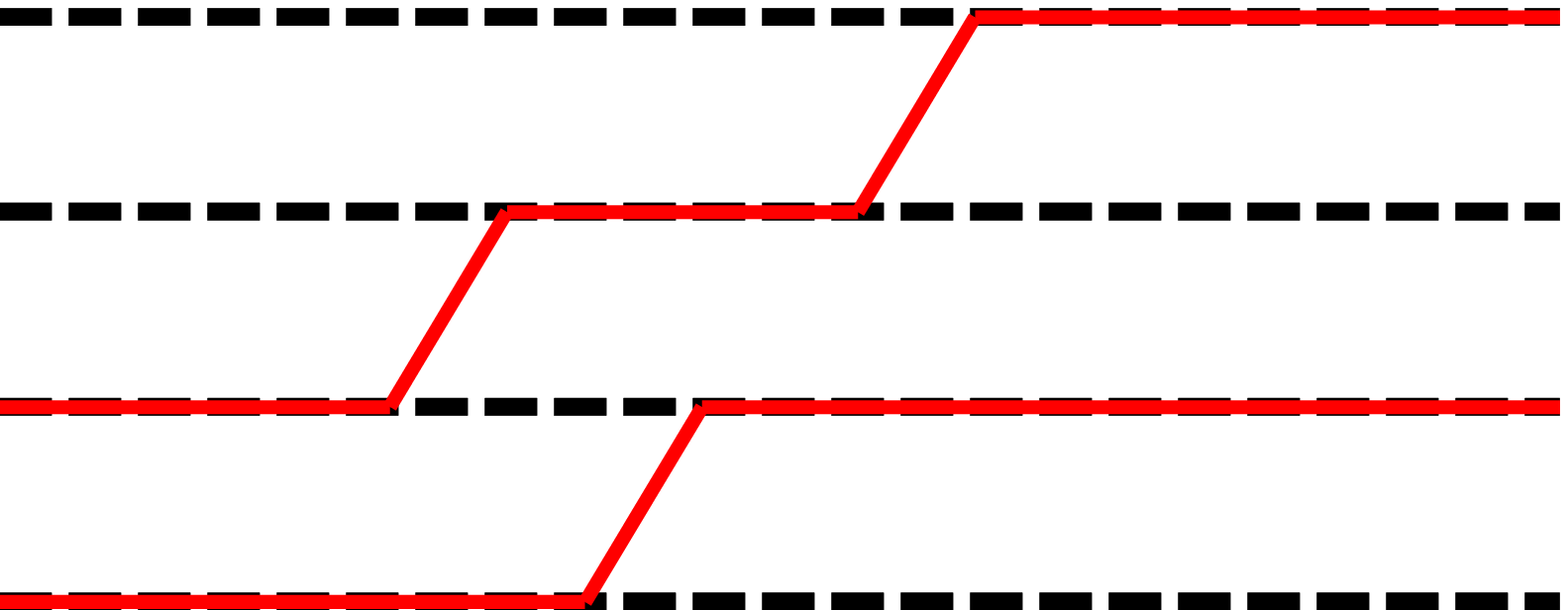} 
& \includegraphics[width=0.33\textwidth]{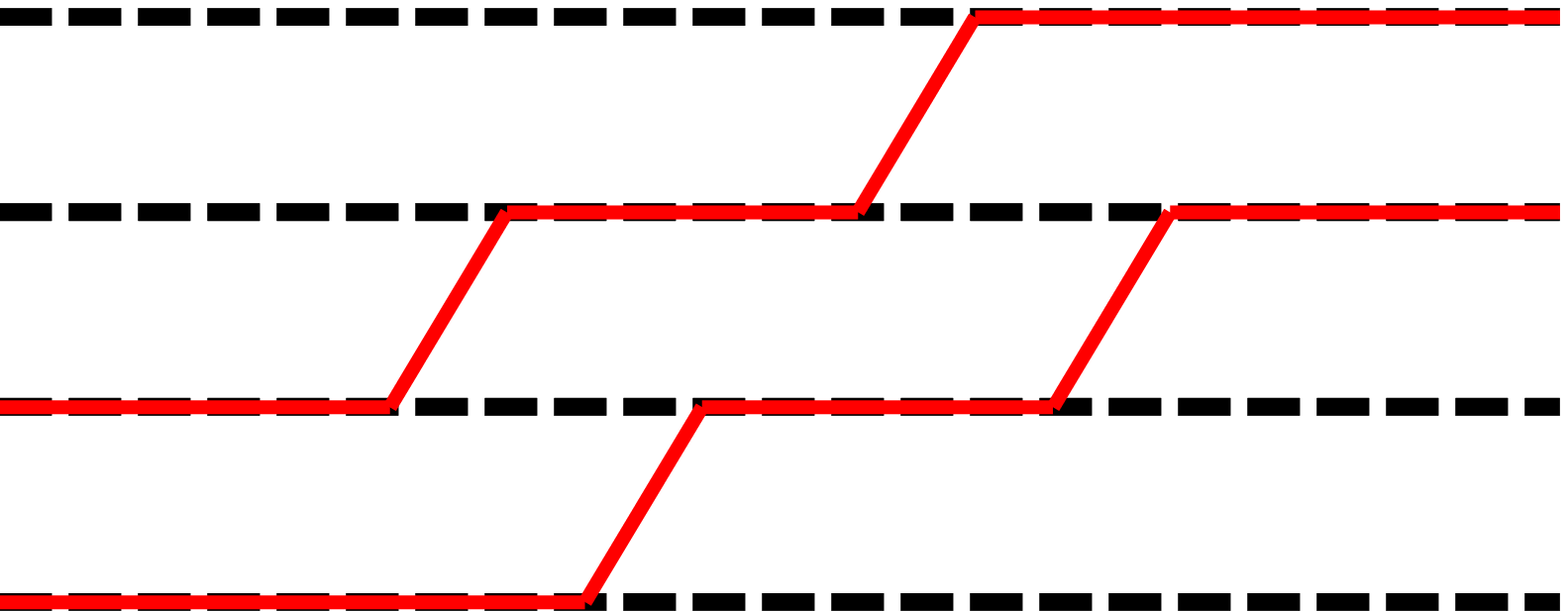}
& \includegraphics[width=0.33\textwidth]{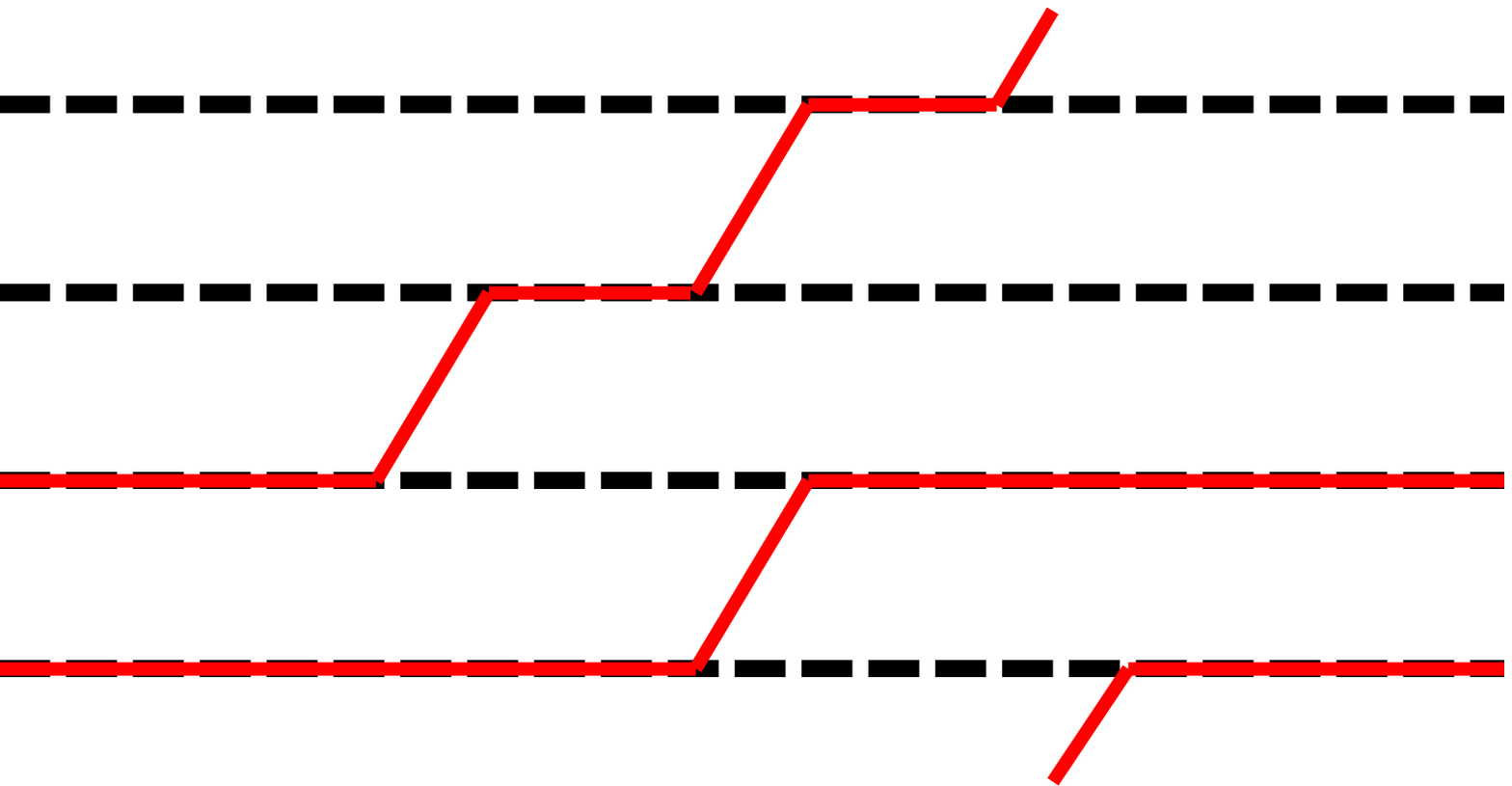}
\\
(d) $(1,2)$ &(e) $(2,2)$ &(f) $(1,3)$
\\
\end{tabular}
\end{center}
\caption{Fractional instantons for two adjacent flavors 
occupied by colors at the left vacuum in $G_{4,2}$, labeled 
by indices (a) $(0,1)$, (b) $(0,2)$, (c) $(1,1)$, (d) $(1,2)$, 
(e) $(2,2)$, and (f) $(1,3)$. 
Diagrams (a) is an elementary fractional instanton, 
and (b)--(e) are composite fractional instantons. 
Diagram (f) is an instanton, but not a composite 
fractional instanton. 
}
\label{fig:Gr42-fractional_adjacent}
\end{figure}


Let us exhibit all possible types of BPS and 
anti-BPS fractional instanton solutions in $Gr_{4,2}$.  
We first consider the case of left vacua with two adjacent 
flavors occupied by colors: namely we consider $f_1=1, f_2=2$. 
There is only one elementary BPS fractional instanton 
with total instanton charge $1/4$, specified by $(k_1,k_2)=(0,1)$, 
as shown in Fig.~\ref{fig:Gr42-fractional_adjacent}(a). 
There are two BPS composite fractional instantons  
with total instanton charge $1/2$, specified by 
$(k_1,k_2)=(1,1)$ and $(0,2)$, as shown in 
Fig.~\ref{fig:Gr42-fractional_adjacent}(b). 
There is only one BPS composite fractional instantons  
with total instanton charge $3/4$, specified by 
$(k_1,k_2)=(1,2)$, as shown in Fig.~\ref{fig:Gr42-fractional_adjacent}(c). 
There are two BPS solutions with 
total instanton charge $1$, specified by $(2,2)$ and $(1,3)$. 
It is surprising and interesting to find that the diagram 
$(2,2)$ shown in Fig.~\ref{fig:Gr42-fractional_adjacent}(e) is representing 
a BPS composite fractional instanton and not an instanton, 
in spite of the unit total instanton charge. 
On the other hand, 
we observe that $(1,3)$ diagram is a genuine 
BPS instanton solution, since the left and right vacua are 
identical as a set of flavors occupied by colors, as shown in 
Fig.~\ref{fig:Gr42-fractional_adjacent}(f). 
We have listed all BPS solutions with the total instanton charge 
less than or equal to unity. 
One can easily check that all other BPS solutions (on this left 
vacuum with two adjacent flavors occupied by colors) are 
composite of BPS fractional instantons with at least one genuine 
BPS instanton.

By the same token, we can list all the anti-BPS solutions as well. 
The anti-BPS elementary fractional instanton with total instanton 
charge $-1/4$ is specified by $(-1,0)$. 
The composite anti-BPS fractional instantons are $(-2,0)$ and 
$(-1,-1)$ with total instanton charge $-1/2$, 
$(-2,-1)$ with total instanton charge $-3/4$, 
$(-2,-2)$ with total instanton charge unity. 
The anti-BPS instanton solution is $(-3,-1)$. 
All other anti-BPS solutions are composite of anti-BPS 
fractional instantons with at least one genuine anti-BPS 
instanton.

\begin{figure}[htbp]
\begin{center}
\begin{tabular}{cc}
   \includegraphics[width=0.33\textwidth]{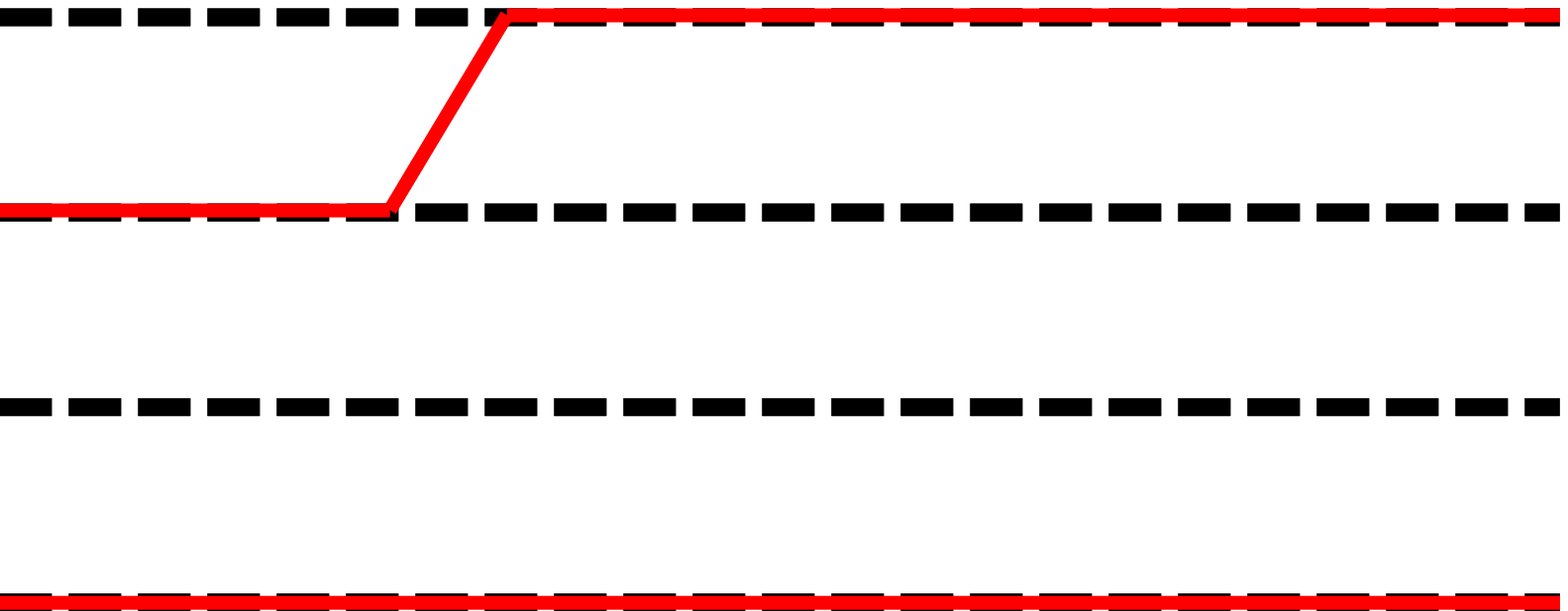} 
& \includegraphics[width=0.33\textwidth]{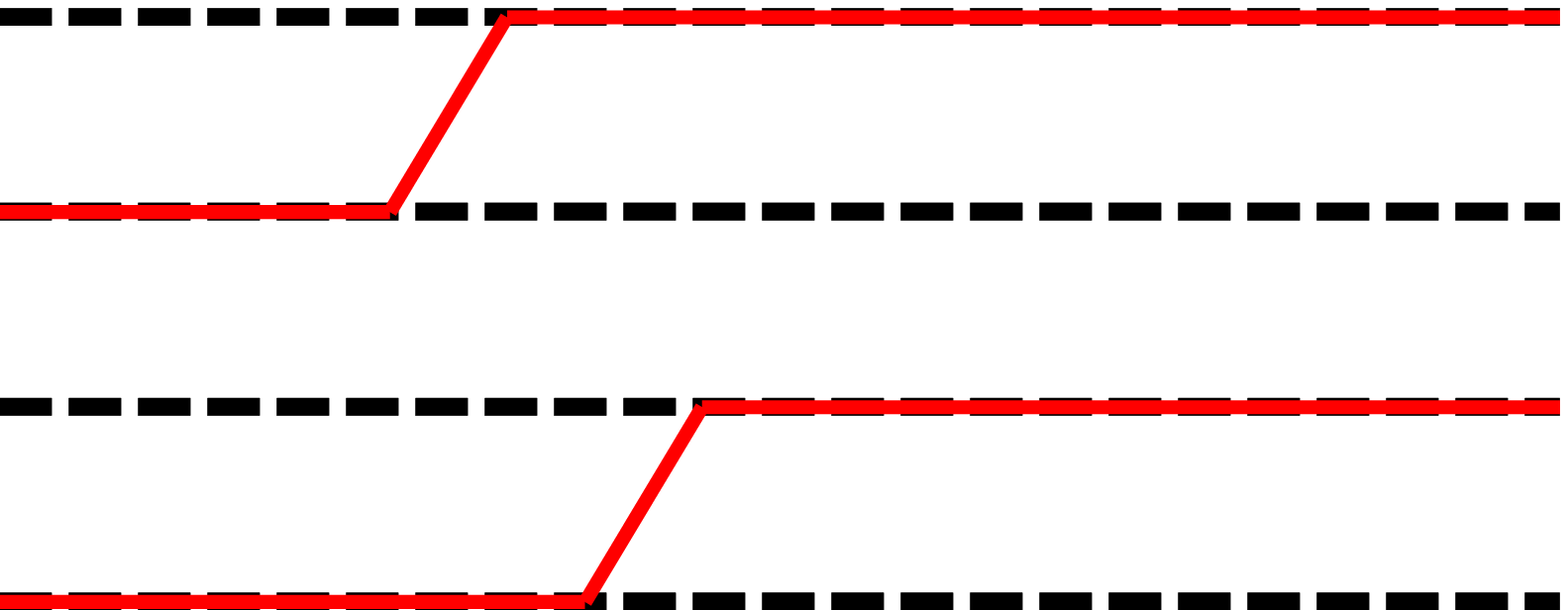}\\
(a) $(0,1)$ & (b) $(1,1)$ \\
  \includegraphics[width=0.33\textwidth]{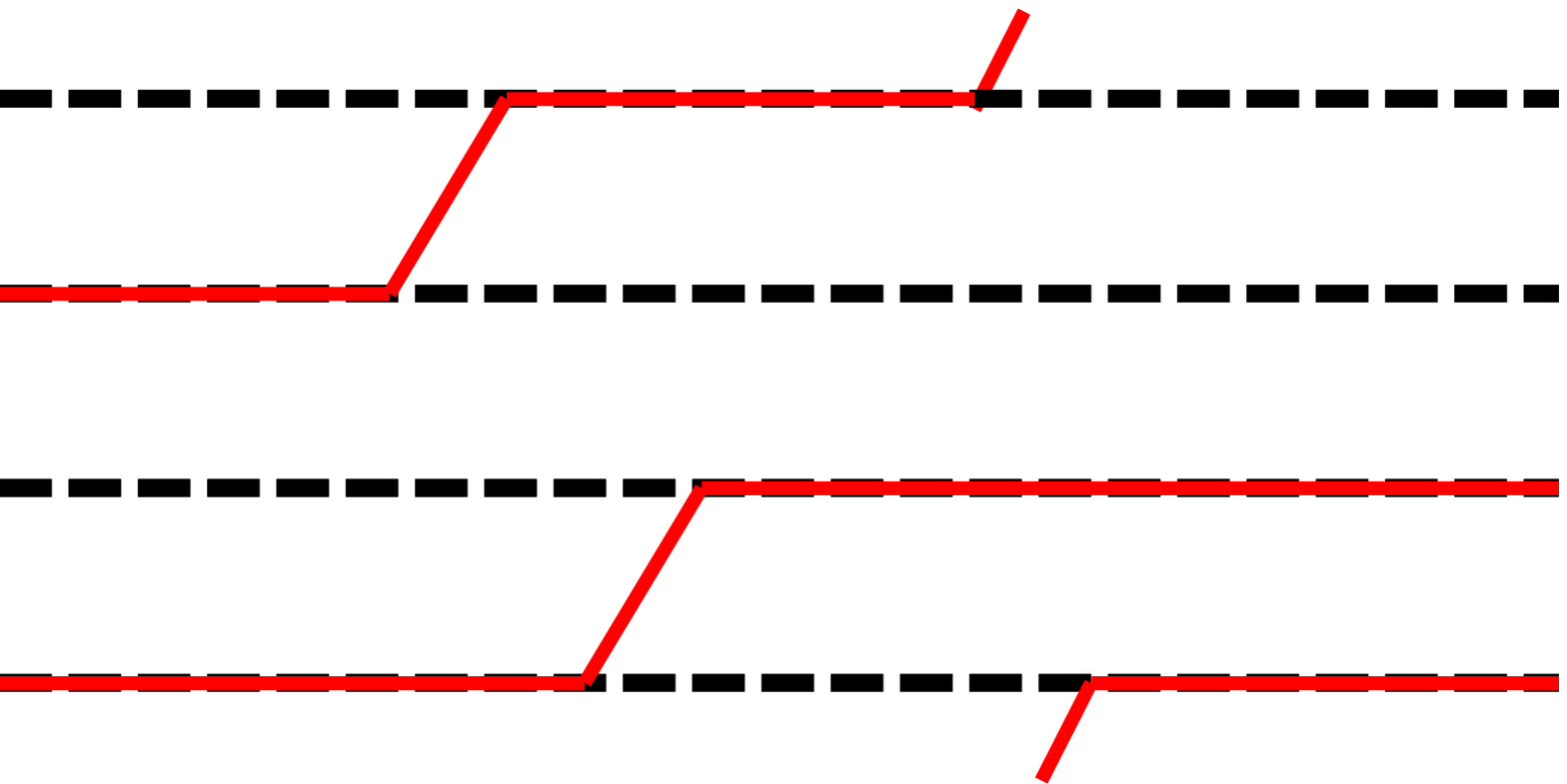}  
& \includegraphics[width=0.33\textwidth]{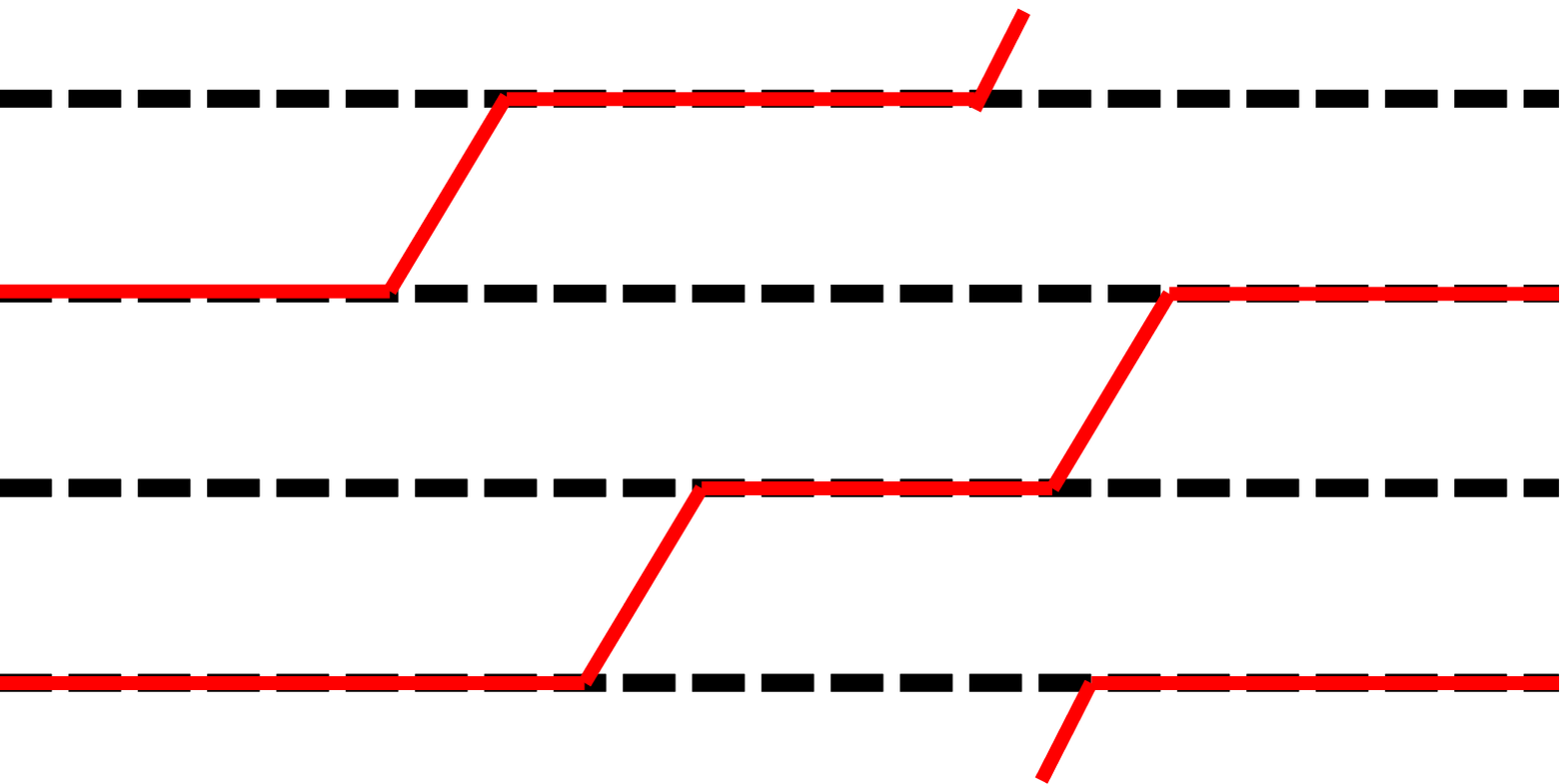}
\\
(c) $(1,2)$ &(d) $(2,2)$ 
\\
\end{tabular}
\end{center}
\caption{Fractional instantons for two non-adjacent flavors 
occupied by colors at the left vacuum in $G_{4,2}$, labeled 
by indices 
(a) $(0,1)$, (b) $(1,1)$, (c) $(1,2)$, and (d) $(2,2)$. 
Diagram (a) is an elementary fractional instanton, 
(b) and (c) are composite fractional instantons, 
(d) is a BPS instanton. 
}
\label{fig:Gr42-fractional_nonadjacent}
\end{figure}

We can also construct all possible BPS (anti-BPS) solutions for 
fractional instantons in the other case of left vacua with 
non-adjacent flavors occupied by colors: namely we consider 
$f_1=1, f_2=3$. 
There is only one type of elementary BPS fractional instanton 
with total instanton charge $1/4$, specified by $(k_1,k_2)=(0,1)$, 
as shown in Fig.~\ref{fig:Gr42-fractional_nonadjacent}(a).  
Let us note that other possible elementary fractional instanton 
$(1,0)$ is of the same type as $(0,1)$ (although physically 
distinct) and we do not list it. 
There is only one type of BPS composite fractional instantons  
with total instanton charge $1/2$, specified by 
$(k_1,k_2)=(1,1)$, as shown in 
Fig.~\ref{fig:Gr42-fractional_nonadjacent}(b).  
There is only one type of BPS composite fractional instantons  
with total instanton charge $3/4$, specified by 
$(k_1,k_2)=(1,2)$, as shown in 
Fig.~\ref{fig:Gr42-fractional_nonadjacent}(c).  
Since $(2,1)$ is of the same type as $(1,2)$ and is not listed. 
There is only one type of BPS solution with total instanton 
charge $1$, specified by $(2,2)$, 
as shown in Fig.~\ref{fig:Gr42-fractional_nonadjacent}(d).  
This $(2,2)$ solution is a genuine 
BPS instanton solution, since the left and right vacua are 
identical as a set of flavors occupied by colors. 
We have listed all BPS solutions with the total instanton 
charge less than or equal to unity. 
One can easily check that all other BPS solutions (on this left 
vacuum with two non-adjacent flavors occupied by colors) are 
composite of BPS fractional instantons with at least one genuine 
BPS instanton. 

Summarizing, we find that BPS fractional instantons in $Gr_{4,2}$ 
model are exhausted by five species on the left vacuum with 
two adjacent flavor occupied by colors in 
Fig.~\ref{fig:Gr42-fractional_adjacent}(a)-(e), 
and three species on the left vacuum with 
two non-adjacent flavor occupied by colors in 
Fig.~\ref{fig:Gr42-fractional_nonadjacent}(a)-(c).

As an evidence for rich varieties of fractional instantons, we 
observe that BPS fractional instantons not reducible to 
composite of instanton and fractional instantons can have 
topological charge of order $N_{\rm C}$ in the case of 
$Gr_{N_{\rm F},N_{\rm C}}$ sigma model with large $N_{\rm F}$. 
By generalizing the $(2,2)$ fractional instanton for adjacent 
vacuum in Fig.~\ref{fig:Gr42-fractional_adjacent}(e), we obtain 
a fractional instanton that is characterized by the vector of 
fractional instanton number 
$(N_{\rm F}-N_{\rm C}, \cdots, N_{\rm F}-N_{\rm C})$ 
constructed on the vacuum with $N_{\rm C}$ adjacent flavors 
occupied by colors in the $Gr_{N_{\rm F},N_{\rm C}}$ sigma model. 
This fractional instanton cannot be reduced to a composite of 
instanton and fractional instantons by moduli deformations, and have 
topological charge $Q=N_{\rm C}(N_{\rm F}-N_{\rm C})/N_{\rm F}$ 
which become $Q\sim N_{\rm C}$ for large $N_{\rm F}$.

As an illustrutive example, we give explicitly the moduli 
matrix $H_0$ for the BPS instanton solution with the set 
$(1,3)$ in Fig.~\ref{fig:Gr42-fractional_adjacent}(f)  
\begin{eqnarray} 
H_{0} = 
\left(
\begin{array}{cccc}
\lambda_1 e^{i\theta_1}e^{-\frac{1}{2}\pi z}, & 1, & 0, & 0 \\
1, & \lambda_3 e^{i\theta_3}e^{-\frac{3}{2}\pi z}, 
& \lambda_4 e^{i\theta_4}e^{-\pi z}, 
& \lambda_5 e^{i\theta_5}e^{-\frac{1}{2}\pi z}  
\end{array}
\right).
\end{eqnarray}
The constituent fractional instantons are located at 
$x^1=\frac{2}{\pi}\log\frac{\lambda_3}{\lambda_4},
\frac{2}{\pi}\log\frac{\lambda_4}{\lambda_5},
\frac{2}{\pi}\log\lambda_1, 
\frac{2}{\pi}\log\lambda_5$. 
This BPS configuration is nothing but an instanton with the unit 
instanton charge, as one can recognize from the fact that the 
set of flavor branes occupied by the color branes in the 
right infinity of this diagram are identical to the corresponding 
set at the left infinity. 
The total instanton charge is of course unity.


As another example, let us write down explicitly the moduli 
matrix for the BPS solution of (composite) fractional instantons 
with the set $(2, 2)$, which is depicted in 
Fig.~\ref{fig:Gr42-fractional_adjacent}(e)  
\begin{eqnarray} 
H_{0}
 = \left(
\begin{array}{cccc}
\lambda_1 e^{i\theta_1}e^{-\pi z}, & 
\lambda_2 e^{i\theta_2}e^{-\frac{1}{2}\pi z}, & 1, & 0 
\\
0, & \lambda_3 e^{i\theta_3}e^{-\pi z}, & 
\lambda_4 e^{i\theta_4}e^{-\frac{1}{2}\pi z}, & 1 
\end{array}
\right). 
\label{eq:H0_22_Gr42}
\end{eqnarray}
Total instanton number of this BPS solution is unity. 
However, one should note that the set of flavors occupied 
by two color branes in the left and right vacua are different. 
Therefore this solution is not an instanton, but is a composite 
soliton of BPS fractional instantons. 

In general, fractional instantons are characterized by 
two adjacent color-flavor locking vacua, which are denoted 
by a set of flavors $(f_i, f_j)$ occupied by the first and 
second colors, respectively in the case of $N_{\rm C}=2$. 
The BPS solution defined by the moduli matrix 
(\ref{eq:H0_22_Gr42}) is a composite soliton of four different 
types of fractional instantons. 
As can be read from the moduli matrix (\ref{eq:H0_22_Gr42}) 
for Fig.~\ref{fig:Gr42-fractional_adjacent}(e), 
the fractional instanton 
connecting vacua $(1,2)\to(1,3)$ is located at 
$\frac{2}{\pi}\log\frac{\lambda_4}{\lambda_5}$, 
the fractional instanton 
connecting vacua $(1,3)\to(2,3)$ is located at 
$\frac{2}{\pi}\log\frac{\lambda_1}{\lambda_2}$, 
the fractional instanton 
connecting vacua $(2,3)\to(2,4)$ is located at 
$\frac{2}{\pi}\log{\lambda_5}$, 
and the fractional instanton 
connecting vacua $(2,4)\to(3,4)$ is located at 
$\frac{2}{\pi}\log{\lambda_2}$. 
In the Fig.~\ref{fig:Gr42-fractional_adjacent}(e), 
the position of the kink on the 
first brane at $\frac{2}{\pi}\log\frac{\lambda_1}{\lambda_2}$ 
is placed to the right of the kink on the second brane at 
$\frac{2}{\pi}\log{\lambda_5}$. 
One should note that their positions can be interchanged 
by taking ${\lambda_5} < \frac{\lambda_1}{\lambda_2}$. 
In that case, the character of fractional instantons at 
these positions change: 
the fractional instanton connecting vacua $(1,3)\to(1,4)$ 
is located at $\frac{2}{\pi}\log{\lambda_2}$, and 
the fractional instanton 
connecting vacua $(1,4)\to(2,4)$ is located at 
$\frac{2}{\pi}\log\frac{\lambda_1}{\lambda_2}$.

\begin{figure}[htbp]
\begin{center}
 \includegraphics[width=1.0\textwidth]{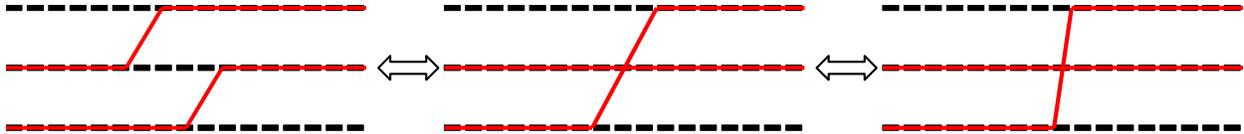}
\end{center}
\caption{
BPS reconnection. 
The moduli space of the configuration after the BPS reconnection 
is a boundary of the moduli space of that before the BPS reconnection.
\label{fig:bps-reconnection}
}
\end{figure}

Similarly to the ${\mathbb C}P^{N_{\rm F}-1}$ model, two BPS 
fractional instantons can be merged together and become a 
compressed fractional instantons with half of the size of the 
individual fractional instanton in the limit, as illustrated in 
Fig.\ref{fig:bps-reconnection}. 
This configuration of the compressed fractional instantons 
can be regarded as a boundary of the moduli space of separated 
fractional instantons. 
Since the compressed kink may be regarded as a reconnection of 
color lines, we call this phenomenon as BPS reconnection.

\begin{figure}[htbp]
\begin{center}
 \includegraphics[width=0.5\textwidth]{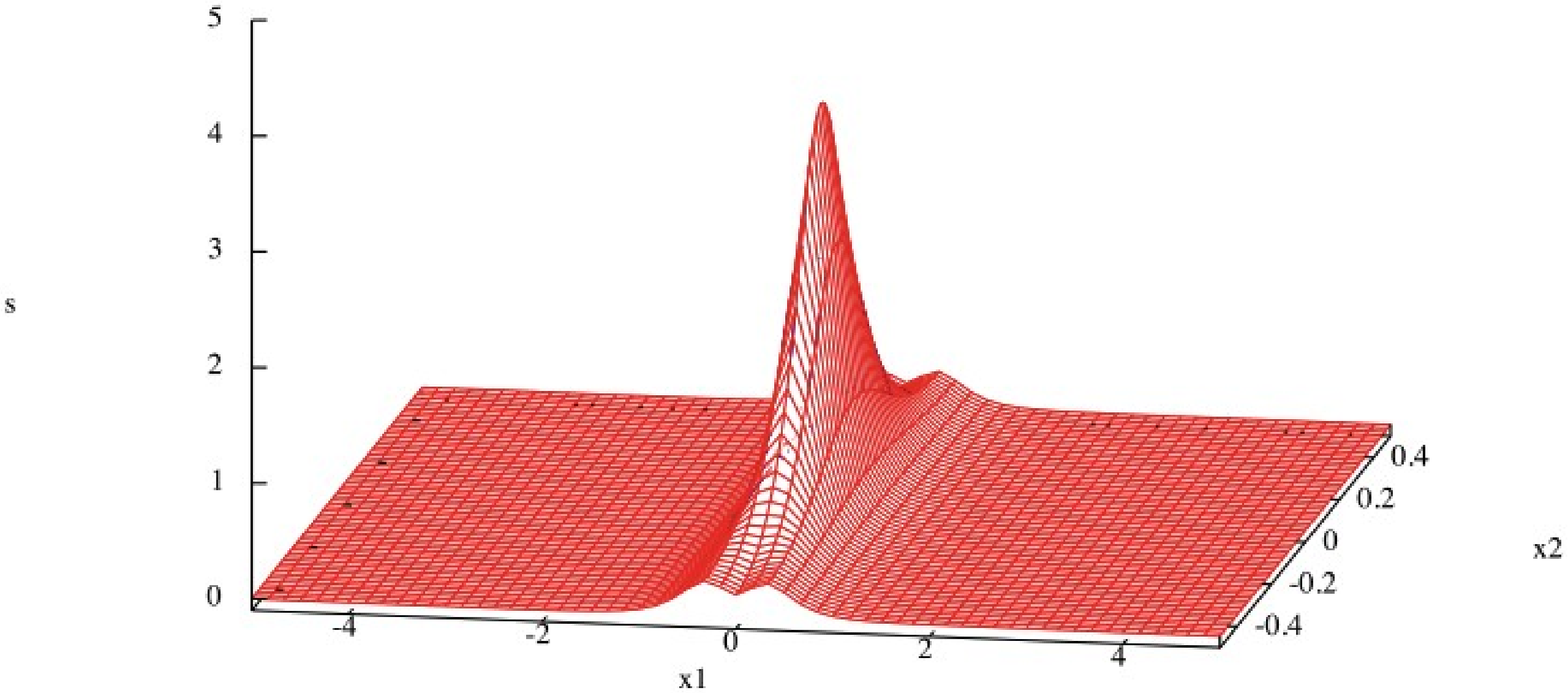}
 \includegraphics[width=0.5\textwidth]{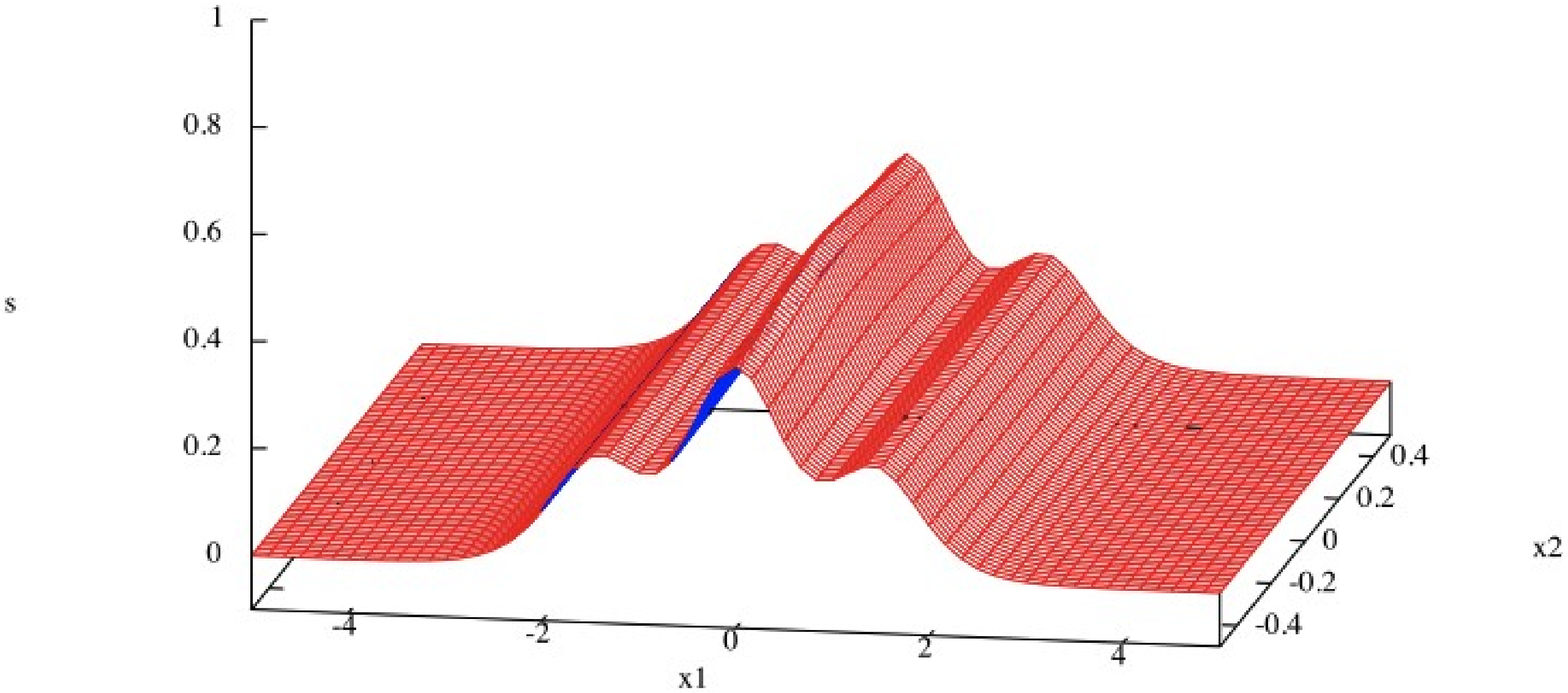}
  \includegraphics[width=0.5\textwidth]{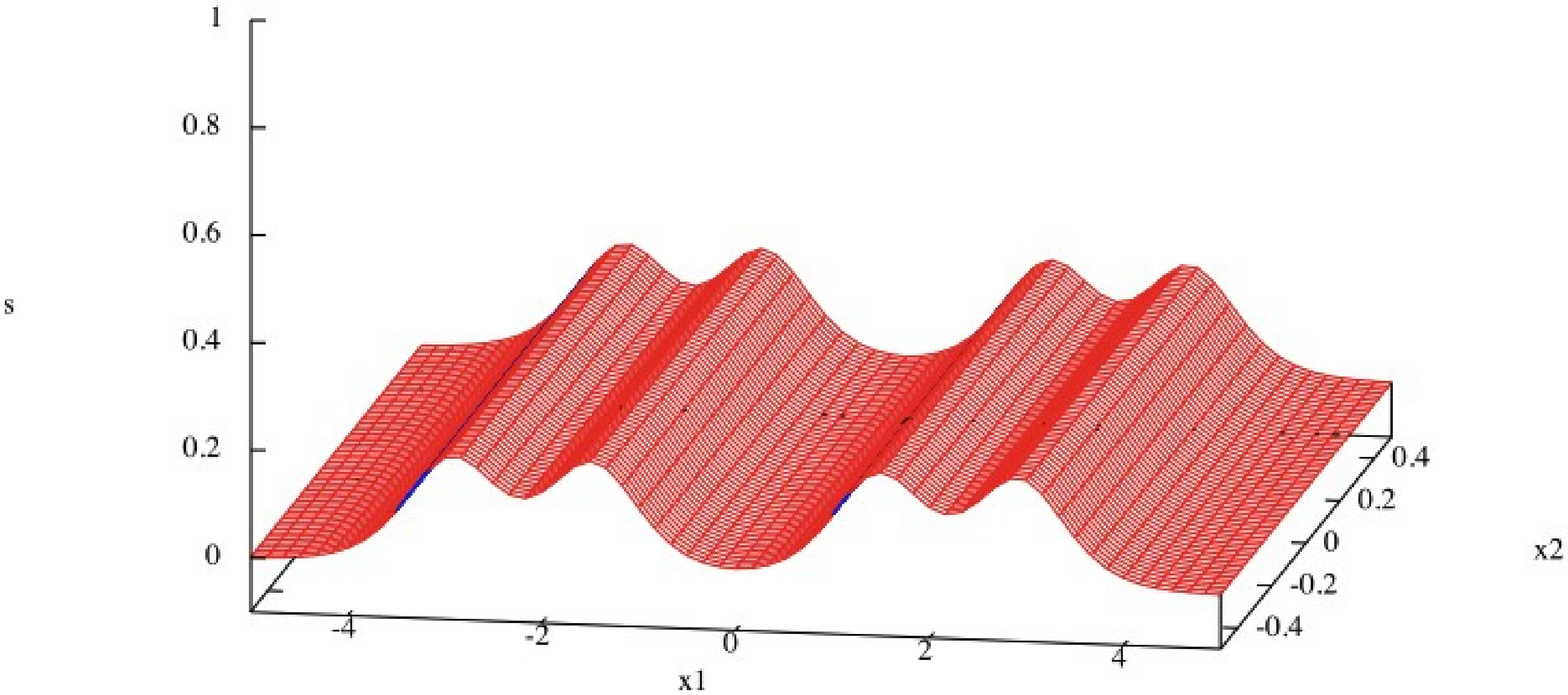}
\end{center}
\caption{Energy density $s({\bf x})$ for the BPS instanton 
configuration of the Grassman sigma model in 
Fig.~\ref{fig:Gr42-fractional_adjacent}(f) for 
$\lambda_{1} =1,\,\lambda_{3}=1,\,\lambda_{4} =1,\,
\lambda_{5}=1$ (top),$\lambda_{1} =1,\,\lambda_{3}=1,\,
\lambda_{4} =10,\,\lambda_{5}=10$ (middle), and 
$\lambda_{1} =10,\,\lambda_{3}=0.1,\,\lambda_{4} =10,\,
\lambda_{5}=100$ (bottom). 
Grassmann sigma model coupling is taken as $v^2=1$. 
}
\label{inst_Gr}
\end{figure}

\subsection{Energy density of BPS instantons in Grassmann Sigma model}
\label{sec:IGS}

We now calculate the energy density of the BPS instanton 
configuration in $Gr_{4,2}$ in Fig.~\ref{fig:Gr42-fractional_adjacent}(f).
For this case, the energy density generically depends on 
$x^{2}$, but this dependence disappear as the separation 
between the fractional instanton constituents gets large. 
We here show how the lump of the instanton is decomposed 
into the four fractional instantons with the $1/4$ topological 
charge.
Fig.~\ref{inst_Gr} depicts the energy densities ${\cal L}(x^1,x^2)$ 
in Eq.~(\ref{eq:mdl:reduced-L}) of 
instanton in the Grassmann sigma model $Gr_{4,2}$ 
(Fig.~\ref{fig:Gr42-fractional_adjacent}(f)) 
in the $x^{1}$ and $x^{2}$ plane for three parameter sets.
The top one is almost equivalent to the instanton configuration 
with topological charge being unity, 
where the $x^{2}$ dependence still remains. 
In the middle one, the instanton starts to be decomposed, 
where the $x^{2}$ dependence is disappearing. 
In the bottom one, the instanton are almost decomposed into 
four fractional instantons, 
where the $x^{2}$ dependence disappears.

\section{Classification of ${\mathbb C}P^{N_{\rm F}-1}$ Bions} \label{sec:CPN}

\subsection{Bions in the ${\mathbb C}P^1$ model: a review}

\begin{figure}[htbp]
\begin{center}
 \includegraphics[width=0.5\textwidth]{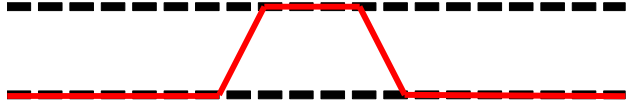}
\end{center}
\caption{A neutral bion in the ${\mathbb C}P^1$ model. 
}
\label{fig:CP1-bion}
\end{figure}

A neutral bion configuration is 
a composite of a fractional instanton and 
fractional anti-instanton with 
the total instanton charge canceled out. 
It is a non-BPS configuration and may not be a solution of 
field equations. 
Let us discuss the ${\mathbb C}P^1$ model first. 
From the solutions in Eq.~(\ref{eq:fractional}) 
and their complex conjugates, 
it is reasonable to consider 
the following ansatz for  the ${\mathbb C}P^{1}$ model 
satisfying a ${\mathbb Z}_{2}$ twisted boundary condition 
(\ref{ZNC}) as \cite{Misumi:2014jua}
\begin{equation}
H_0
 = \left(\lambda_{1}e^{i\theta_{1}}e^{-\pi z}
+\lambda_{2}e^{i\theta_{2}}e^{\pi\bar{z}}, 1 \right)\,. 
 \label{BBozero}
\end{equation}
As shown in Fig.\ref{fig:CP1-bion}, the fractional instanton 
is located at $\frac{1}{\pi}\log\lambda_1$, 
and the fractional anti-instanton is at 
$\frac{1}{\pi}\log\frac{1}{\lambda_2}$. 
As the separation becomes negative 
$\lambda_1\lambda_2 \to \infty$ (with $\lambda_1/\lambda_2$ held fixed), 
they are compressed together 
and eventually becomes a vacuum $H_0\to (1,0)$. 

The energy with arbitrary separation 
between fractional instanton and anti-instanton 
was calculated in Ref.~\cite{Misumi:2014jua}.

No charged bion is possible in 
the ${\mathbb C}P^1$ model.  

\subsection{Bions in the ${\mathbb C}P^{N_{\rm F}-1}$ model}

\begin{figure}[htbp]
\begin{center}
\begin{tabular}{cc}
 \includegraphics[width=0.5\textwidth]{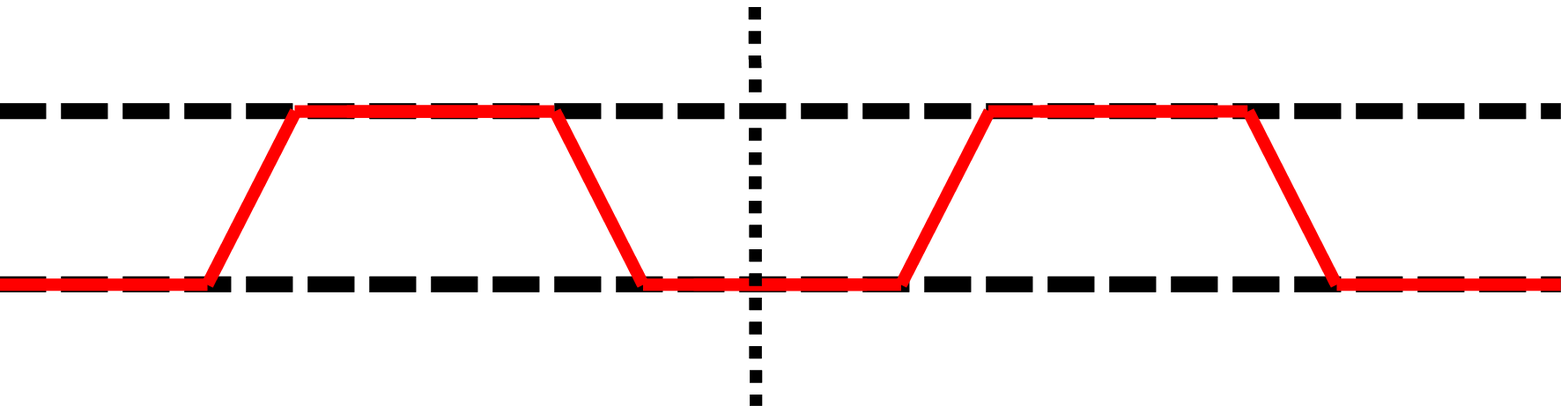}
&
 \includegraphics[width=0.5\textwidth]{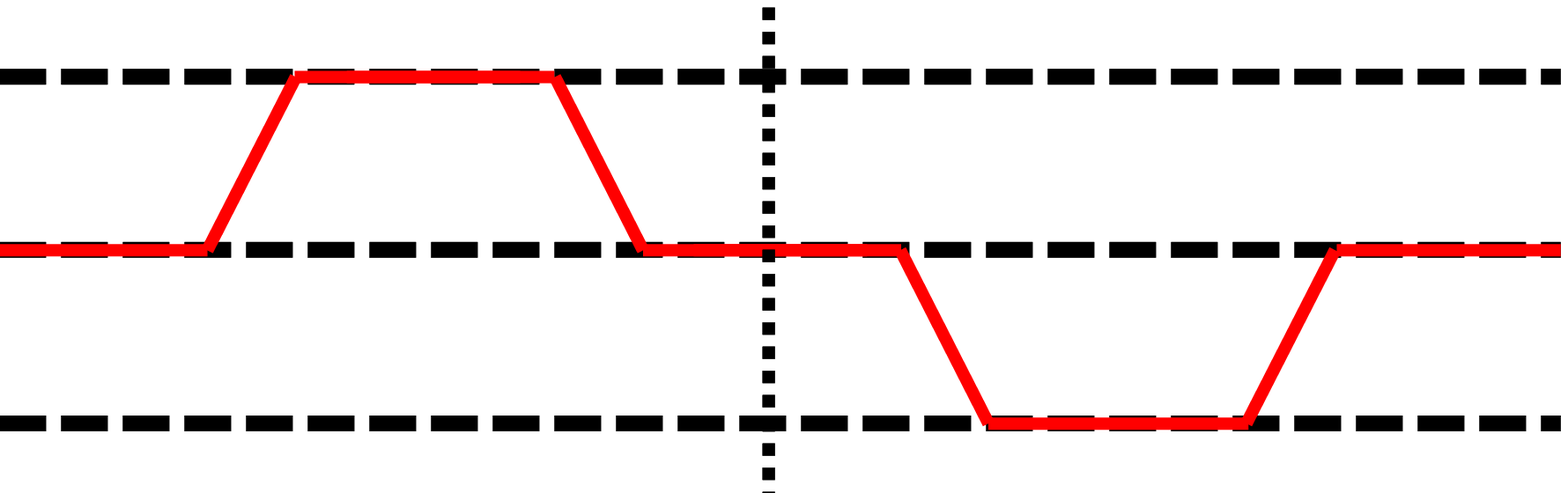}\\
(a) &  (b) \\~\\
 \includegraphics[width=0.5\textwidth]{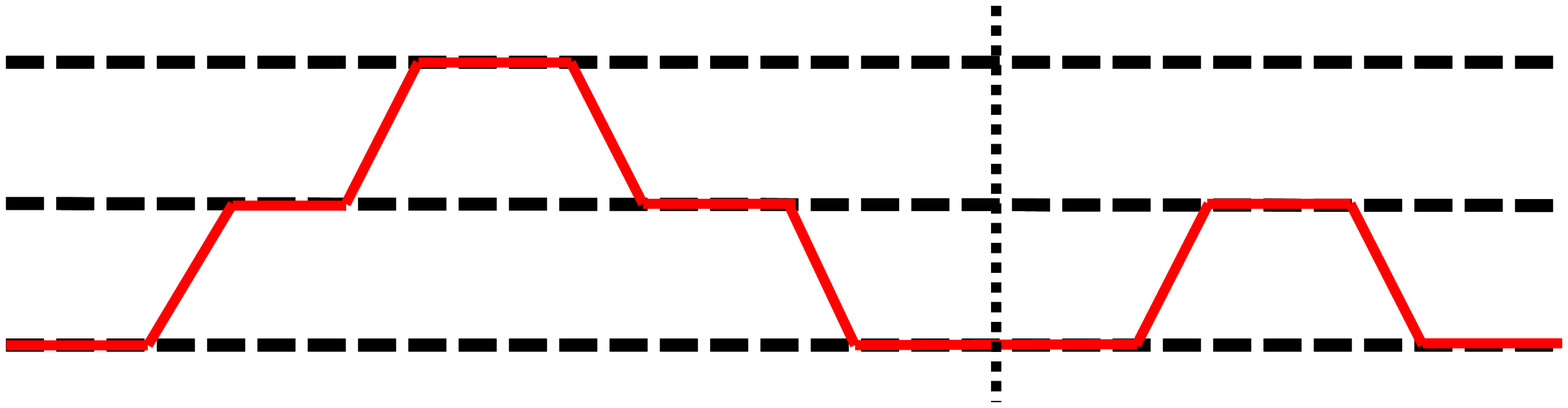}
&
 \includegraphics[width=0.5\textwidth]{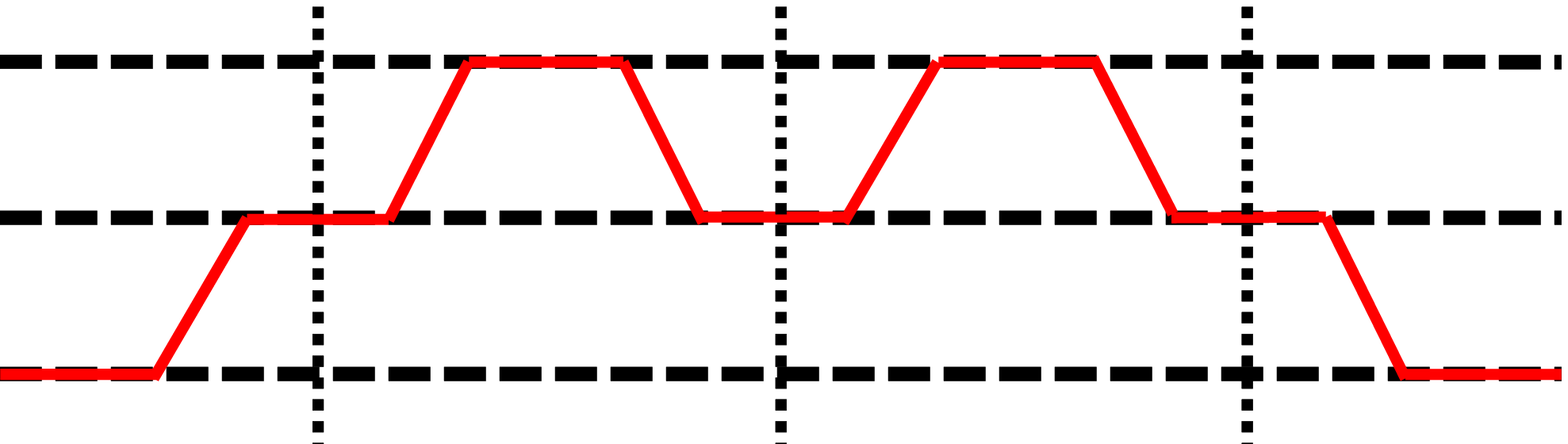}\\
(c) & (d)
\end{tabular}
\end{center}
\caption{Reducible neutral bions in the ${\mathbb C}P^{N_{\rm F}-1}$ model.  
}
\label{fig:CPN-reduciblebion}
\end{figure}

\begin{figure}[htbp]
\begin{center}
\begin{tabular}{cc}
 \includegraphics[width=0.4\textwidth]{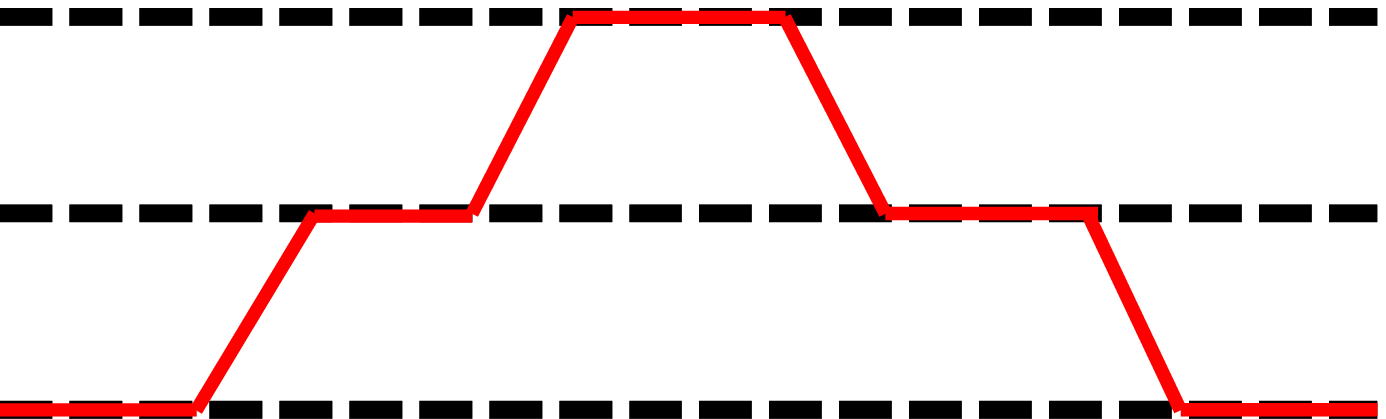}
& \includegraphics[width=0.4\textwidth]{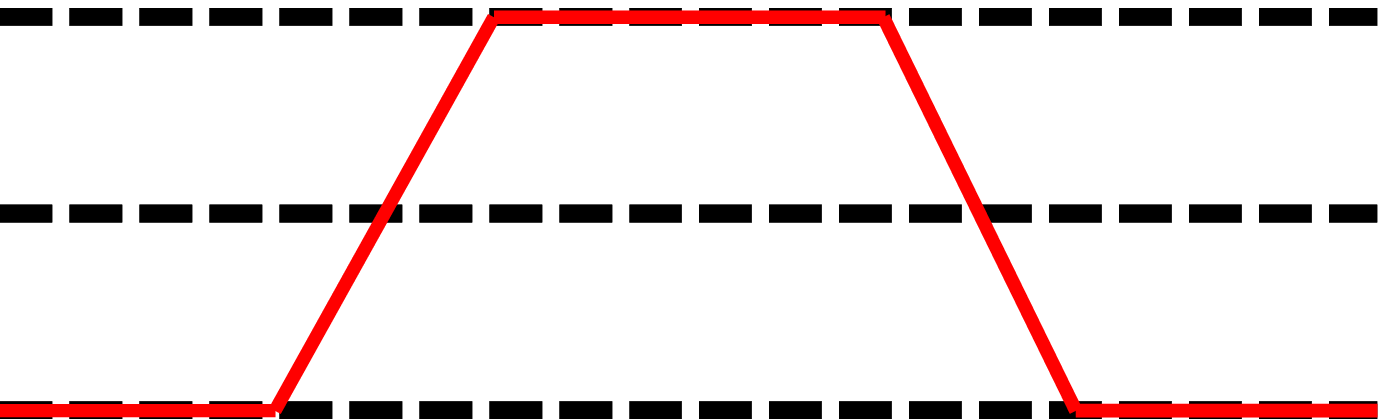}\\
(a)  & (b)
\end{tabular}
\end{center}
\caption{Irreducible and composite neutral bions  
in the ${\mathbb C}P^2$ model.  
(a) generic position, (b) compressed.
}
\label{fig:CPN-bion}
\end{figure}

Let us consider
neutral bions in the ${\mathbb C}P^{N_{\rm F}-1}$ model. 
A configuration is said to be 
{\it reducible} (or {\it irreducible}) when 
it can (or cannot) 
be decomposed into multiple neutral bions 
by changing moduli parameters. 
Examples of reducible and irreducible neutral 
bions are shown in Figs.~\ref{fig:CPN-reduciblebion} 
and \ref{fig:CPN-bion}, respectively.
The configurations
 in Fig.~\ref{fig:CPN-reduciblebion}(a)--(c)
can be decomposed along the dotted line into 
two neutral bions, so that they are all reducible. 
We say a configuration to be reducible even when 
its subconfiguration can be 
decomposed into multiple neutral parts.  
An example of such a configuration is given in 
Fig.~\ref{fig:CPN-reduciblebion}(d), 
where the two regions between the dotted lines 
are decomposed neutral bions in
Fig.~\ref{fig:CPN-reduciblebion}(a). 

Irreducible neutral bions can be characterized in terms 
constituent BPS fractional instanton. 
We define the topological charge of neutral bions 
by the topological charge of constituent BPS fractional 
instanton, namely by the number of fractional instantons, 
which is the same with that of fractional anti-instantons 
from its neutrality. 
The total energy is proportional to the topological charge.

For instance, an irreducible bion in Fig.~\ref{fig:CP1-bion} 
is a pair of an elementary fractional instanton and an 
elementary fractional anti-instanton. 
The simplest example of neutral bion in the ${\mathbb C}P^2$ 
model consisting of composite fractional instantons 
is shown in Fig.~\ref{fig:CPN-bion}. 
We consider to produce t
This diagram may be regarded as a result a pair creation 
of elementary fractional instanton and anti-instanton on 
the diagram in Fig.~\ref{fig:CP1-bion}. 
One of characteristic features of composite fractional 
instantons is that the constituent fractional instantons 
can merge into a compressed fractional instanton, as shown in 
Fig.~\ref{fig:CPN-wall}. 
For instance, two fractional (anti-)instantons 
can merge to form a compressed fractional (anti-)instanton, 
as shown in Fig.~\ref{fig:CPN-bion}.

No charged bion is possible in the ${\mathbb C}P^{N_{\rm F}-1}$ 
model.

\section{Classification of Bions in Grassmann sigma models} 
\label{sec:Gr}

\subsection{Neutral bions in the Grassmann sigma models}
\label{subsec:neutralBionGr}

\begin{figure}[htbp]
\begin{center}
 \includegraphics[width=0.7\textwidth]{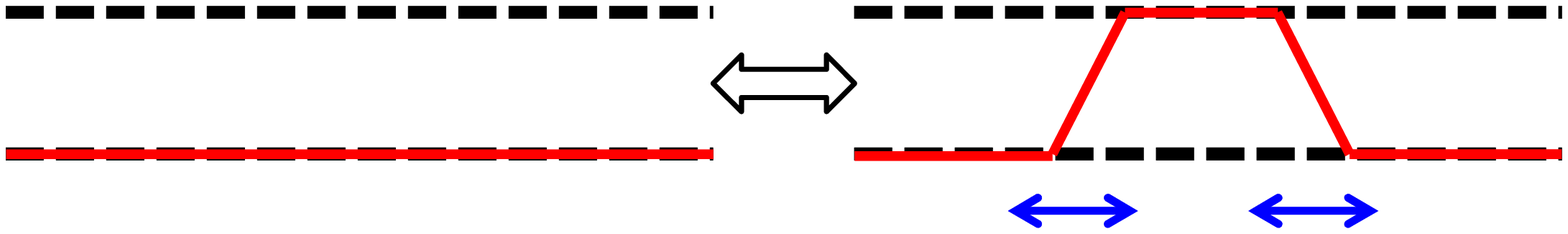}\\
(a)\\
  ~\\
 \includegraphics[width=0.7\textwidth]{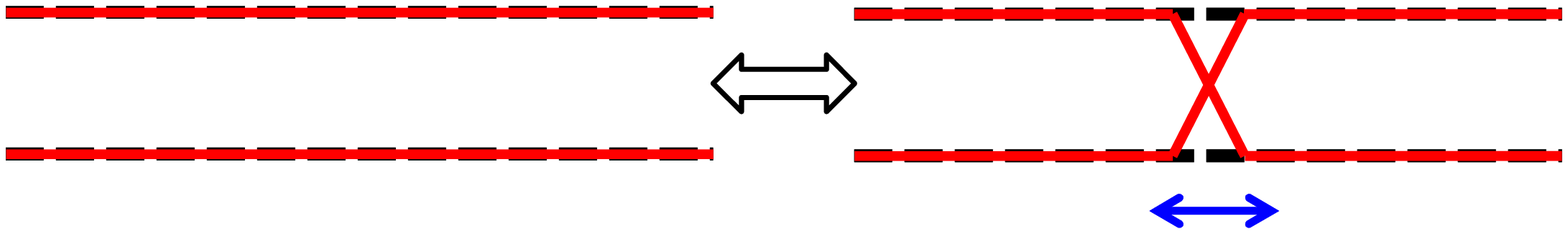}\\
(b)
\end{center}
\caption{Fundamental procedures to create neutral bions. 
(a) A pair creation of fractional instanton and fractional 
anti-instanton.
(b) Crossing/non-BPS reconnection of two color branes.   
(a) creates neutral bions but (b) does {\it not}.
}
\label{fig:bion-creation}
\end{figure}

In order to classify bions in  the Grassmann sigma models systematically, 
we may consider two fundamental procedures 
 to create neutral bions, as shown in Fig.~\ref{fig:bion-creation};
(a) a pair creation of fractional instanton and anti-instanton 
and (b) a crossing of two color branes, which may be 
called non-BPS crossing.
The pair creation (a) in Fig.\ref{fig:bion-creation}  
increases the numbers of 
fractional (anti-)instantons by one, 
and consequently the topological charge by 
$1/\NF$. 
The pair creation procedure can be used 
for constructing neutral bions in the ${\mathbb C}P^{N_{\rm F}-1}$ model.  
For instance, 
the neutral bions in 
Figs.~\ref{fig:CPN-reduciblebion} and 
(\ref{fig:CPN-bion})   
can be all constructed by repeating the pair creations two and four times,
respectively.  
As we see later, we can connect the right hand side of (a) 
 in Fig.\ref{fig:bion-creation} with finite energy continuously 
to left hand side of (a) representing the vacuum configuration 
with vanishing energy by changing parameters of field configuration. 

On the other hand, we can write down a moduli matrix 
representing the non-BPS crossing shown in the right 
hand side of (b) in Fig.\ref{fig:bion-creation} as 
\begin{equation}
H_0
 = \left(
\begin{array}{cc}
\lambda_{2}e^{i\theta_{2}}e^{\pi\bar{z}}, &1 \\
\lambda_{1}e^{i\theta_{1}}e^{-\pi z}, & 1 
\end{array}
\right)\,. 
 \label{nonBPScrossing}
\end{equation}
We find the energy density to vanish identically, indicating 
that the right hand side of (b) is actually the same as the 
left-hand side, namely the vacuum itself \footnote{
The matrix is nonsingular det$H_0\not=0$ except 
possibly at isolated points. Then it is easy to see that $H$ 
given in our fundamental formula in Eq.~(\ref{eq:H_formula}) 
gives $H H^\dagger =v^2{\bf 1}=H^\dagger H$. Hence the energy 
density in Eq.~(\ref{eq:mdl:reduced-L}) vanish identically. 
}. 
Therefore we do not use the non-BPS crossing in our approach 
to generate bion configurations in Grassmann sigma model. 
It is conceivable that non-BPS crossing needs to be considered 
if we consider finite gauge coupling and/or quantum effects 
properly.

\begin{figure}[htbp]
\begin{center}
 \includegraphics[width=0.6\textwidth]{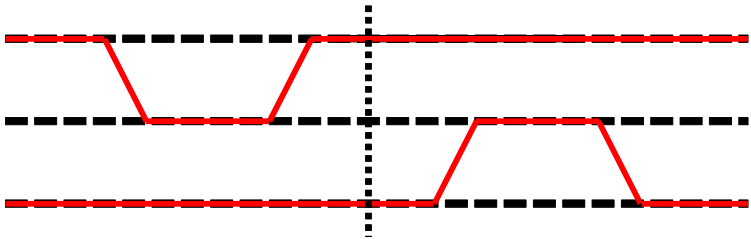}
\end{center}
\caption{
A reducible neutral bion for the Grassmann sigma models. 
This is typical reducible bion in addition to the configurations in 
Fig.~\ref{fig:CPN-reduciblebion} (a) and (b), which exist for 
the ${\mathbb C}P^{N_{\rm F}-1}$ model. 
This configuration is a dual to that of 
Fig.~\ref{fig:CPN-reduciblebion} (b).
}
\label{fig:composite}
\end{figure}

The definition of the reducibility is the same with the last 
subsection for 
the ${\mathbb C}P^{N_{\rm F}-1}$ model.
In this paper, we classify irreducible neutral bions. 
In addition to the configurations in 
Fig.~\ref{fig:CPN-reduciblebion} (a) and (b)
the configuration in Fig.~\ref{fig:composite}
 is a reducible neutral bion 
because each of them can be split into the left and right parts.
The configuration in Fig.~\ref{fig:composite} is a Seiberg dual 
of that of Fig.~\ref{fig:CPN-reduciblebion} (b).
We do not consider these reducible configurations 
when we classify irreducible neutral bions.

\begin{figure}[htbp]
\begin{center}
 \includegraphics[width=0.8\textwidth]{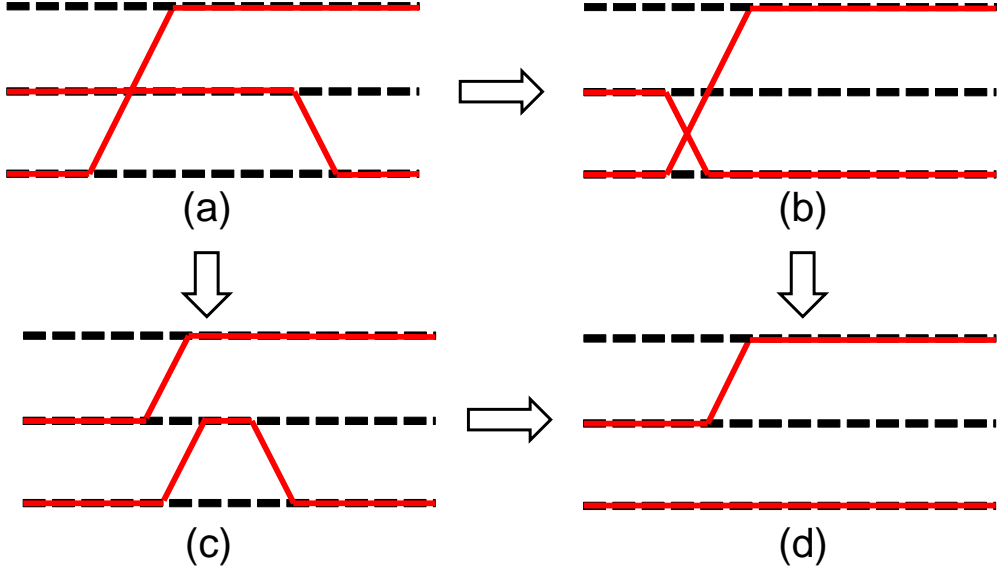}\\
\end{center}
\caption{
Annhilation of a compressed kink and a single anti-kink.
(a) is the initial configuration of 
a compressed kink and a single anti-kink, 
and (d) is the final configuration with one kink.  
The non-BPS crossing (b) represents a direct collision 
of an anti-kink with a compressed kink, and can be 
understood through the BPS reconnection (a) $\to$ (c) to be 
equivalent to the pair annihilation of a kink and an anti-kink. 
}
\label{fig:equivalence}
\end{figure}
Although we do not consider non-BPS crossing in this work, 
we here present a case that the non-BPS crossing and reconnection 
reduce to the other procedure that we use, namely pair 
annihilation, as shown in Fig.~\ref{fig:equivalence}. 
Instead of considering the reconnection from the non-BPS 
crossing (b) in Fig.~\ref{fig:equivalence} to the vacuum (d) 
(in the two lower color lines),  
the non-BPS crossing (b) can be safely deformed by deformations 
keeping BPS properties to (a) and (c), and then finally (c) to (d) 
by pair-annihilation process. 
Therefore we find that the non-BPS crossing and reconnection 
can be reduced to the pair-annihilation process in some region 
of parameter space by deformations keeping BPS conditions. 
This result supports our strategy not to use 
non-BPS crossing to enumerate bion configurations.


When we repeat to insert pair creations in 
Fig.~\ref{fig:bion-creation} (a),  
we do not insert the second pair 
outside the first pair, 
resulting in a zigzag configuration 
as in Fig.~\ref{fig:CPN-reduciblebion}(a),
which is reducible.
Instead, we allow inserting 
the second pair creations
between the first pair of fractional instanton and anti-instanton,   
as in Fig.~\ref{fig:CPN-bion} (a).
If we insert the second pair with the direction 
opposite to the first one, we again have  
 a zigzag configuration as in Fig.~\ref{fig:CPN-reduciblebion} (a), 
which is reducible.

For each color brane, $k$ BPS fractional instantons 
are placed on the left (right) and 
$k$ fractional anti-instantons are placed on the right (left) 
to cancel the instanton charge in total. 
We label it by $k$ ($-k$).
Therefore, irreducible neutral bions in $G_{\NF,\NC}$
can be labeled by a set of $\NC$ integers:
\beq
 (k_1,k_2,\cdots,k_{\NC}),  \quad   Q = \sum_{a=1}^{\NC} k_a /\NF, 
\eeq
where $(k_1,\cdots,k_{N_{\rm C}})$ and $Q$ give constituent 
fractional instanton numbers and the total instanton charge 
of the BPS fractional instantons corresponding to the left 
half of the bion.

\begin{figure}[htbp]
\begin{center}
 \includegraphics[width=0.7\textwidth]{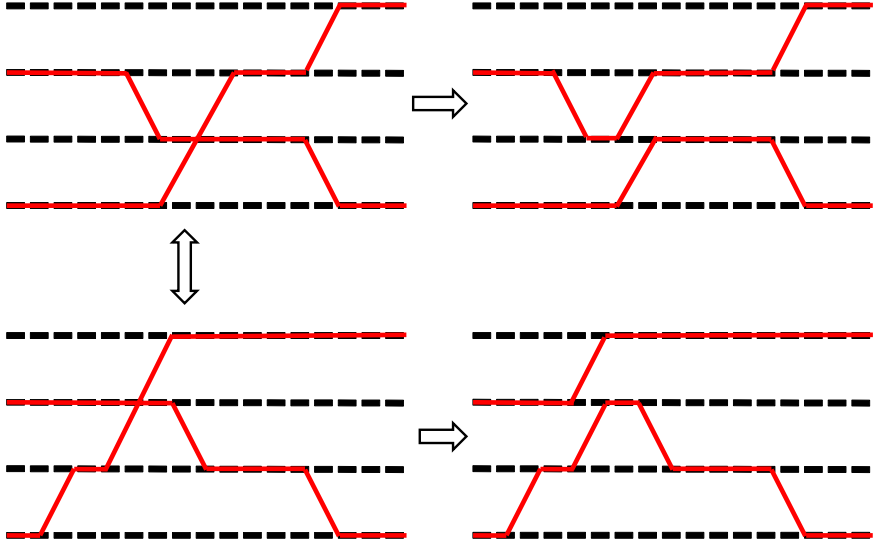}\\
\end{center}
\caption{
The upper left and down left configurations, 
which are connected each other by a smooth deformation, 
become the upper right and down right configurations, respectively,  
after BPS reconnections, which are disconnected from each other.
}
\label{fig:BPSrec-moduli}
\end{figure}
The BPS reconnection in Fig.~\ref{fig:bps-reconnection} 
is always possible, 
but a configuration after the reconnection is a compressed limit 
of the configuration before the reconnection. 
The moduli space of configurations after the BPS reconnection 
is a boundary of the moduli space of configurations before 
the BPS reconnection.
So we do not regard configurations with one color brane crossing 
another color brane to be independent configuration.
However, note that two moduli which are not connected to each other 
may be connected by a BPS reconnection, 
as shown in an example in Fig.~\ref{fig:BPSrec-moduli}.
A BPS reconnection 
is always allowed to exhaust all possible configurations, 
and we have to use it to understand the whole structure of the 
moduli space.

Now, we are ready to classify all possible irreducible neutral bions 
in the Grassmann sigma models.
As an example, we take $G_{4,2}$ sigma model, which is 
the simplest Grassmann sigma model. 
Considering the above arguments, we can obtain all irreducible 
neutral bions by combining the BPS fractional instantons with the 
anti-BPS fractional instantons listed in 
Sec.\ref{subsec:frac_inst_grassmann}. 

\begin{figure}[htbp]
\begin{center}
\begin{tabular}{ccc}
   \includegraphics[width=0.33\textwidth]{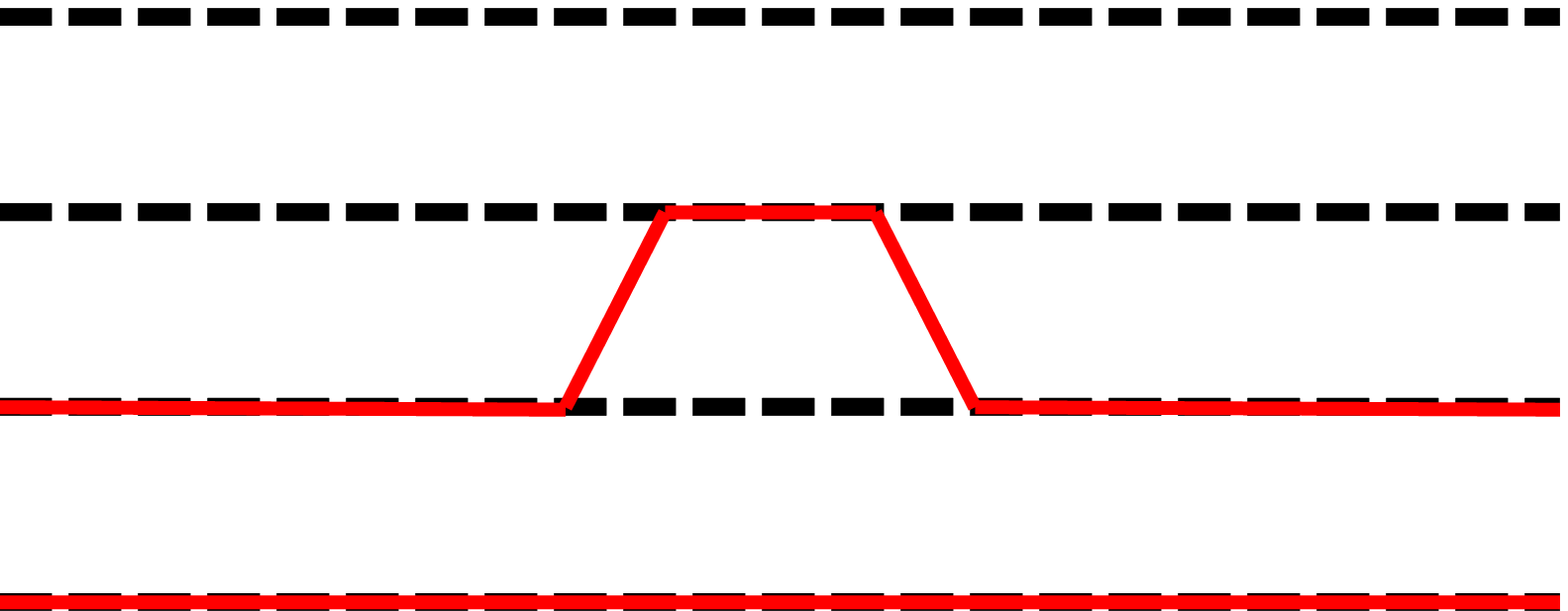} 
& \includegraphics[width=0.33\textwidth]{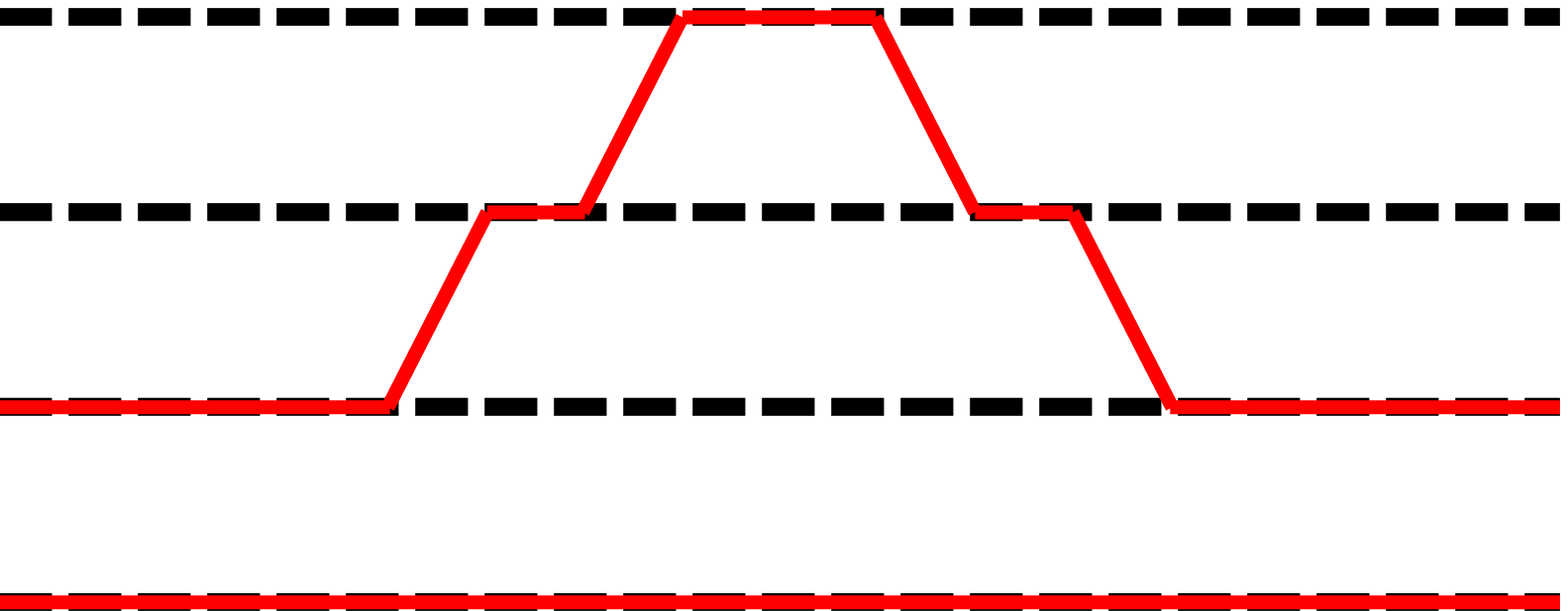}
& \includegraphics[width=0.33\textwidth]{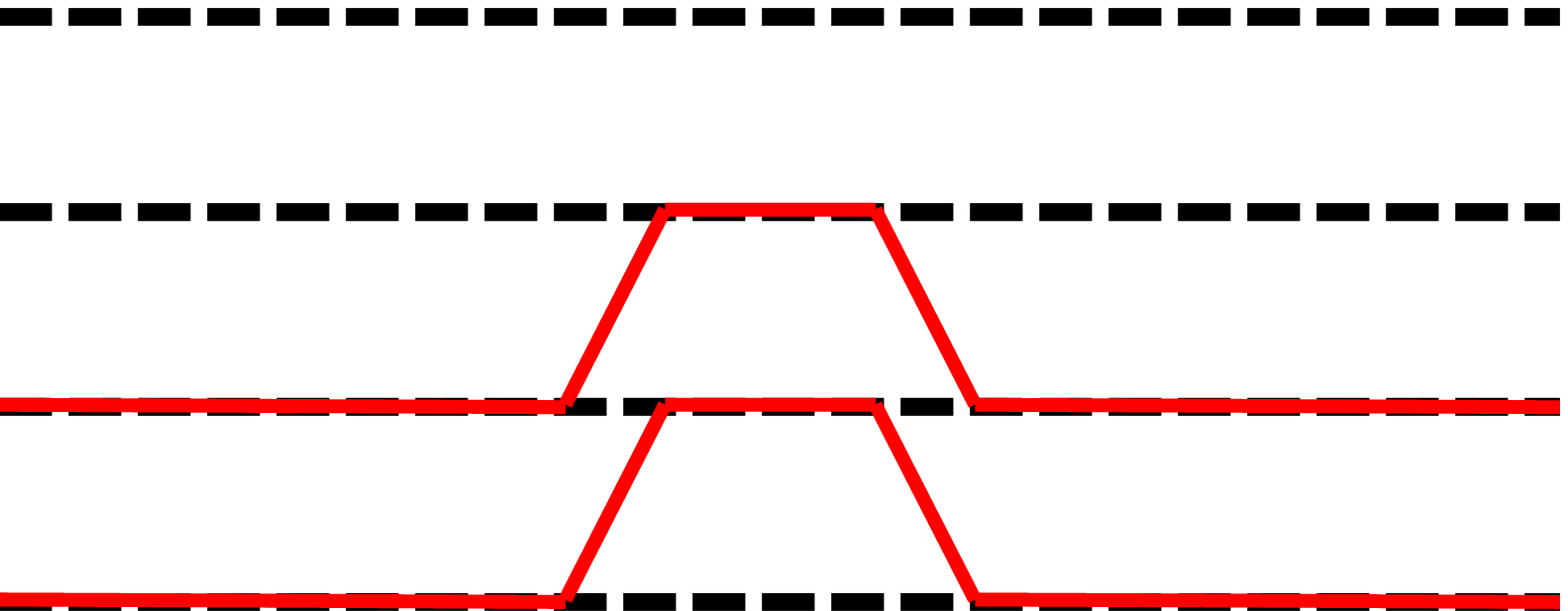}
\\
(a) $(0,1)$ &(b) $(1,1)$ &(c) $(0,2)$\\
   \includegraphics[width=0.33\textwidth]{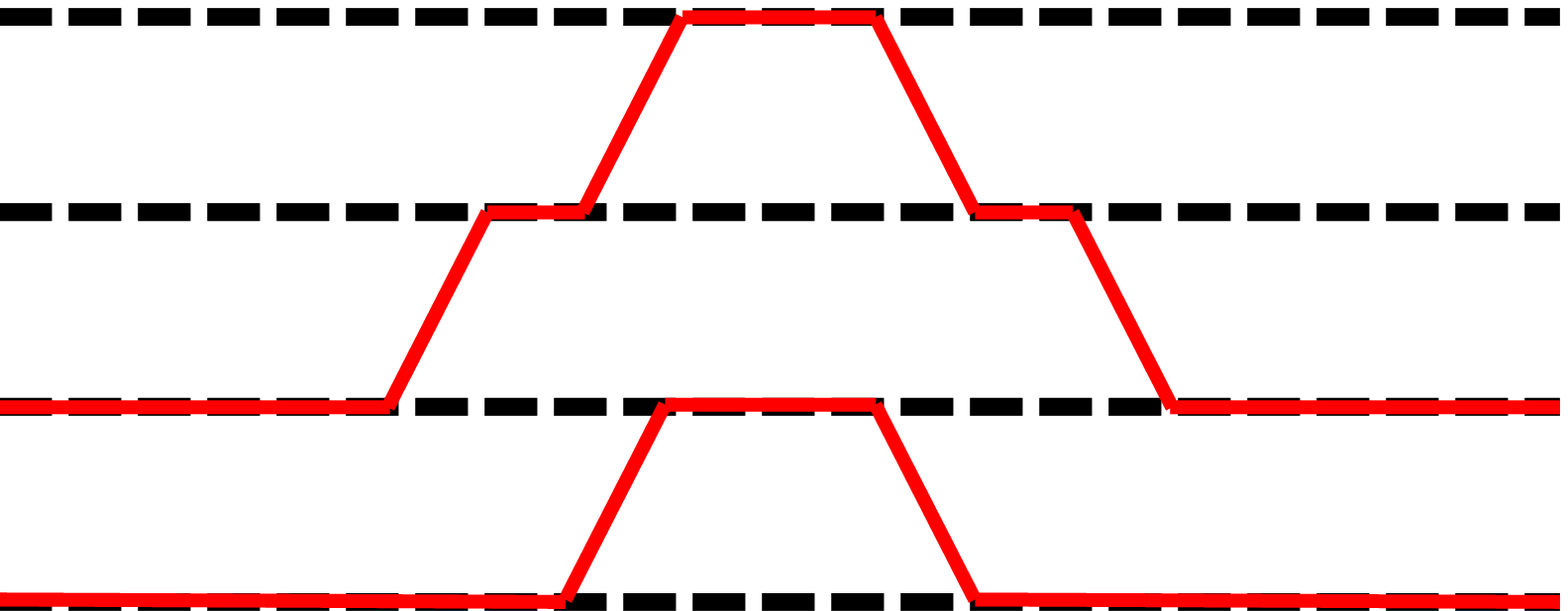} 
& \includegraphics[width=0.33\textwidth]{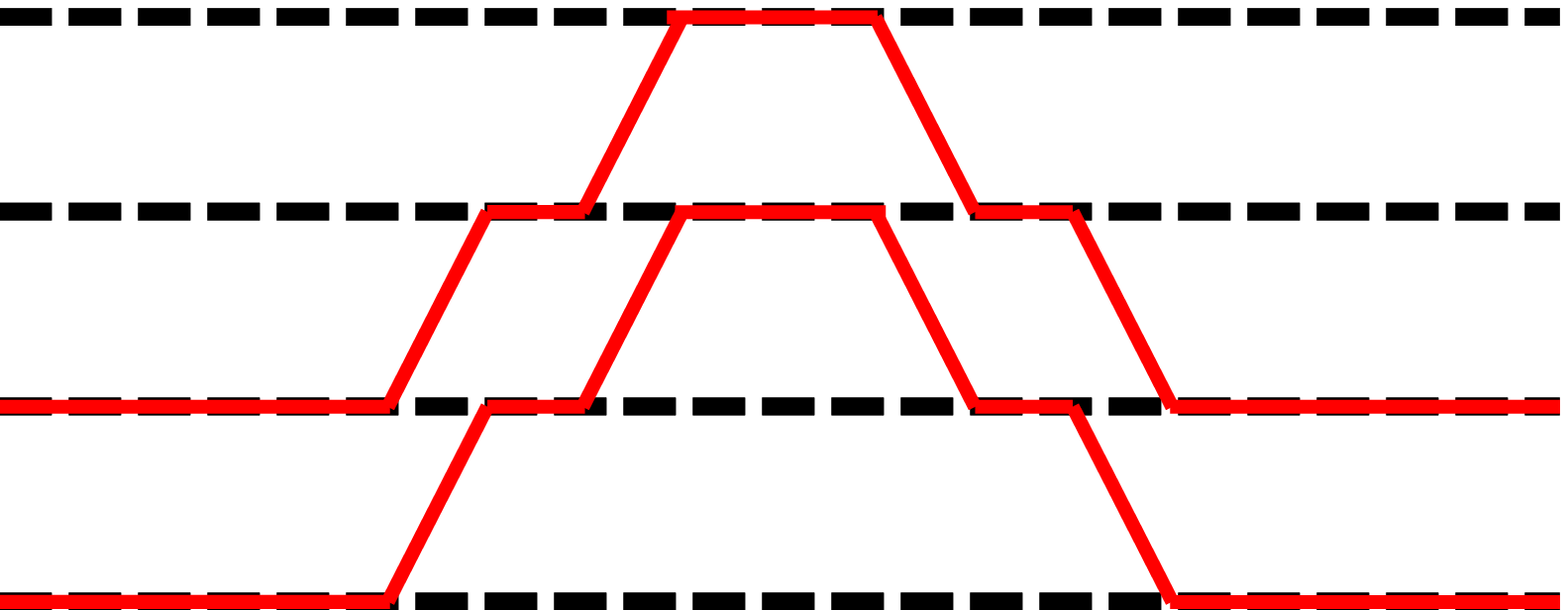}
& 
\\
(d) $(1,2)$ &(e) $(2,2)$ & \\
\end{tabular}
\end{center}
\caption{Neutral bions in $G_{4,2}$, labeled by indices 
(a) $(0,1)$, (b) $(0,2)$, (c) $(1,1)$, (d) $(1,2)$, 
(e) $(2,2)$. 
(a)--(e) are elementary neutral bions in $G_{4,2}$. 
}
\label{fig:Gr42-bion_adjacent_flavor}
\end{figure}

Let us first consider the case of left vacuum being two 
adjacent flavors occupied by color branes. 
On this vacuum, there are five diagrams of the BPS fractional 
instantons in Fig.~\ref{fig:Gr42-fractional_adjacent}(a)-(e). 
The neutral bions obtained from these fractional instantons are 
listed in Fig.~\ref{fig:Gr42-bion_adjacent_flavor}(a)-(e), 
respectively. 

\begin{figure}[htbp]
\begin{center}
 \includegraphics[width=0.55\textwidth]{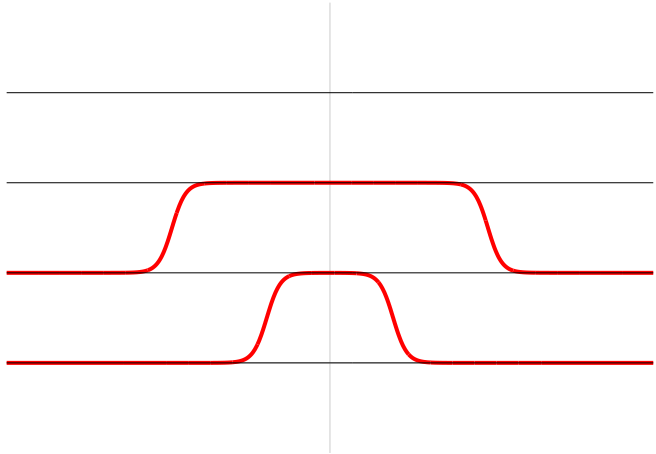}
\end{center}
\caption{Brane configuration calculated from the ansatz 
(\ref{eq:11neutral_42model}) for $\lambda_{1} =10^{-2},\,\lambda_{2}=10^{-2},\,\lambda_{3}=10^{-5}, \,\lambda_{4}=10^{-5},\,N_{\rm F}=4$.}
\label{brane_nb}
\end{figure}

We will write down the moduli matrix of these neutral bions 
explicitly for a more general situation of $Gr_{N_{\rm F}, 2}$ 
sigma model. 
By creating a pair of fractional instanton and anti-instanton 
on the second color brane in the vacuum, we have 
Fig.\ref{fig:Gr42-bion_adjacent_flavor}(a) with a BPS fractional 
instantons $(0,1)$ for the left half of the diagram. 
The moduli matrix for this $(0,1)$ neutral bion for the model 
$Gr_{N_{\rm F}, 2}$ is given by 
\begin{equation}
H_0
 = \left(
\begin{array}{ccccc}
1, & 0, & 0, & 0, & \cdots \\
0, & \lambda_{1}e^{i\theta_{1}}e^{-\frac{2\pi}{N_{\rm F}} z} 
+\lambda_{3}e^{i\theta_{3}}e^{\frac{2\pi}{N_{\rm F}}\bar{z}}, 
& 1, & 0, & \cdots 
\end{array}
\right)\,. 
 \label{eq:01neutral_42model}
\end{equation}
Similarly to the ${\mathbb C}P^{N_{\rm F}-1}$ case, 
fractional instanton is situated where the magnitude of 
two neighboring elements in each row (each color) become equal. 
In Eq.~(\ref{eq:01neutral_42model}), the fractional instanton 
is located at $x^1=\frac{N_{\rm F}}{2\pi}\log\lambda_1$, and fractional 
anti-instanton is at 
$x^1=\frac{N_{\rm F}}{2\pi}\log\frac{1}{\lambda_3}$. 
By creating a pair of fractional instanton and anti-instanton 
between the pair of fractional instanton and anti-instanton 
on the second brane in Fig.\ref{fig:Gr42-bion_adjacent_flavor}(a), 
we obtain Fig.\ref{fig:Gr42-bion_adjacent_flavor}(b) with a 
BPS fractional instantons $(0,2)$ for the left half of the diagram. 
The moduli matrix for this $(0,2)$ neutral bion for the model 
$Gr_{N_{\rm F}, 2}$ is given by 
\begin{equation}
H_0
 = \left(
\begin{array}{cccccc}
1, & 0, & 0, & 0, & 0, & \cdots \\
0, & \lambda_{1}e^{i\theta_{1}}e^{-\frac{2\pi}{N_{\rm F}} 2z} 
+\lambda_{5}e^{i\theta_{5}}e^{\frac{2\pi}{N_{\rm F}}2\bar{z}}, 
 & \lambda_{2}e^{i\theta_{2}}e^{-\frac{2\pi}{N_{\rm F}} z} 
+\lambda_{4}e^{i\theta_{4}}e^{\frac{2\pi}{N_{\rm F}}\bar{z}}, 
& 1, & 0, & \cdots 
\end{array}
\right)\,. 
 \label{eq:02neutral_42model}
\end{equation}
The fractional instantons are located at 
$x^1=\frac{N_{\rm F}}{2\pi}\log\lambda_1, 
\frac{N_{\rm F}}{2\pi}\log\frac{1}{\lambda_2}$, 
and fractional anti-instantons are at 
$x^1=\frac{N_{\rm F}}{2\pi}\log\frac{1}{\lambda_4}, 
\frac{N_{\rm F}}{2\pi}\log\frac{\lambda_4}{\lambda_5}$. 
By creating a pair of fractional instanton and 
anti-instanton on the first color brane in 
Fig.\ref{fig:Gr42-bion_adjacent_flavor}(a), we obtain 
Fig.\ref{fig:Gr42-bion_adjacent_flavor}(c) with a BPS 
fractional instantons $(1,1)$ for the left half of the diagram. 
The moduli matrix for this $(1,1)$ neutral bion for the model 
$Gr_{N_{\rm F}, 2}$ is given by 
\begin{equation}
H_0
 = \left(
\begin{array}{ccccc}
\lambda_{1}e^{i\theta_{1}}e^{-\frac{2\pi}{N_{\rm F}} z} 
+\lambda_{2}e^{i\theta_{2}}e^{\frac{2\pi}{N_{\rm F}}\bar{z}}, 
& 1, & 0, & 0, & \cdots \\
0, & \lambda_{3}e^{i\theta_{3}}e^{-\frac{2\pi}{N_{\rm F}} z} 
+\lambda_{4}e^{i\theta_{4}}e^{\frac{2\pi}{N_{\rm F}}\bar{z}}, 
 & 1, & 0, & \cdots 
\end{array}
\right)\,. 
 \label{eq:11neutral_42model}
\end{equation}
The fractional instantons are located at 
$x^1=\frac{N_{\rm F}}{2\pi}\log\lambda_3, 
\frac{N_{\rm F}}{2\pi}\log\lambda_1$, and fractional 
anti-instantons are at 
$x^1=\frac{N_{\rm F}}{2\pi}\log\frac{1}{\lambda_2},
\frac{N_{\rm F}}{2\pi}\log\frac{1}{\lambda_4}$. 
To visualize the brane diagram for the bion $(1,1)$ given 
by the moduli matrix ansatz (\ref{eq:11neutral_42model}), 
we computed the relative weight of absolute value square of 
each flavor components of moduli matrix for each row 
corresponding to the parameter set $\lambda_{1} =10^{-2},\,
\lambda_{2}=10^{-2},\,\lambda_{3}=10^{-5}, \,
\lambda_{4}=10^{-5},\,N_{\rm F}=4$, and plotted in 
Fig.~\ref{brane_nb}. 
Since $\Sigma$ is almost diagonal when fractional instantons 
are far apart, the relative weight becomes indistinguishable 
with the diagonal elements of $\Sigma$ given in 
Eq.~(\ref{eq:defsigma}). 
This result nicely agrees with the schematic picture in 
Fig.~\ref{fig:Gr42-bion_adjacent_flavor}(c). 

By further creating a pair of fractional instanton and anti-instanton 
between the pair of fractional instanton and anti-instanton 
in the second color brane in Fig.~\ref{fig:Gr42-bion_adjacent_flavor}(c), 
we obtain Fig.~\ref{fig:Gr42-bion_adjacent_flavor}(d) with a 
BPS fractional instantons $(1,2)$ for the left half of the diagram. 
The moduli matrix for this $(1,2)$ neutral bion for the model 
$Gr_{N_{\rm F}, 2}$ is given by 
\begin{eqnarray}
&&H_0  = 
\\
\!\!\!\!&&\!\!\left(
\begin{array}{cccccc
}
\lambda_{1}e^{i\theta_{1}}e^{-\frac{2\pi}{N_{\rm F}} z} 
+\lambda_{3}e^{i\theta_{3}}e^{\frac{2\pi}{N_{\rm F}}\bar{z}}, 
& 1, & 0, & 0, & 0, & \cdots \\
0, & \lambda_{4}e^{i\theta_{4}}e^{-\frac{2\pi}{N_{\rm F}} 2z} 
+\lambda_{8}e^{i\theta_{8}}e^{\frac{2\pi}{N_{\rm F}}2\bar{z}}, 
& \lambda_{5}e^{i\theta_{5}}e^{-\frac{2\pi}{N_{\rm F}} z} 
+\lambda_{7}e^{i\theta_{7}}e^{\frac{2\pi}{N_{\rm F}}\bar{z}}, 
 & 1, & 0, & \cdots 
\end{array}
\right)\,. 
\nonumber
 \label{eq:12neutral_42model}
\end{eqnarray}
The fractional instantons are located at 
$x^1=\frac{N_{\rm F}}{2\pi}\log\frac{\lambda_4}{\lambda_5},
\frac{N_{\rm F}}{2\pi}\log\lambda_5, 
\frac{N_{\rm F}}{2\pi}\log\lambda_1$, and fractional 
anti-instantons are at 
$x^1=\frac{N_{\rm F}}{2\pi}\log\frac{1}{\lambda_3},
\frac{N_{\rm F}}{2\pi}\log\frac{1}{\lambda_7},
\frac{N_{\rm F}}{2\pi}\log\frac{\lambda_7}{\lambda_8}$. 
If we further create  a pair of fractional instanton and 
anti-instanton between the innermost pair of fractional 
instanton and anti-instanton in the first color brane in 
Fig.\ref{fig:Gr42-bion_adjacent_flavor}(d), we obtain 
Fig.\ref{fig:Gr42-bion_adjacent_flavor}(e) with a BPS 
fractional instantons $(2,2)$ for the left half of the diagram. 
The moduli matrix for this $(2,2)$ neutral bion for the model 
$Gr_{N_{\rm F}, 2}$ is given by 
\begin{eqnarray}
H_0 = 
\left(
\begin{array}{cc
}
\lambda_{1}e^{i\theta_{1}}e^{-\frac{2\pi}{N_{\rm F}} 2z} 
+\lambda_{5}e^{i\theta_{5}}e^{\frac{2\pi}{N_{\rm F}}2\bar{z}}, 
& 0 \\
 \lambda_{2}e^{i\theta_{2}}e^{-\frac{2\pi}{N_{\rm F}} z} 
+\lambda_{4}e^{i\theta_{4}}e^{\frac{2\pi}{N_{\rm F}}\bar{z}}, 
& \lambda_{6}e^{i\theta_{6}}e^{-\frac{2\pi}{N_{\rm F}} 2z} 
+\lambda_{10}e^{i\theta_{10}}e^{\frac{2\pi}{N_{\rm F}}2\bar{z}}\\
1, & \lambda_{7}e^{i\theta_{7}}e^{-\frac{2\pi}{N_{\rm F}} z} 
+\lambda_{9}e^{i\theta_{9}}e^{\frac{2\pi}{N_{\rm F}}\bar{z}}\\
0& 1 \\
 0, & 0\\
\vdots & \vdots 
\end{array}
\right)^T\,. 
 \label{eq:22neutral_42model}
\end{eqnarray}
The fractional instantons are located at 
$x^1=\frac{N_{\rm F}}{2\pi}\log\frac{\lambda_6}{\lambda_7},
\frac{N_{\rm F}}{2\pi}\log\frac{\lambda_1}{\lambda_2},
\frac{N_{\rm F}}{2\pi}\log\lambda_7, 
\frac{N_{\rm F}}{2\pi}\log\lambda_2$, and fractional 
anti-instantons are at 
$x^1=\frac{N_{\rm F}}{2\pi}\log\frac{1}{\lambda_4},
\frac{N_{\rm F}}{2\pi}\log\frac{1}{\lambda_9},
\frac{N_{\rm F}}{2\pi}\log\frac{\lambda_4}{\lambda_5},
\frac{N_{\rm F}}{2\pi}\log\frac{\lambda_9}{\lambda_{10}}$.

\begin{figure}[htbp]
\begin{center}
\begin{tabular}{ccc}
   \includegraphics[width=0.33\textwidth]{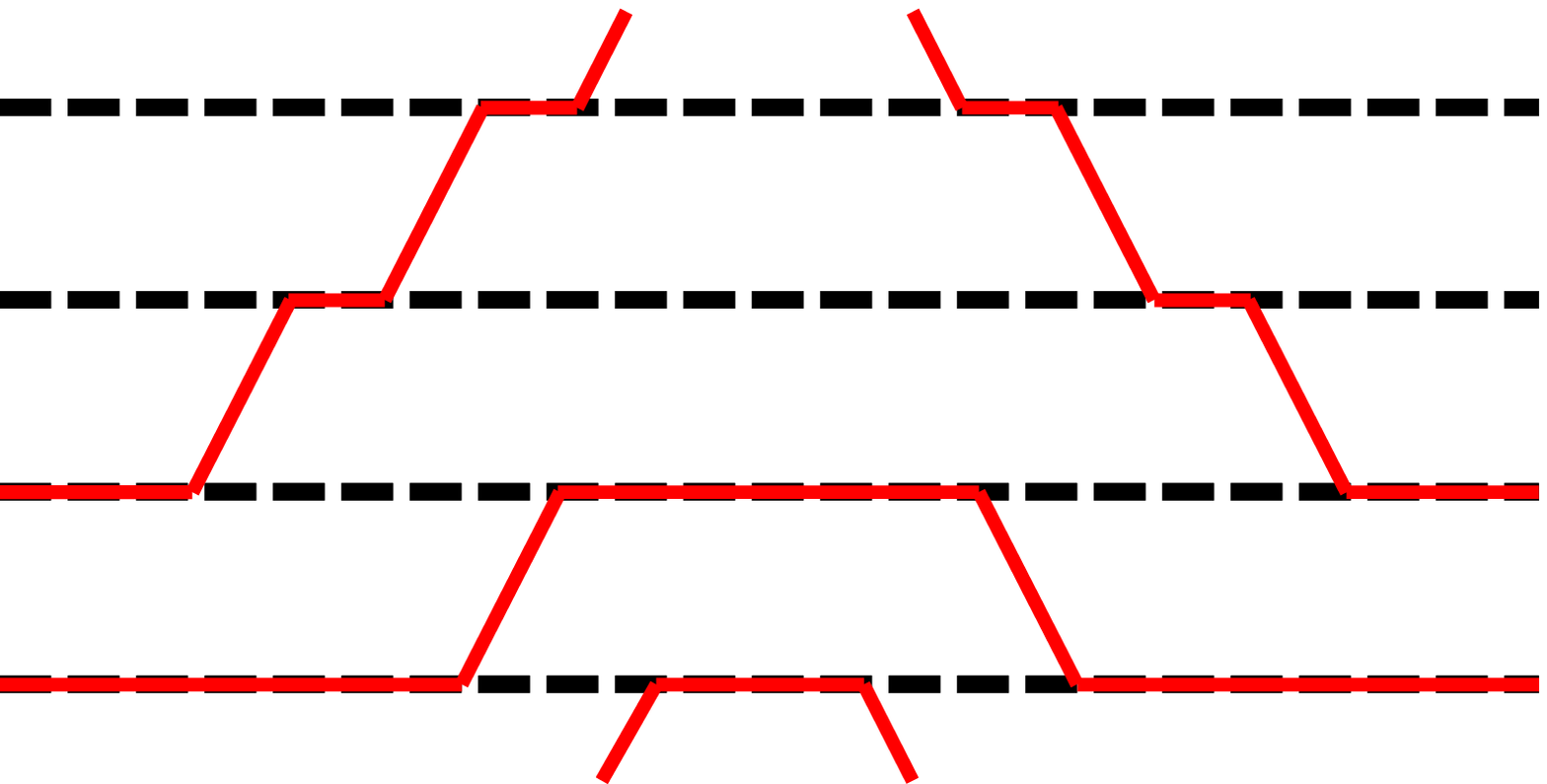}
& \includegraphics[width=0.33\textwidth]{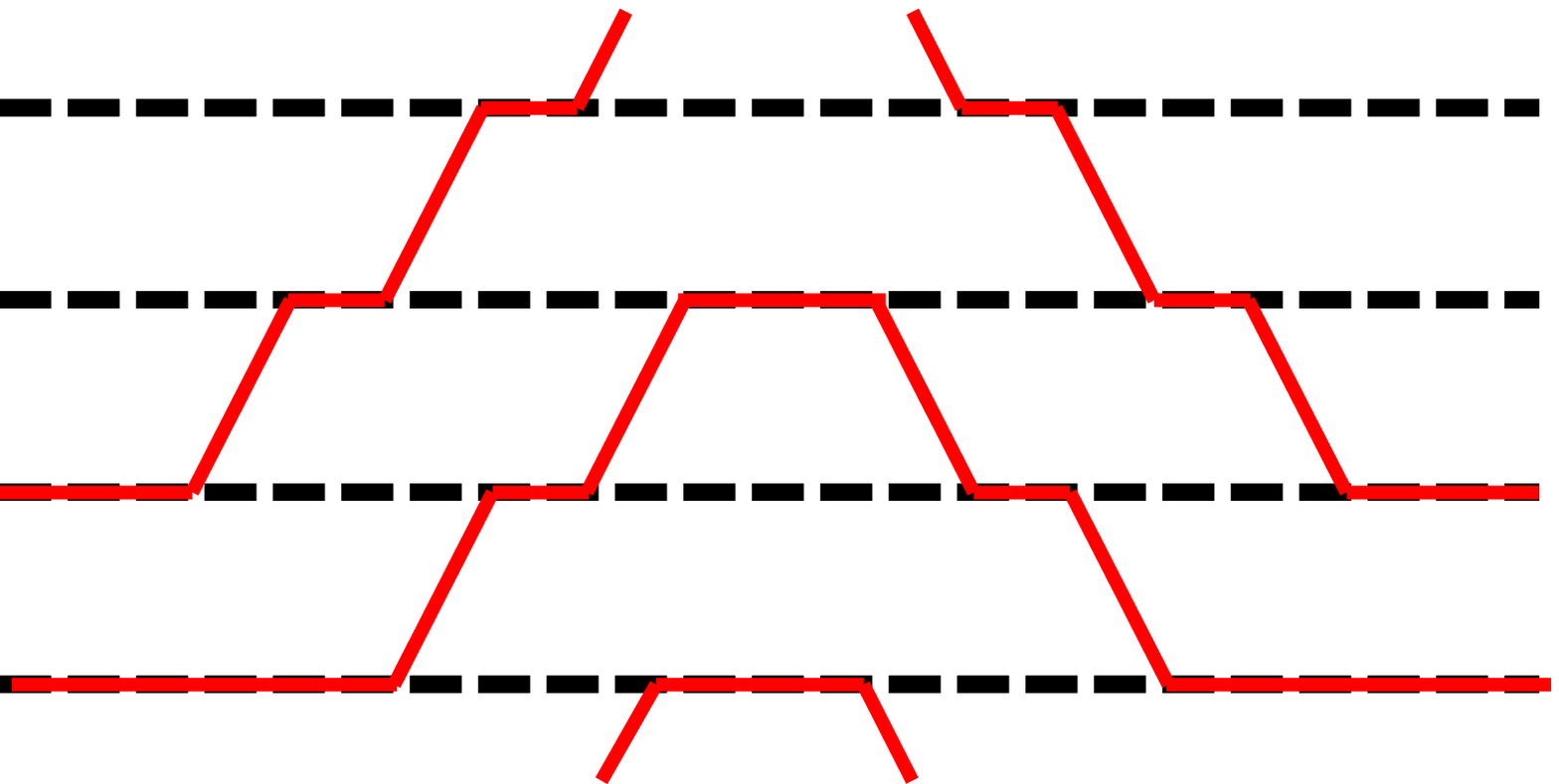} 
& \includegraphics[width=0.33\textwidth]{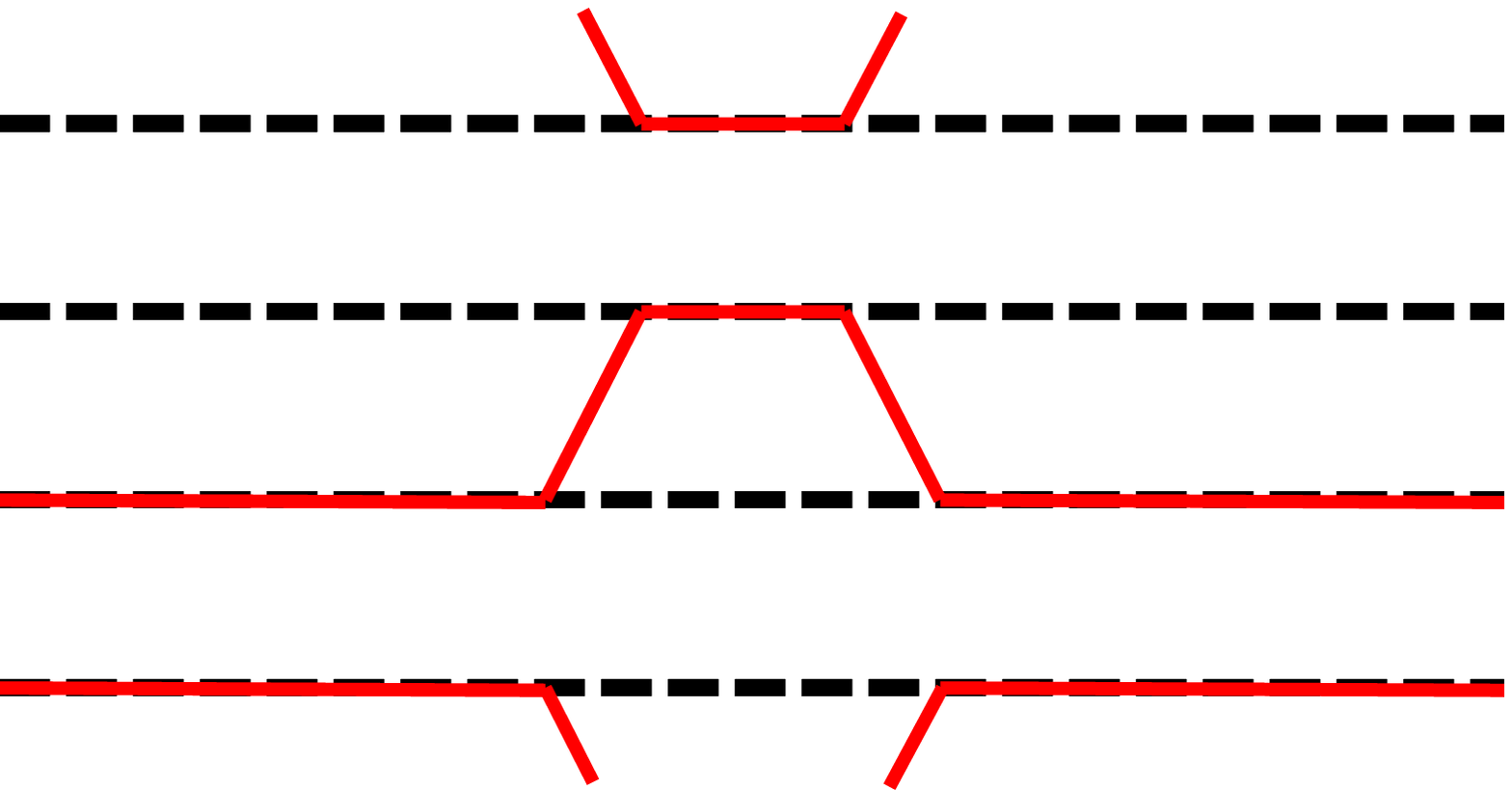}
\\
(a) $(1,3)$ &(b) $(2,3)$ &(c) $(-1,1)$\\
   \includegraphics[width=0.33\textwidth]{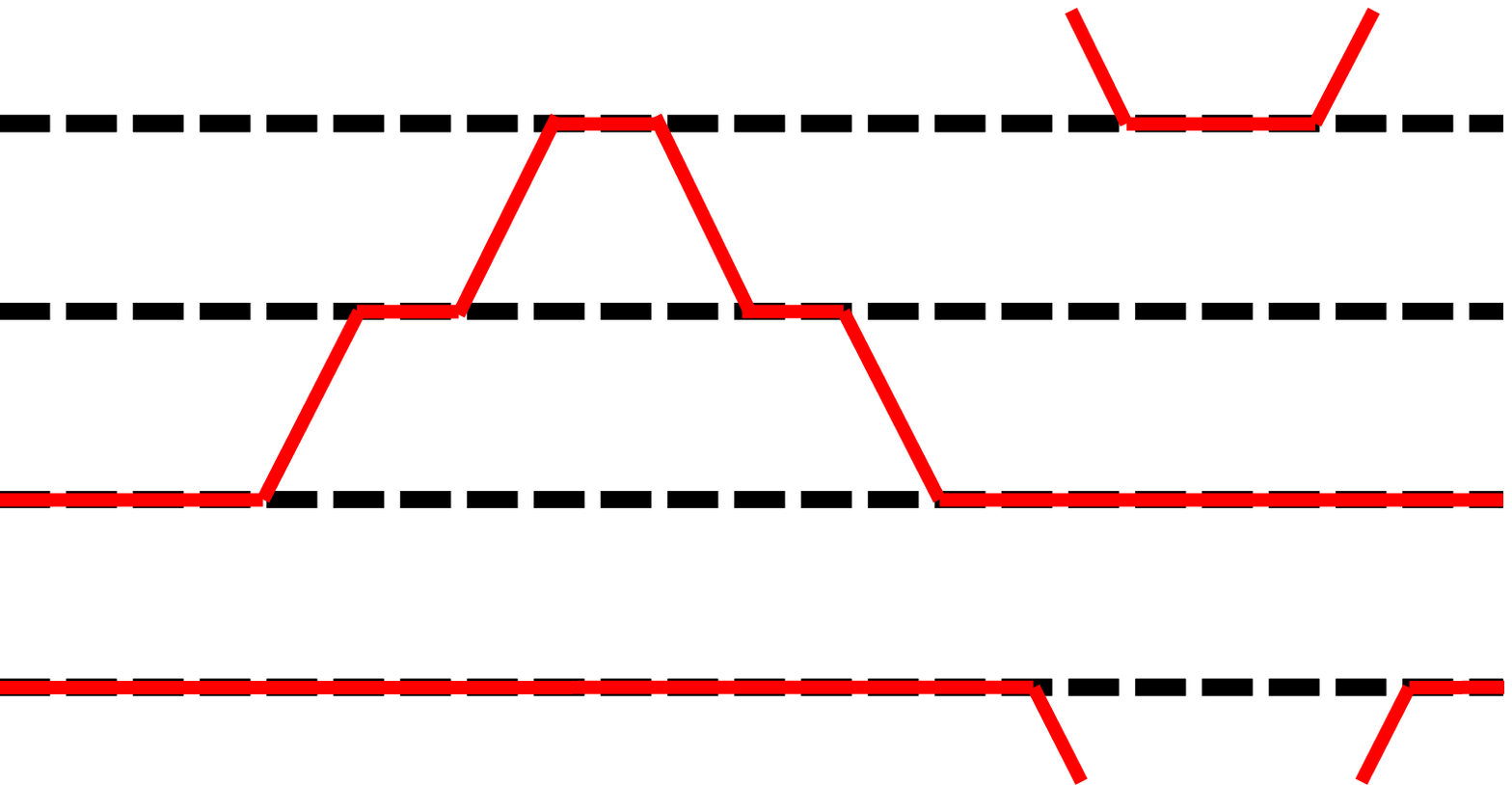}
& \includegraphics[width=0.33\textwidth]{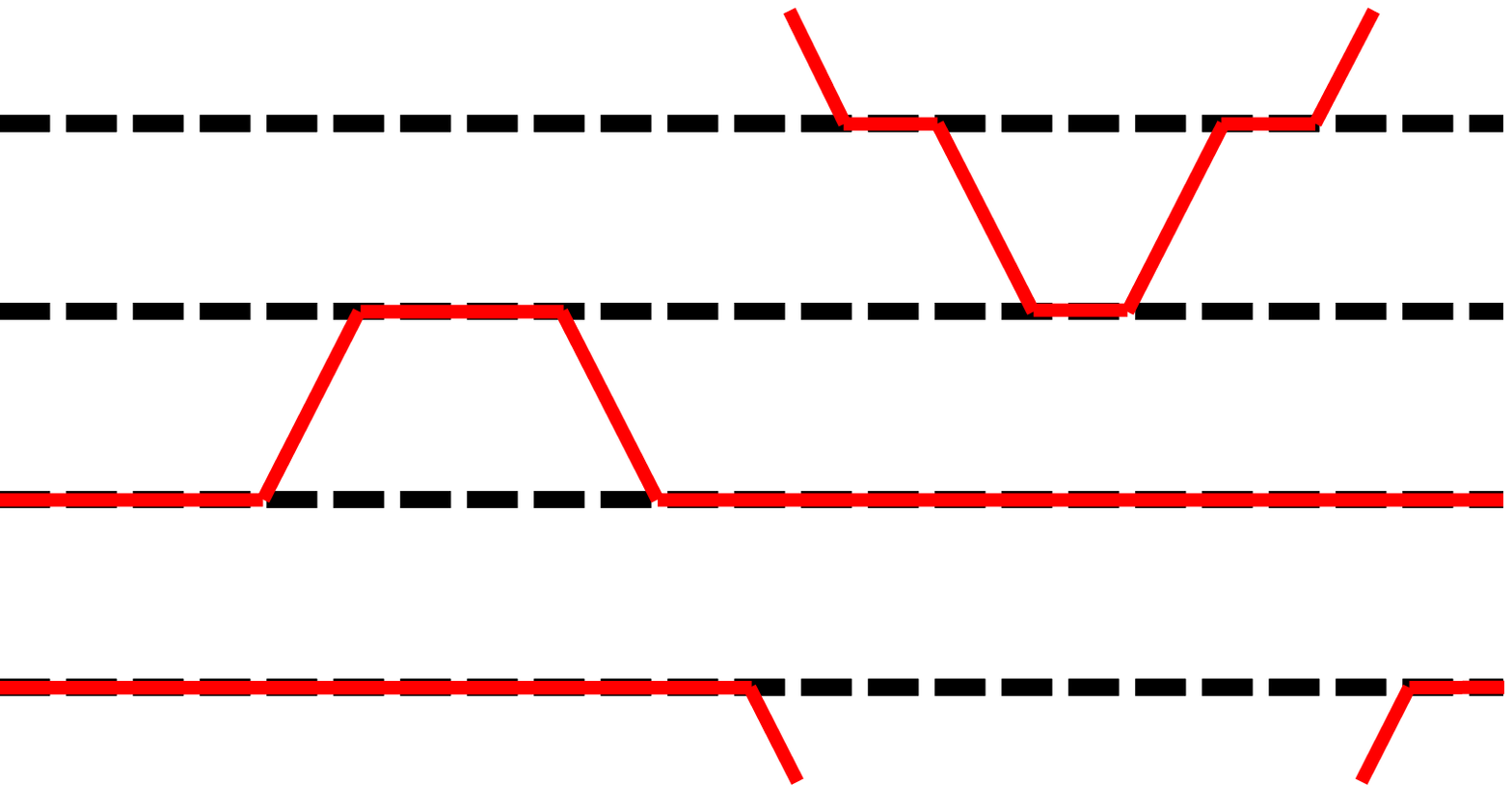} 
& \\
(d) $(-1,2)$ &(e) $(-2,1)$ & \\
\end{tabular}
\end{center}
\caption{Neutral bions in $G_{4,2}$, labeled by indices 
(a) $(1,3)$, (b) $(2,3)$, (c) $(-1,1)$, (d) $(-1,2)$, and  
(e) $(-2,1)$, are not neutral bions in $G_{4,2}$. 
(a) $(1,3)$ is a pair of instanton and anti-instanton 
and is not a bion.
(b) $(2,3)$ is not an elementary bion anymore 
because of the decomposition
$(2,3) = (1,3) + (1,0)$, where the former is 
an instanton-anti-instanton pair. 
(c) is composite.
(d) and (e) are not elementary neutral bion which can 
be understood in dual pictures.
}
\label{fig:Gr42-bion_non-bion}
\end{figure}
There are other dagrams as a composite of BPS solutions and 
anti-BPS solutions with the set of higher fractional instanton 
numbers $(k_1,k_2)$, but they contain at least one instanton 
and are not bions anymore, or reducible diagrams.  
For instance, if we create  a pair of fractional instanton and 
anti-instanton between the innermost pair of fractional 
instanton and anti-instanton in the second color brane of $(1,2)$ 
in Fig.~\ref{fig:Gr42-bion_adjacent_flavor}(d), we obtain 
Fig.~\ref{fig:Gr42-bion_non-bion}(a) 
with a BPS configuration $(1,3)$ for the left half of the diagram. 
Since this left half of the brane diagram is the BPS instanton 
$(1,3)$ in Fig.~\ref{fig:Gr42-fractional_adjacent}(f), 
Fig.~\ref{fig:Gr42-bion_non-bion}(a) represents 
a pair of instanton and anti-instanton 
with the unit instanton charges. 
The moduli matrix for this $(1,3)$ instanton-anti-instanton 
pair for the model 
$Gr_{N_{\rm F}, 2}$ is given by 
\begin{eqnarray}
H_0 = 
\left(
\begin{array}{cc
}
\lambda_{1}e^{i\theta_{1}}e^{-\frac{2\pi}{N_{\rm F}} z} 
+\lambda_{3}e^{i\theta_{3}}e^{\frac{2\pi}{N_{\rm F}}\bar{z}}, 
& 0 \\
 1, & \lambda_{4}e^{i\theta_{4}}e^{-\frac{2\pi}{N_{\rm F}} 3z} 
+\lambda_{10}e^{i\theta_{10}}e^{\frac{2\pi}{N_{\rm F}}3\bar{z}} 
\\
0, & \lambda_{5}e^{i\theta_{5}}e^{-\frac{2\pi}{N_{\rm F}} 2z} 
+\lambda_{9}e^{i\theta_{9}}e^{\frac{2\pi}{N_{\rm F}}2\bar{z}}
\\
 0, &\lambda_{6}e^{i\theta_{6}}e^{-\frac{2\pi}{N_{\rm F}} z} 
+\lambda_{8}e^{i\theta_{8}}e^{\frac{2\pi}{N_{\rm F}}\bar{z}}
\\
 0, & 1 
\\
0, & 0 \\
\vdots & \vdots 
\end{array}
\right)^T \,. 
 \label{eq:13neutral_42model}
\end{eqnarray}
The fractional instantons are located at 
$x^1=\frac{N_{\rm F}}{2\pi}\log\frac{\lambda_6}{\lambda_7},
\frac{N_{\rm F}}{2\pi}\log\frac{\lambda_1}{\lambda_2},
\frac{N_{\rm F}}{2\pi}\log\lambda_7, 
\frac{N_{\rm F}}{2\pi}\log\lambda_2$, and fractional 
anti-instantons are at 
$x^1=\frac{N_{\rm F}}{2\pi}\log\frac{1}{\lambda_4},
\frac{N_{\rm F}}{2\pi}\log\frac{1}{\lambda_9},
\frac{N_{\rm F}}{2\pi}\log\frac{\lambda_4}{\lambda_5},
\frac{N_{\rm F}}{2\pi}\log\frac{\lambda_9}{\lambda_{10}}$.

The brane diagram in Fig.~\ref{fig:Gr42-bion_non-bion}(b) is 
characterized by $(2,3)$, and is a reducible neutral bion diagram 
because of the decomposition 
$(2,3) = (1,3) + (1,0)$, namely it is a composite of an 
instanton-anti-instanton pair $(1,3)$ and a feactional instanton 
$(1,0)$. 
Fig.~\ref{fig:Gr42-bion_non-bion}(c) is characterized by $(-1,1)$, 
and is constructed by placing 
in the left half of the diagram the exact non-BPS charged bion 
solution in Fig.~\ref{fig:Gr42-bion_adjacent_flavor}(g).  
This is a reducible neutral bion diagram. 
In general, all the exact non-BPS solutions that we can construct 
in moduli matrix formalism is the case of composite of 
non-interacting BPS and anti-BPS fractional instantons. 
Therefore neutral bions constructible from these non-BPS exact 
solutions are always reducible. 
The brane diagram in Fig.~\ref{fig:Gr42-bion_non-bion} (d) 
and (e) are characterized by $(-1,2)$, and $(-2,1)$, 
respectively. They are reducible neutral bion diagrams 
because they contain a configuration in Fig.~\ref{fig:composite}.

Lastly, let us consider the other case of left vacuum being 
two non-adjacent flavor branes occupied by color branes. 
We can obtain neutral bions by combining BPS fractional 
instantons in Fig.~\ref{fig:Gr42-fractional_nonadjacent}(a)-(d) 
with corresponding anti-BPS fractional instantons. 
\begin{figure}[htbp]
\begin{center}
\begin{tabular}{cc}
   \includegraphics[width=0.33\textwidth]{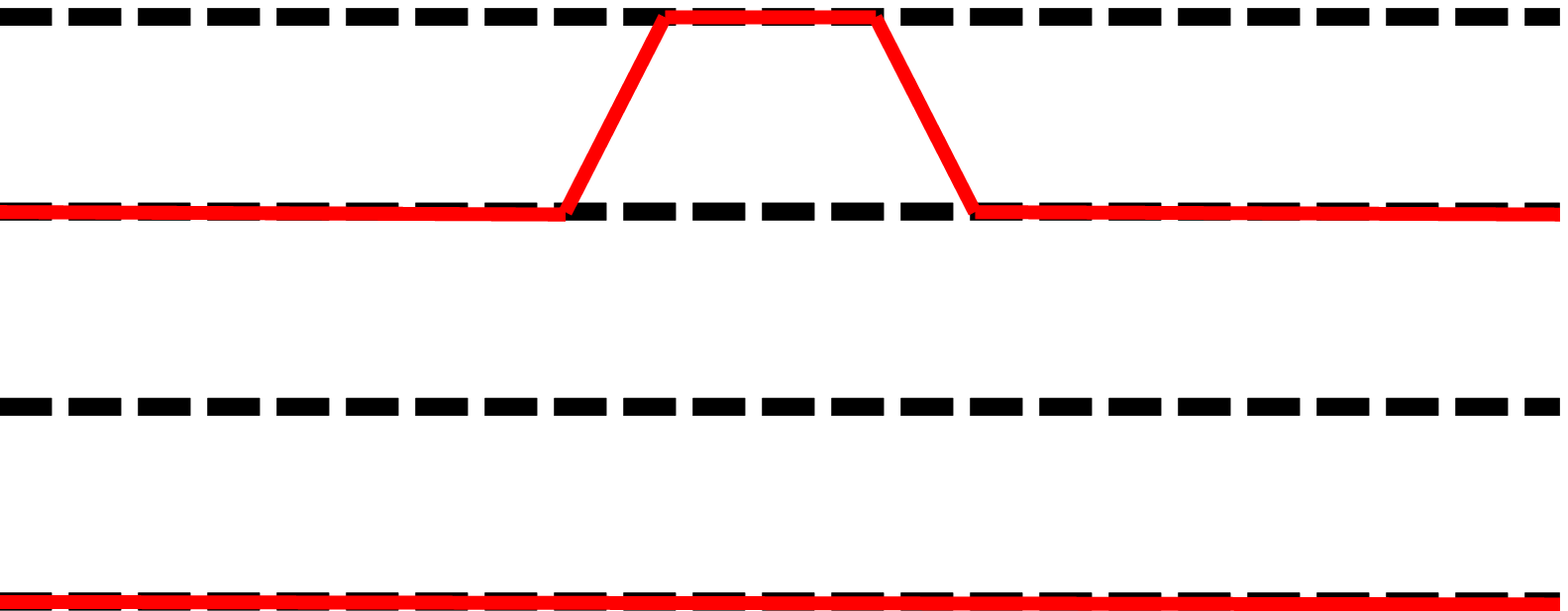}
& \includegraphics[width=0.33\textwidth]{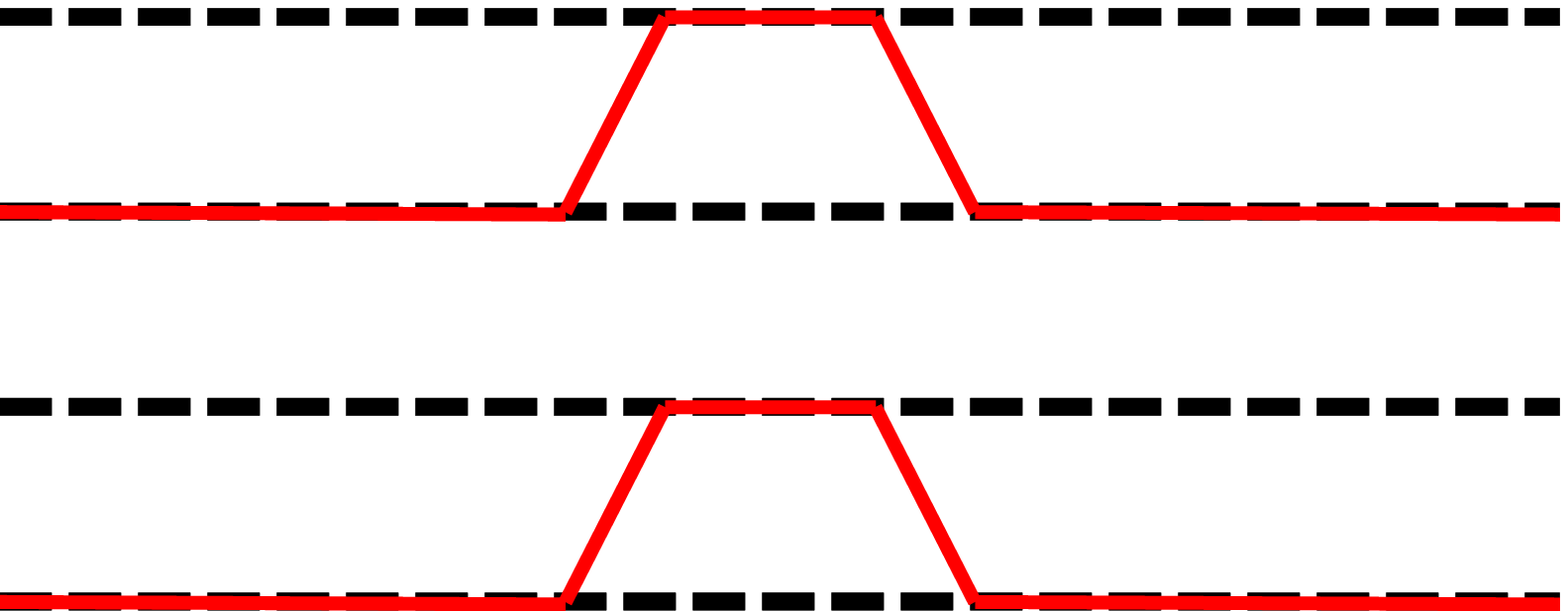}\\
(a)  $(0,1)$ & (b) $(1,1)$\\
   \includegraphics[width=0.33\textwidth]{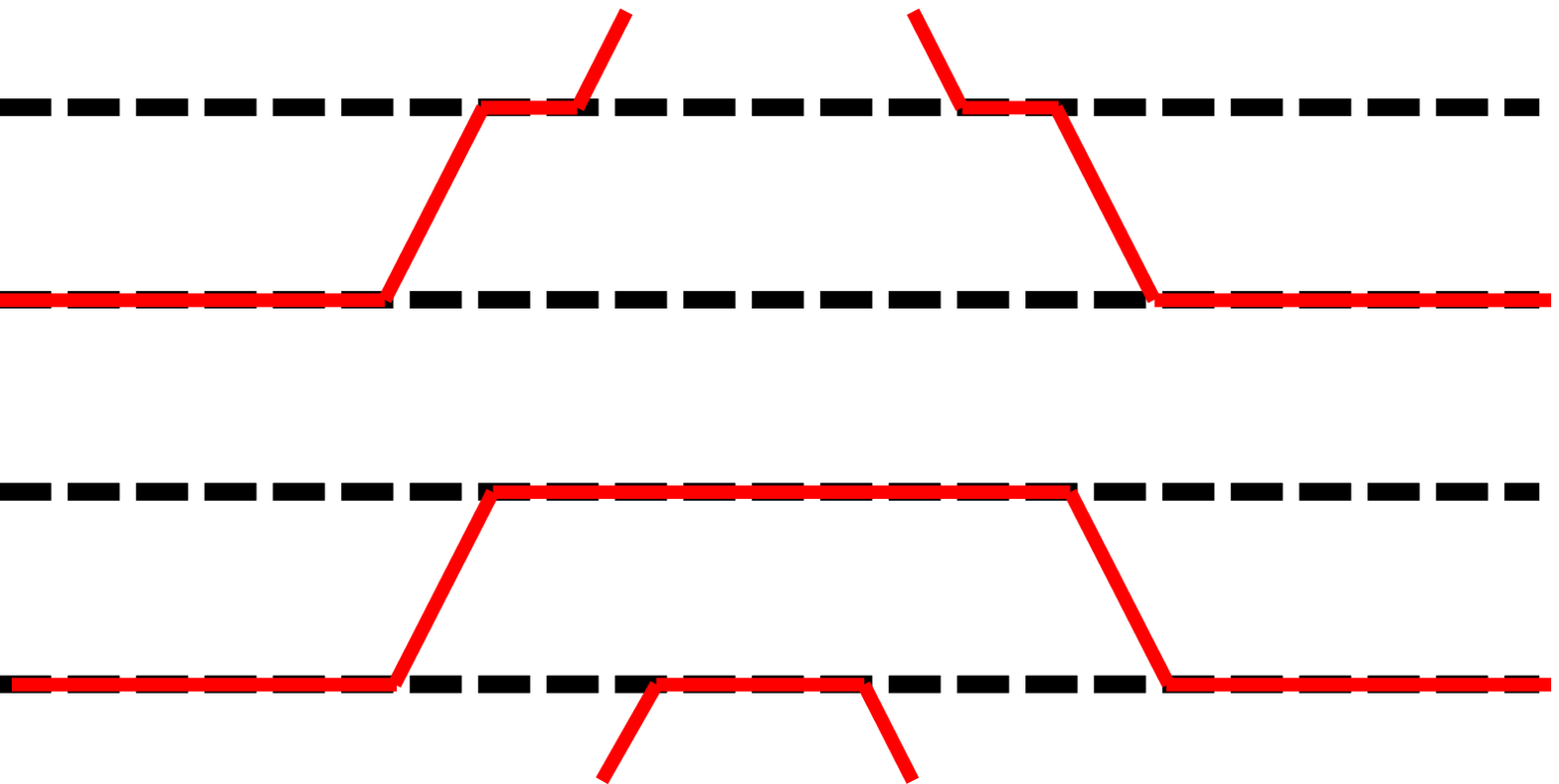} 
& \includegraphics[width=0.33\textwidth]{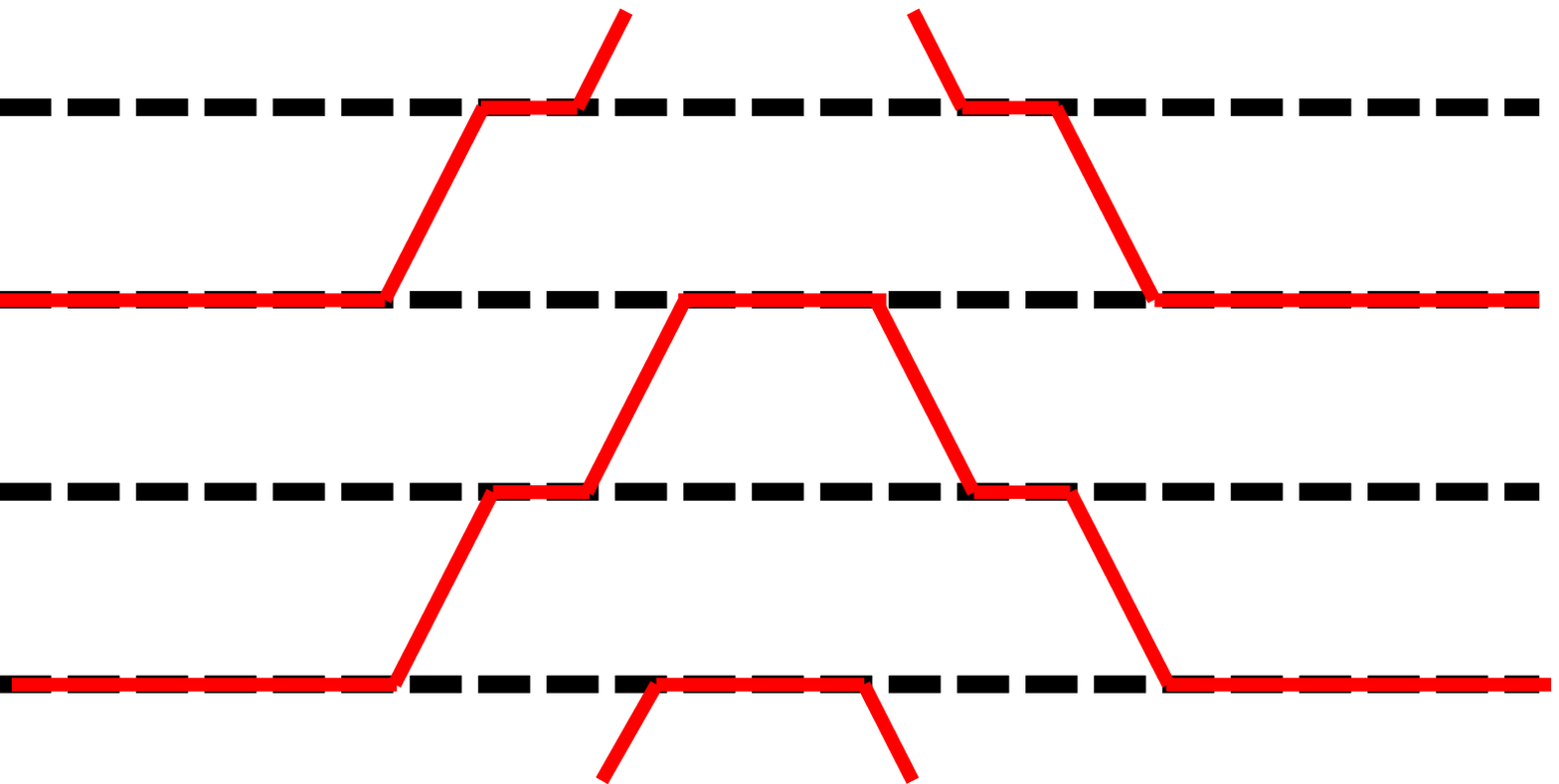}
\\
(c) $(1,2)$ & (d) $(2,2)$  \\
\end{tabular}
\end{center}
\caption{Neutral bions in $G_{4,2}$ for the left vacuum with 
non-adjacent flavors occupied by colors, labeled by indices 
(a) $(0,1)$, (b) $(1,1)$, (c) $(1,2)$,  and  
(d) $(2,2)$. 
(a) $(0,1)$ is an elementary neutral bion in $G_{4,2}$. 
(b) $(1,1)$ is a reducible diagram. 
(c) $(1,2)$ is a neutral bion. 
(d) $(2,2)$ is a composite of a pair of an instanton and an 
anti-instanton and is not a neutral bion anymore. 
}
\label{fig:Gr42-bion_nonadjacent}
\end{figure}
The elementary neutral bion in Fig.\ref{fig:Gr42-bion_nonadjacent}(a) 
with an elementary BPS fractional instantons $(0,1)$ for the 
left half of the diagram is given by the moduli matrix 
\begin{equation}
H_0
 = \left(
\begin{array}{cccccc}
1, & 0, & 0, & 0, & 0 &\cdots \\
0, & 0, & \lambda_{1}e^{i\theta_{1}}e^{-\frac{2\pi}{N_{\rm F}} z} 
+\lambda_{3}e^{i\theta_{3}}e^{\frac{2\pi}{N_{\rm F}}\bar{z}}, 
& 1, & 0, & \cdots 
\end{array}
\right)\,. 
 \label{eq:01neutral_42model-na}
\end{equation}
The fractional instanton is located at 
$x^1=\frac{N_{\rm F}}{2\pi}\log\lambda_1$, and fractional 
anti-instanton is at 
$x^1=\frac{N_{\rm F}}{2\pi}\log\frac{1}{\lambda_3}$. 
Two color lines of $(1,1)$ in Fig.\ref{fig:Gr42-bion_nonadjacent}(b) 
do not share any common flavor lines. This is the case of 
composite of two non-interacting elementary fractional 
instantons whose moduli matrix can be given as 
\begin{equation}
H_0
 = \left(
\begin{array}{cccccc}
\lambda_{3}e^{i\theta_{3}}e^{-\frac{2\pi}{N_{\rm F}} z} 
+\lambda_{4}e^{i\theta_{4}}e^{\frac{2\pi}{N_{\rm F}}\bar{z}}, 
& 1, & 0, & 0, & 0 &\cdots \\
0, & 0, & \lambda_{1}e^{i\theta_{1}}e^{-\frac{2\pi}{N_{\rm F}} z} 
+\lambda_{2}e^{i\theta_{2}}e^{\frac{2\pi}{N_{\rm F}}\bar{z}}, 
& 1, & 0, & \cdots 
\end{array}
\right)\,. 
 \label{eq:11neutral_42model-na}
\end{equation}
The neutral bion in Fig.\ref{fig:Gr42-bion_nonadjacent}(c) 
with a BPS fractional instantons $(1,2)$ for the left half 
of the diagram is given by the moduli matrix 
\begin{equation}
H_0
 = \left(
\begin{array}{cc
}
\lambda_{1}e^{i\theta_{1}}e^{-\frac{2\pi}{N_{\rm F}} z} 
+\lambda_{3}e^{i\theta_{3}}e^{\frac{2\pi}{N_{\rm F}}\bar{z}}, 
& 0 \\
1, & 0 \\
0 & \lambda_{4}e^{i\theta_{4}}e^{-\frac{2\pi}{N_{\rm F}} 2z} 
+\lambda_{8}e^{i\theta_{8}}e^{\frac{2\pi}{N_{\rm F}}2\bar{z}} \\
0, & \lambda_{5}e^{i\theta_{5}}e^{-\frac{2\pi}{N_{\rm F}} z} 
+\lambda_{7}e^{i\theta_{7}}e^{\frac{2\pi}{N_{\rm F}}\bar{z}} \\
0, & 1 \\
0, & 0, \\
\vdots & \vdots 
\end{array}
\right)^T\,. 
 \label{eq:12neutral_42model-na}
\end{equation}
The fractional instantons are located at 
$x^1=\frac{N_{\rm F}}{2\pi}\log\frac{\lambda_4}{\lambda_5},
\frac{N_{\rm F}}{2\pi}\log\lambda_5, 
\frac{N_{\rm F}}{2\pi}\log\lambda_1$, and fractional 
anti-instantons are at 
$x^1=\frac{N_{\rm F}}{2\pi}\log\frac{1}{\lambda_3},
\frac{N_{\rm F}}{2\pi}\log\frac{1}{\lambda_7},
\frac{N_{\rm F}}{2\pi}\log\frac{\lambda_7}{\lambda_8}$. 
One can think of another possibility of $(2,1)$, but this 
case is equivalent to $(1,2)$ case and is not listed here.

Summarizing, all possible types of irreducible neutral bion 
configurations for $G_{4,2}$ are exhaustively listed as eight 
possibilities in Figs.~\ref{fig:Gr42-bion_adjacent_flavor}(a)-(e), 
and \ref{fig:Gr42-bion_nonadjacent}(a)-(c). 
With the same procedures, we can exhaust 
all possible irreducible  neutral bions in 
the general $G_{\NF,\NC}$ model.

\begin{figure}[htbp]
\begin{center}
 \includegraphics[width=0.4\textwidth]{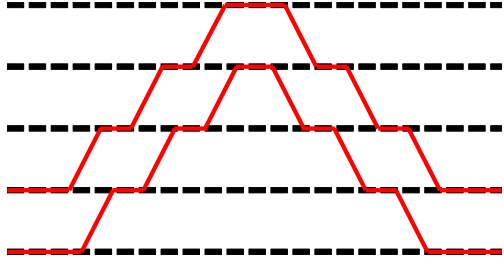}
\end{center}
\caption{An irreducible neutral bion with the instanton charge larger than one 
in $G_{5,2}$. 
}
\label{fig:gr52}
\end{figure}

In this smallest Grassmann sigma model $Gr_{4,2}$, we encountered a 
neutral bion whose constituent fractional instanton 
has unit instanton charge but is different from genuine 
instanton, as shown in Fig.~\ref{fig:Gr42-bion_adjacent_flavor}(e). 
For larger Grassmann sigma models, 
there exist neutral bions with the instanton charge 
even larger than one. 
In Fig.~\ref{fig:gr52}, we show a neutral bion in  $G_{5,2}$, 
half of which has the instanton charge $6/5$ greater than one.
This cannot be decomposed into an instanton and the rest.
This kind of phenomena arises due to an increasingly large number of 
different species of fractional instantons as 
$N_{\rm F}, N_{\rm C}$ increases. 

\subsection{Charged bions in the Grassmann sigma models}

Charged bions have no instanton charge in total but 
non-zero vector 
of fractional 
instanton numbers for the whole field configuration. 
Although the ${\mathbb C}P^{N_{\rm F}-1}$ models do not 
admit charged bions, the Grassmann sigma models do.

For a given charged bion configuration, 
it is always possible to insert neutral bions.
Namely, as classification of charged bions,  
we remove all neutral bions to 
obtain ``purely charged bions."
Insertion of neutral bions in general gives physically distinct 
contributions, but we can choose the representative of 
cohomology classes as purely charged bions. 
Thus we obtain a constraint for purely charged bions that 
each color brane must be either BPS or anti-BPS. 
This drastically simplifies the classification.
We label purely charged bions by a vector of the number 
of fractional instantons $n_a$ for the $a$-th color brane 
\beq 
   [n_1,n_2,\cdots, n_{\NC}]
\eeq 
 with the constraint 
\beq
 0 = \sum_{a=1}^{\NC} n_a ,
\eeq
where we use the notation $[\;\; ]$ to distinguish it from 
the vector $(\;\; )$ specifing the BPS fractional instantons and 
neutral bions made from them.

\begin{figure}[htbp]
\begin{center}
\begin{tabular}{cc}
 \includegraphics[width=0.4\textwidth]{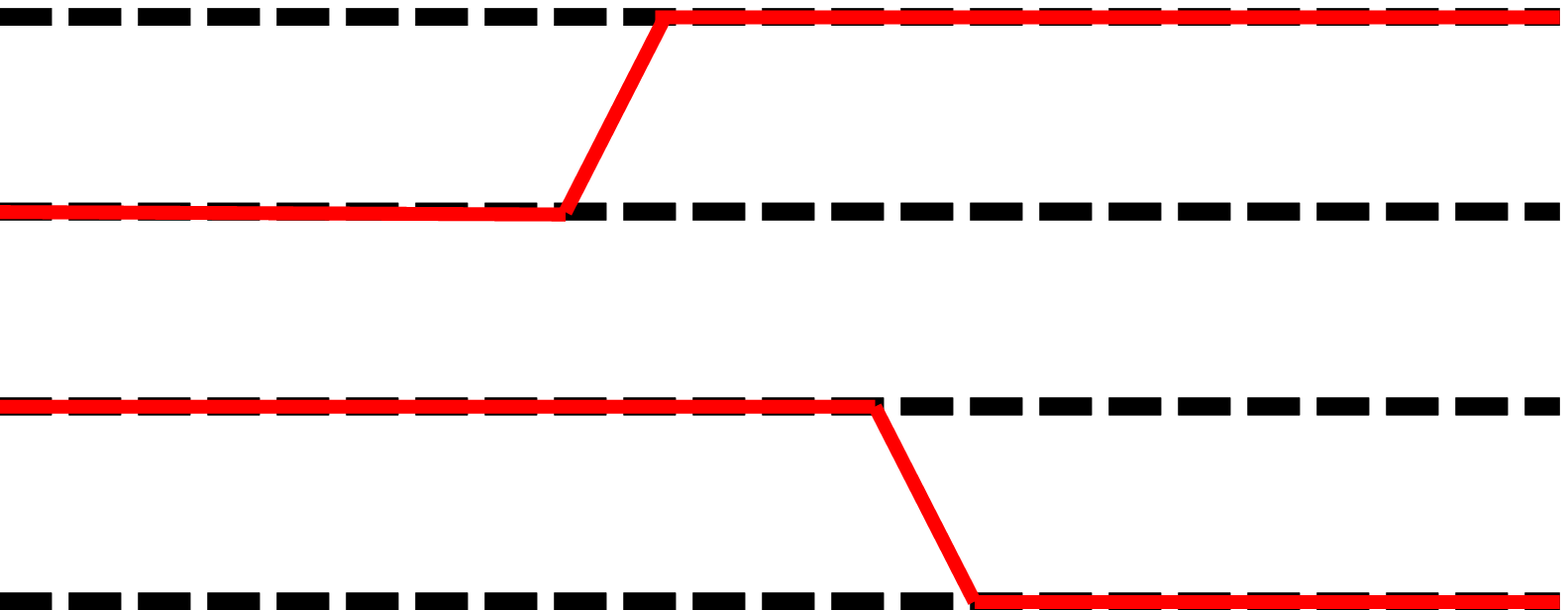}
& \includegraphics[width=0.4\textwidth]{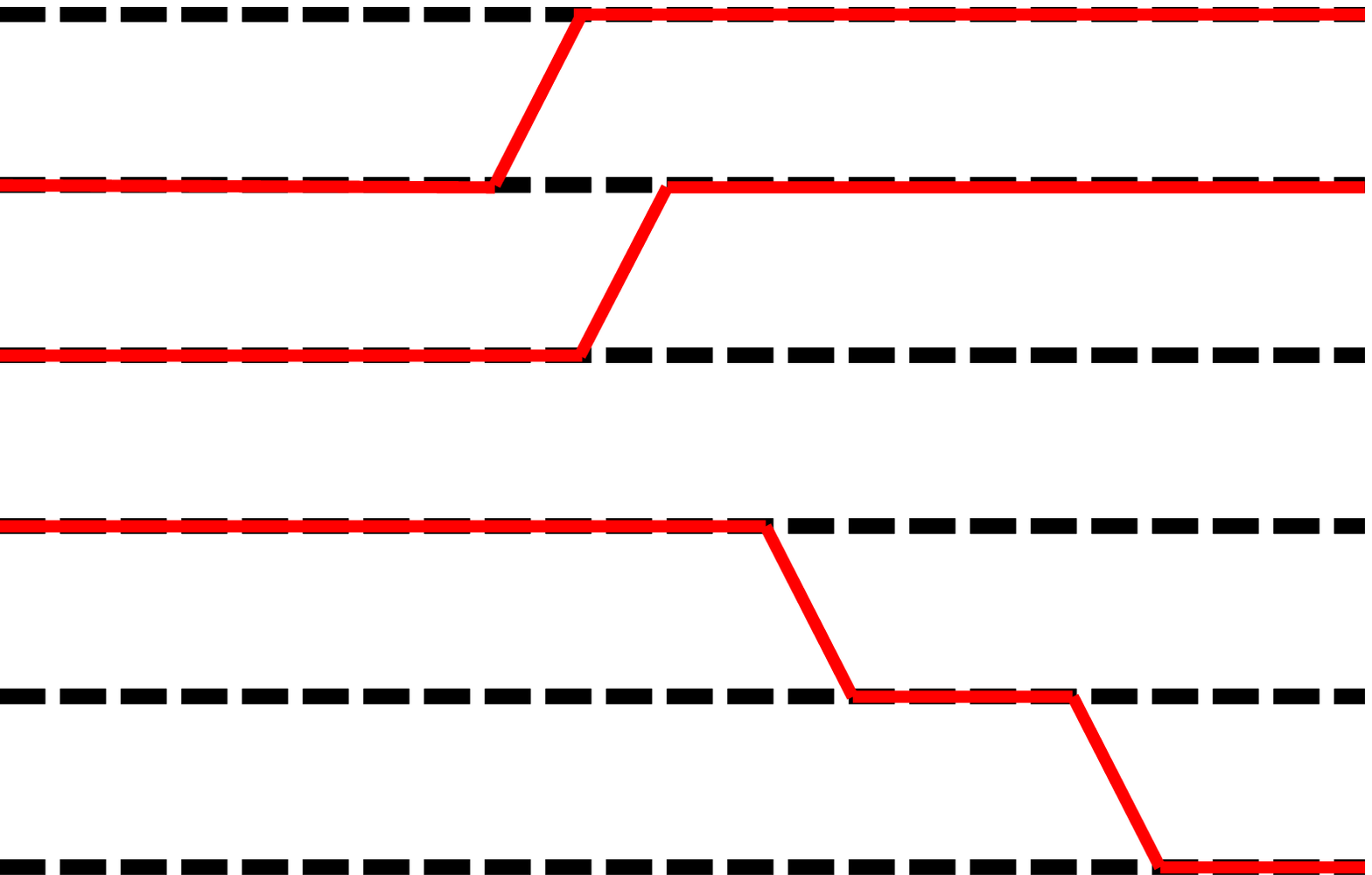}
\\
(a) & (b) \\
\end{tabular}
\end{center}
\caption{Non-BPS exact solutions of charged bions in the 
Grassmann sigma models. 
(a) Only one possible charged bion in $G_{4,2}$. 
(b) An example of a charged bion in $G_{6,3}$. 
}
\label{fig:charged-bion}
\end{figure}

For $G_{4,2}$, there is only one possible charged bion 
$[1,-1]$ on the left vacuum with adjacent flavors occupied by colors, 
shown in Fig.~\ref{fig:charged-bion} (a). 
If one wants to increase the vector of fractional instanton numbers 
to $[2,-2]$, a non-BPS crossing is inevitable, and no new charged bion 
can be generated. 
No charged bion is possible for the left vacuum with 
non-adjacent flavors occupied by colors. 

\begin{figure}[htbp]
\begin{center}
 \includegraphics[width=0.55\textwidth]{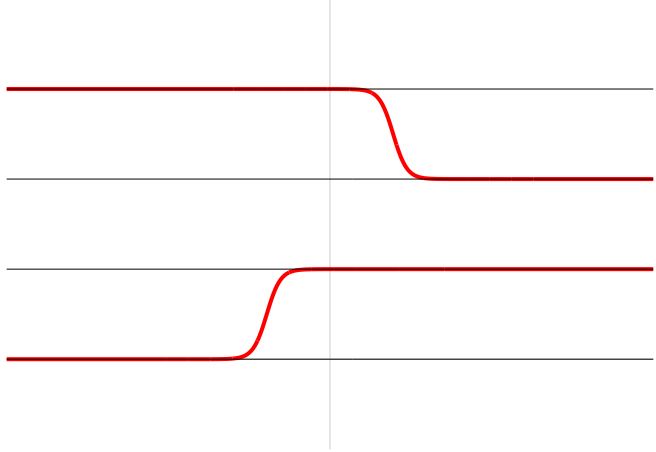}
\end{center}
\caption{
The brane 
configuration calculated from the ansatz (\ref{eq:charged_bion}) 
of the exact non-BPS solution of the charge bion 
for $\lambda_{1} =10^{2},\,\lambda_{2}=10^{2},\,N_{\rm F}=4$.}
\label{brane_cb}
\end{figure}

We note that BPS (nonotonically increasing) color line and 
anti-BPS (monotonically decreasing) color line in this 
charged bion share no 
flavors in common. 
This feature is generic for charged bions, since we do not 
allow color lines to cross. 
This fact implies that fractional instantons residing on the set 
of BPS color lines and the set of anti-BPS color lines are 
non-interacting, namely they do not exert any static 
force. 
In such circumstances, we have observed already that our moduli 
matrix formalism allows non-BPS exact solutions of field 
equations \cite{Eto:2004vy}. 
The moduli matrix of the exact solution can be given by 
\begin{equation}
H_0
 = \left(
\begin{array}{ccccc}
1, 
& \lambda_{1}e^{i\theta_{1}}e^{\frac{2\pi}{N_{\rm F}} z} , & 0, 
& 0, & \cdots \\
0, &0, 
 & 1, & \lambda_{2}e^{i\theta_{2}}e^{-\frac{2\pi}{N_{\rm F}} \bar{z}} , 
& \cdots 
\end{array}
\right)\,. 
 \label{eq:charged_bion}
\end{equation}
To visualize the brane picture more exactly, we compute 
$\Sigma$ defined in Eq.~(\ref{eq:defsigma}) for this charged 
bion. 
Since $\Sigma$ turned out to be diagonal (reflecting the fact that 
two fractional instantons on different color lines are noninteracting), 
we can exactly compute $\Sigma$ from the moduli matrix 
(\ref{eq:charged_bion}). 
The plot of $\Sigma$ in Fig.~\ref{brane_cb} with the parameter 
set $\lambda_{1} =10^{2},\,\lambda_{2}=10^{2},\,N_{\rm F}=4$ 
nicely realizes the brane picture that is schematically drawn in 
Fig.~\ref{fig:charged-bion} (a). 

For larger Grassmann sigma models, 
there are more combinations. 
An example of charged bion in $G_{6,3}$ 
is shown in Fig.~\ref{fig:charged-bion} (b).

\section{Interaction of Grassmann Bions}\label{sec:int}

\subsection{Neutral bions}

In this section, we calculate the (normalized) energy density 
$s(x^1)$ and topological-charge density $q(x^1)$ defined as 
\begin{equation}
s(x^1)=\int dx^2 \frac{1}{2\pi v^2}{\cal L}(x^1,x^2), 
\quad
Q=\int dx^1 q(x^1), 
\end{equation}
for the Grassman-model-specific neutral bion configurations 
with $(1,1)$ in Eq.~(\ref{eq:11neutral_42model}) 
and in Fig.~\ref{fig:Gr42-bion_adjacent_flavor}(c) in 
Sec.\ref{subsec:neutralBionGr}. 
We take, the separation between the instanton (at 
$(N_{\rm F}/2\pi) \log\lambda_{1}$) and anti-instanton 
(at $-(N_{\rm F}/2\pi) \log\lambda_{2}$), 
as a single parameter $\tau=-(N_{\rm F}/2\pi)\log \lambda_1\lambda_2$. 
We from now refer to the instanton at 
$\pm(N_{\rm F}/2\pi) \log\lambda_{i}$ just as $\lambda_{i}$. 
In most of numerical study, we fix $\lambda_{3}$ and 
$\lambda_{4}$ as $\lambda_{3}=\lambda_{4}=10^{5}$, then 
vary $\lambda_{1}$ and $\lambda_{2}$ with the symmetric 
condition $\lambda_{1}=\lambda_{2}=\lambda$, unless stated 
otherwise.

\begin{figure}[htbp]
\begin{center}
 \includegraphics[width=0.4\textwidth]{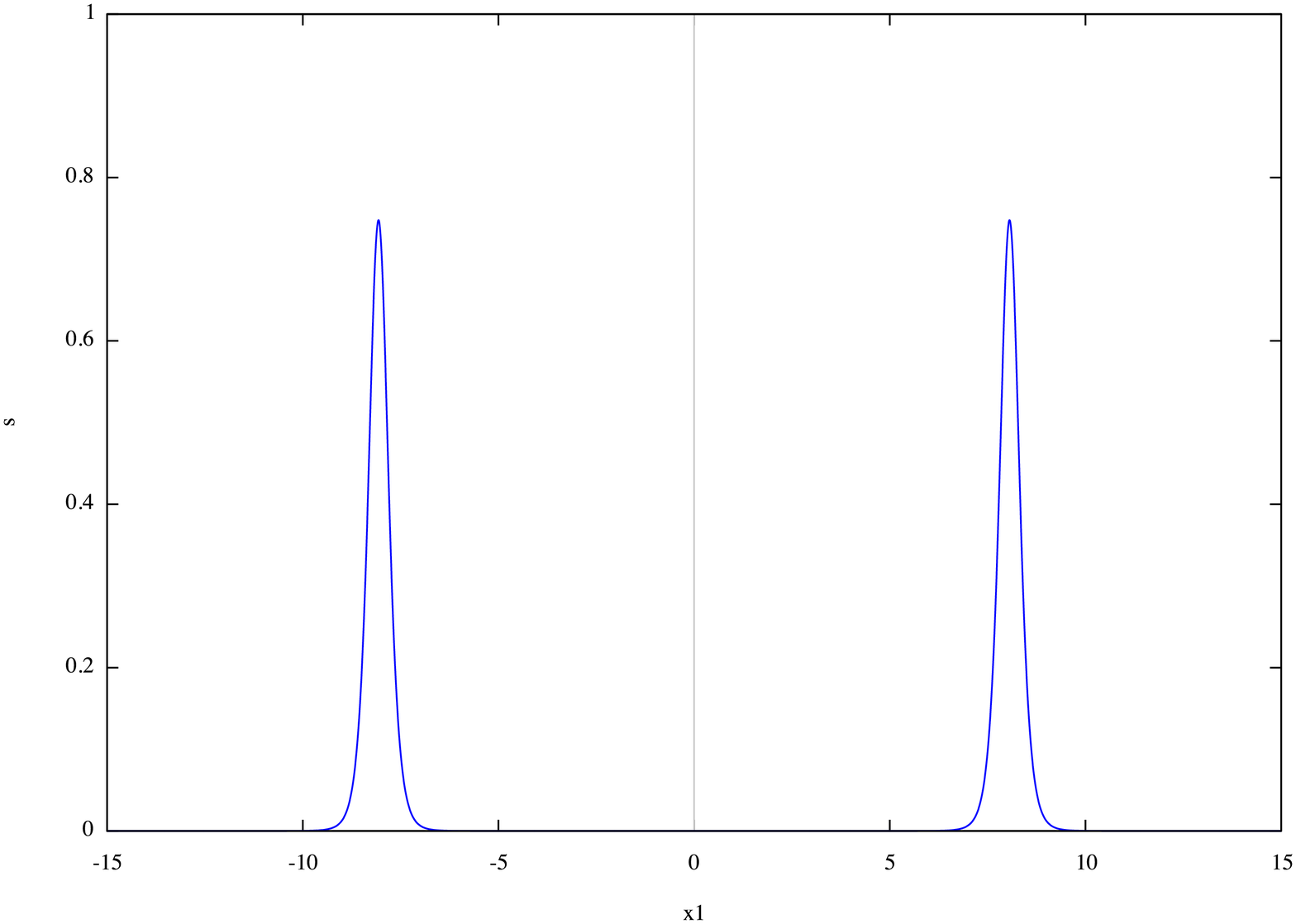}
 \includegraphics[width=0.4\textwidth]{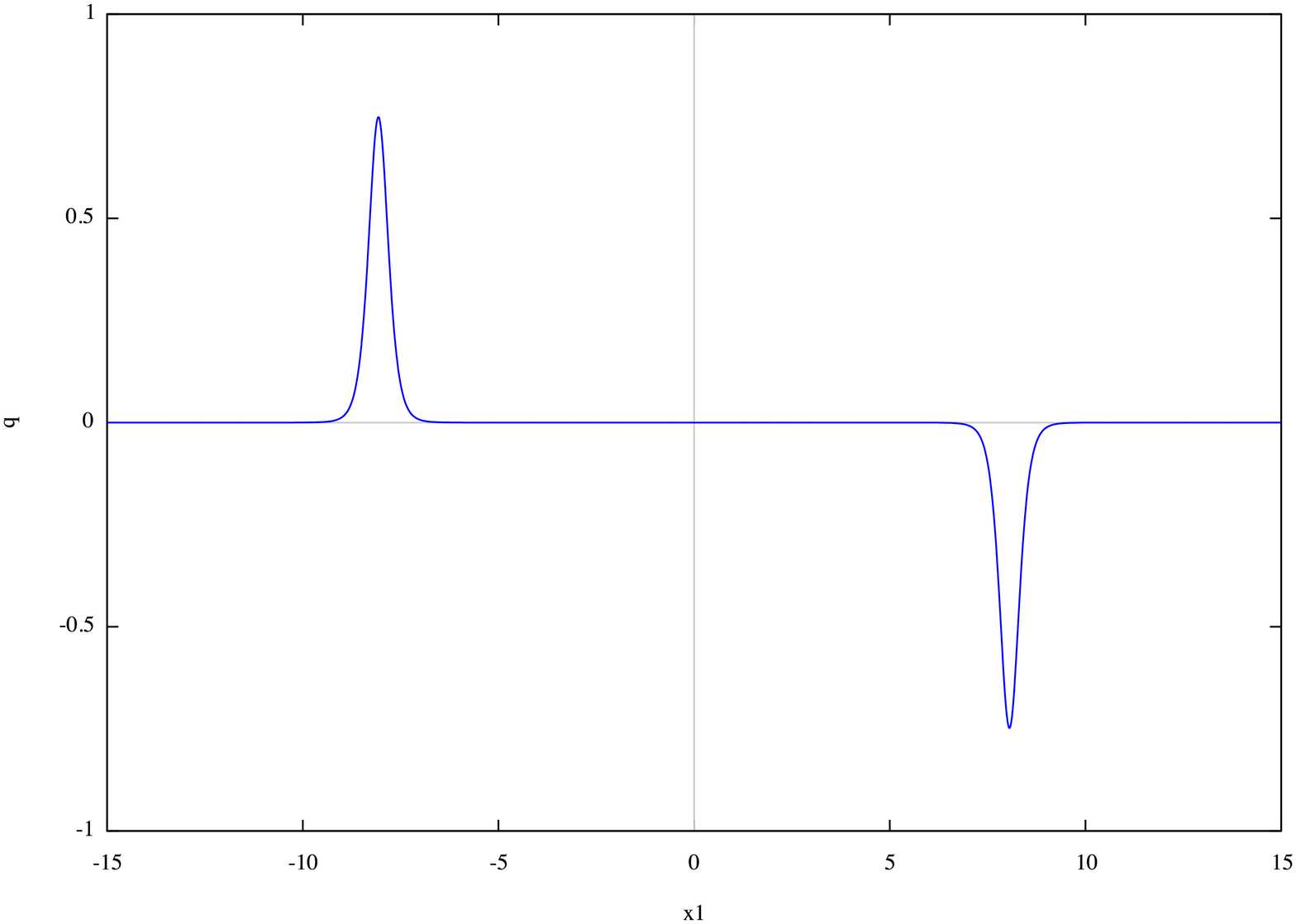}
\end{center}
\caption{Energy density $s(x^{1})$ (left) and topological charge density $q(x^{1})$ (right) for
the configuration of Eq.~(\ref{eq:11neutral_42model}) for $\lambda_{1} =10^{-6},\,\lambda_{2}=10^{-6},\,\lambda_{3}=10^{-5}, \,\lambda_{4}=10^{-5},\,N_{\rm F}=4$.}
\label{sq66554}
\end{figure}

\begin{figure}[htbp]
\begin{center}
 \includegraphics[width=0.4\textwidth]{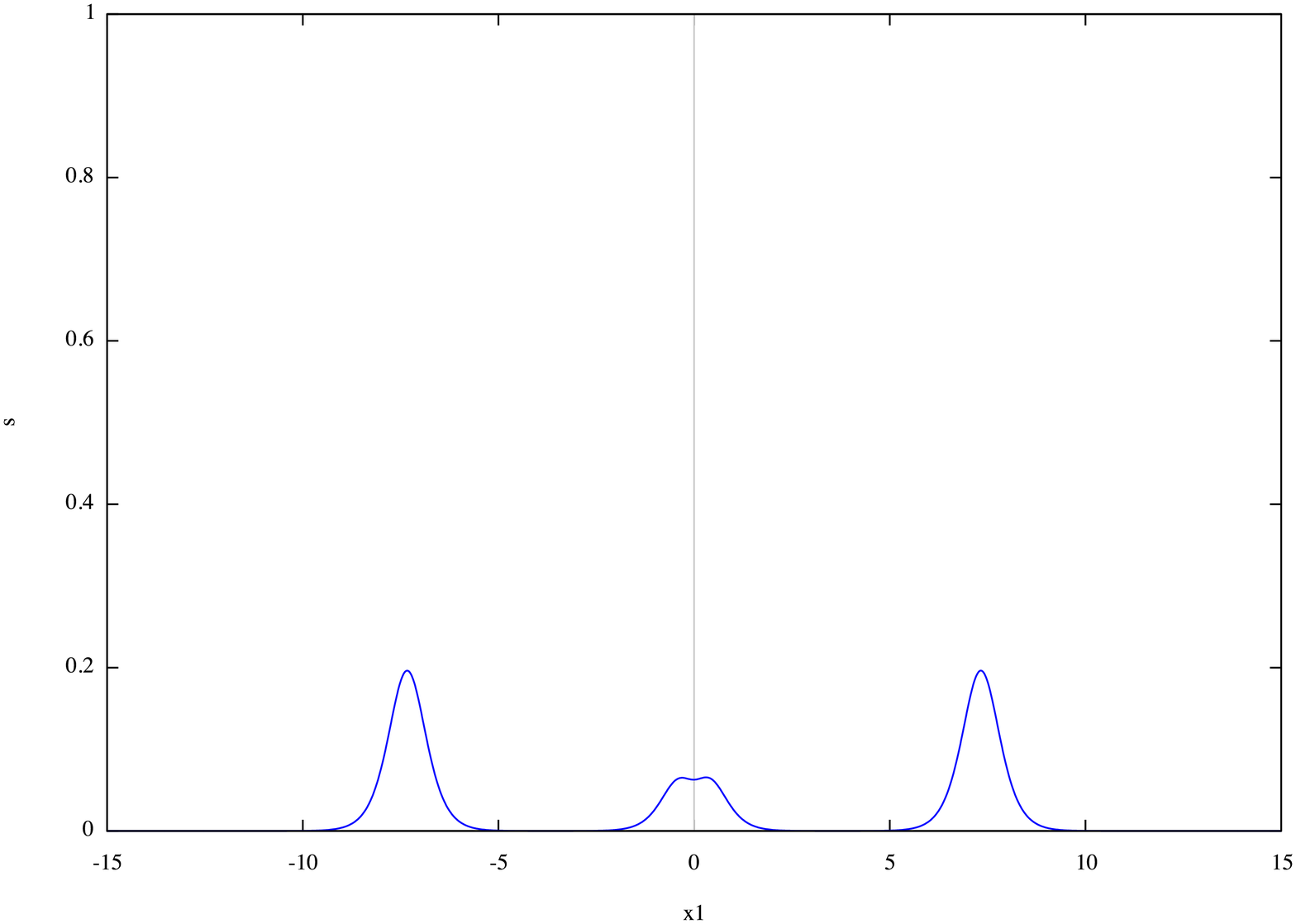}
 \includegraphics[width=0.4\textwidth]{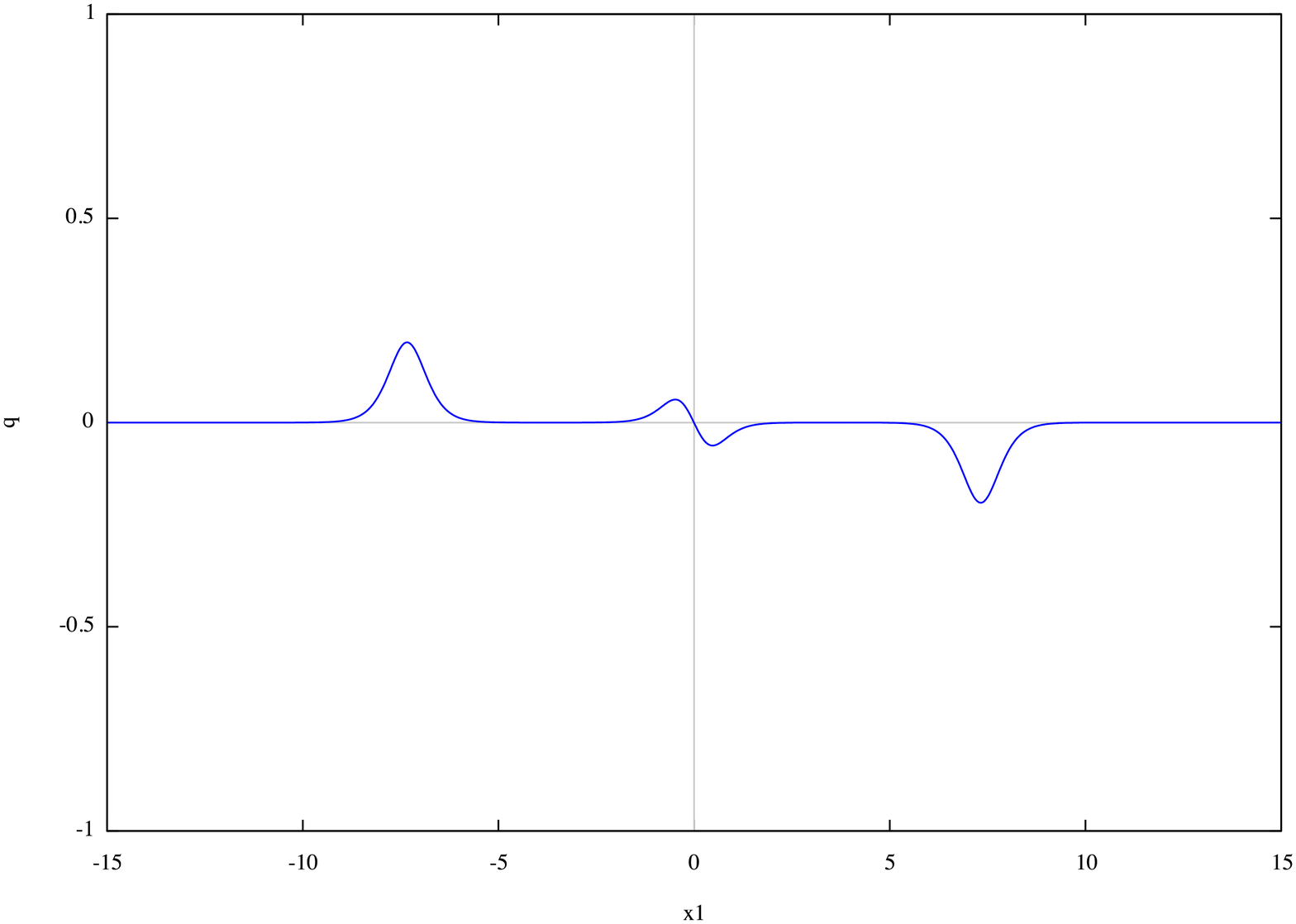}
\end{center}
\caption{Energy density $s(x^{1})$ (left) and topological charge 
density $q(x^{1})$ (right) for the configuration of 
Eq.~(\ref{eq:11neutral_42model}) for $\lambda_{1} =1,\,
\lambda_{2}=1,\,\lambda_{3}=10^{-5}, \,\lambda_{4}=10^{-5},\,
N_{\rm F}=4$.}
\label{sq00554}
\end{figure}

Before discussing the results of numerical evaluation, we make 
some comments:First, we take the phase parameter $\theta_{i}=0$ 
($i=1,2,3,4$) for simplicity.
Indeed, we find that nonzero $\theta_{i}$ just increase the 
total energy as with the neutral bion in the $CP^{N-1}$ model 
\cite{Misumi:2014jua}, thus $\theta_{i}=0$ are a natural 
choice of phase parameters.
Second, we do not show the $x^{2}$ axis in the figure of the 
energy and charge densities since the energy and charge 
densities of the BPS fractional instanton, neutral bions and 
charged bions are independent of $x^{2}$, unlike the case of 
the BPS instanton as shown in Sec.\ref{sec:IGS}. 
\begin{figure}[htbp]
\begin{center}
 \includegraphics[width=0.4\textwidth]{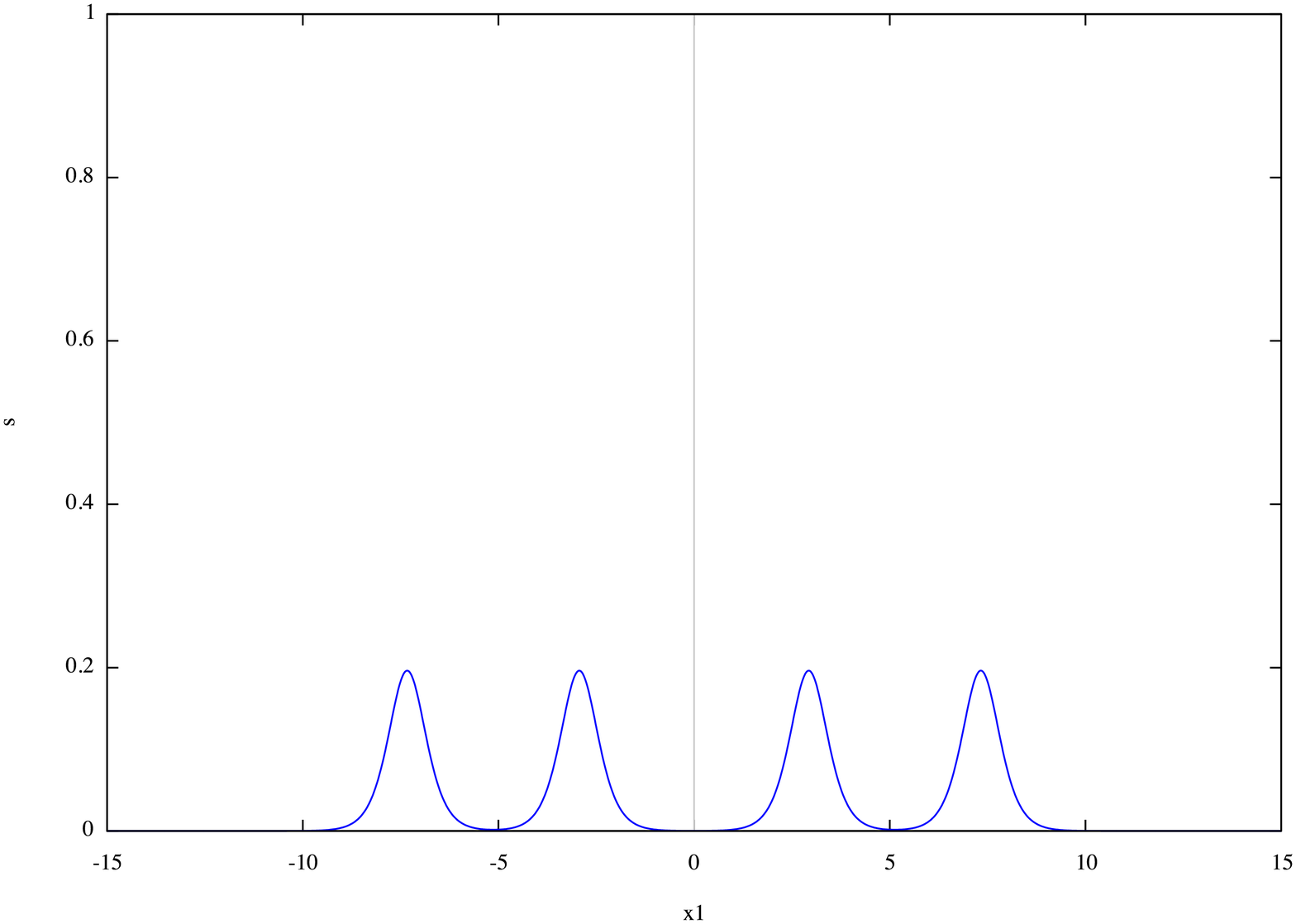}
 \includegraphics[width=0.4\textwidth]{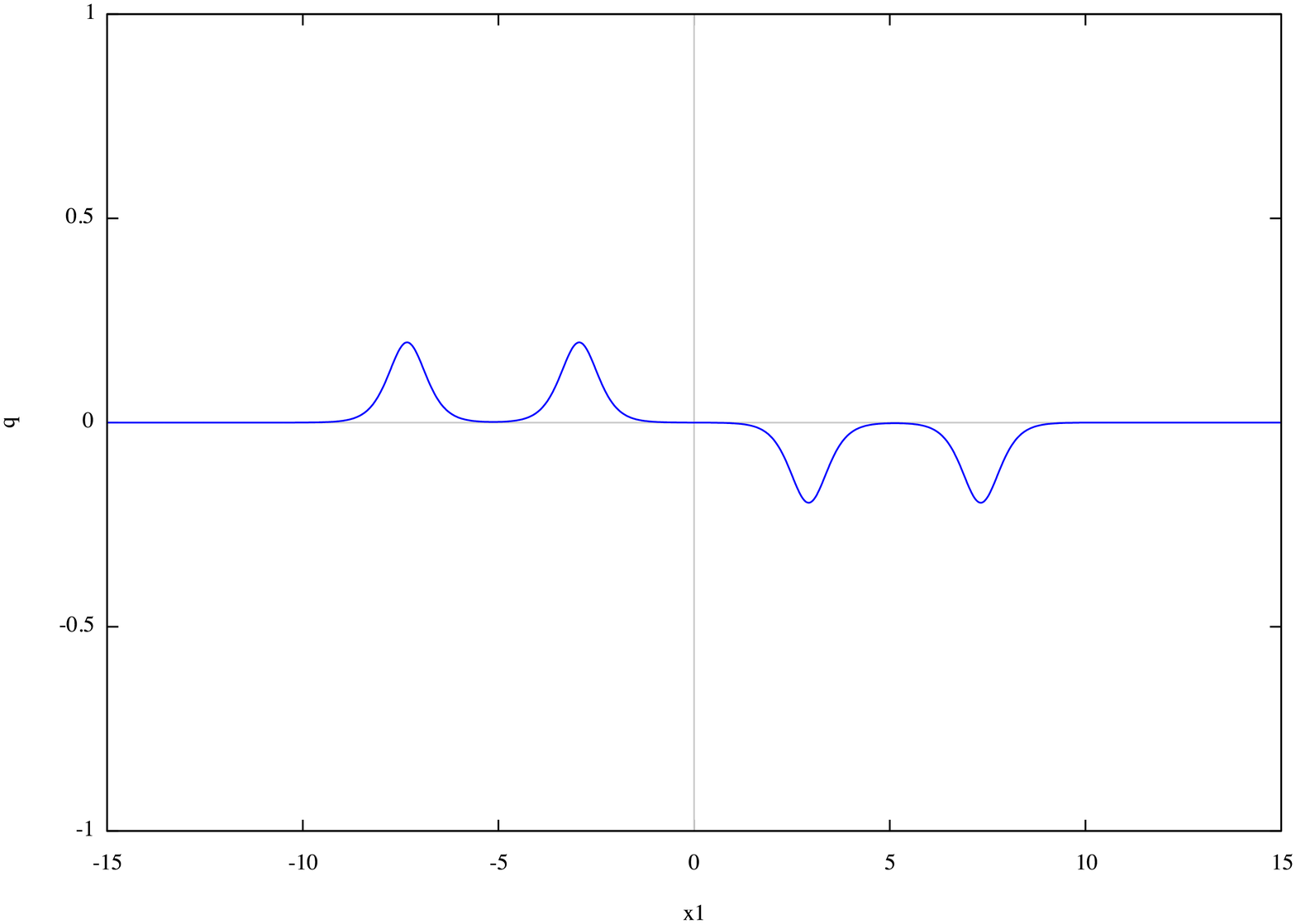}
\end{center}
\caption{Energy density $s(x^{1})$ (left) and topological charge 
density $q(x^{1})$ (right) for the neutral bion $(1,1)$ in 
Eq.~(\ref{eq:11neutral_42model}) and in 
Fig.~\ref{fig:Gr42-bion_adjacent_flavor}(c)  
for $\lambda_{1} =10^{-2},\,\lambda_{2}=10^{-2},\,
\lambda_{3}=10^{-5}, \,\lambda_{4}=10^{-5},\,N_{\rm F}=4$.
}
\label{sq22554}
\end{figure}
To figure out the properties of this configuration, we depict 
energy and charge densities for three sets of parameters,
\begin{equation}
\lambda_{1} =10^{-2},\,\lambda_{2}=10^{-2},\,
\lambda_{3}=10^{-5}, \,\lambda_{4}=10^{-5},\,N_{\rm F}=4,
\end{equation}
in Figs.~\ref{sq22554}, 
\begin{equation}
\lambda_{1} =10^{-6},\,\lambda_{2}=10^{-6},\,
\lambda_{3}=10^{-5}, \,\lambda_{4}=10^{-5},\,N_{\rm F}=4, 
\end{equation}
in Figs.~\ref{sq66554}, 
\begin{equation}
\lambda_{1} =1,\,\lambda_{2}=1,\,\lambda_{3}=10^{-5}, \,
\lambda_{4}=10^{-5},\,N_{\rm F}=4, 
\end{equation}
 and in Figs.~\ref{sq00554}, respectively. 

For $1\ll \lambda < 10^5$, this configuration is composed of 
four kinks, two BPS fractional instantons ($S=1/4$, $Q=1/4$) 
and BPS fractional anti-instantons ($S=1/4$, $Q=-1/4$), 
which are separately located as shown in Fig.~\ref{sq22554}.
For $ \lambda \gtrsim 10^5$, the two instantons are merged 
into one compressed fractional instanton ($S=1/2$, $Q=1/2$) 
while the two anti-instantons are merged into one compressed 
fractional anti-instanton ($S=1/2$, $Q=-1/2$) as shown in 
Fig.~\ref{sq66554}.
It is notable that the size of the fractional (anti-)instantons 
becomes a half smaller ($\sim 4\to\sim 2$) when they are 
compressed with another (anti-)instantons, which is 
consistent with the argument in Sec.\ref{sec:ZN} 
(size of instanton$=1/\Delta m=L/S$). 
 In this case, the total energy is unchanged. 
 For $ \lambda \lesssim 1 $, the instanton and the 
anti-instanton characterized by $\lambda_{1}$ and $\lambda_{2}$ 
respectively are annihilated and disappear, which ends up with 
 one instanton ($S=1/4$, $Q=1/4$) and one anti-instanton 
($S=1/4$, $Q=-1/4$) 
 as shown in Fig.~\ref{sq00554}. 
In this case, the total energy reduces from $S=1/4\times4=1$ 
to $S=1/4\times2=1/2$.

\begin{figure}[htbp]
\begin{center}
 \includegraphics[width=0.5\textwidth]{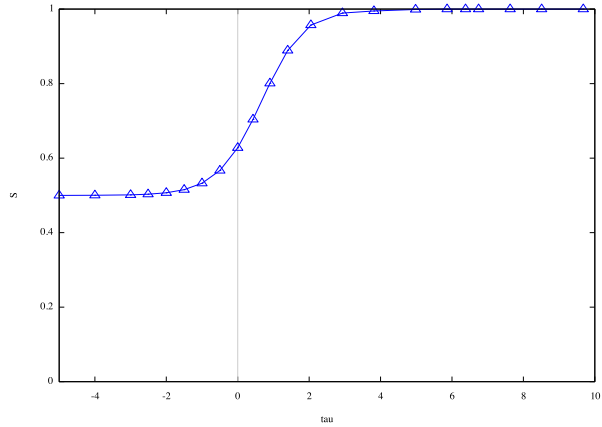}
\end{center}
\caption{The total energy as a function of $\tau=-(N_{\rm F}/\pi)\log \lambda$ with 
$\lambda=\lambda_{1}=\lambda_{2}$ for $\lambda_{3}=\lambda_{4}=10^{-5}\,$ 
with $N_{\rm F}=4$ fixed.}
\label{S_tau12}
\end{figure}
Next, we calculate the parameter dependence of the total energy 
for the neutral bion configuration (\ref{eq:11neutral_42model}).
As a characteristic case, 
we again vary $\lambda=\lambda_{1}=\lambda_{2}$ with 
$\lambda_{3}=\lambda_{4}=10^{-5}$ fixed.
The result is given in Fig.~\ref{S_tau12}. 
$\tau\sim10$ corresponds to the compressed-kink cases as 
Fig.~\ref{sq66554} while $\tau=1$ corresponds to the 
pair-annihilation case between the instanton and anti-instanton 
as Fig.~\ref{sq00554}.
The total energy is changed from $S=4\times1/4=1$ to 
$S=2\times1/4=1/2$ as $\tau$ gets smaller. 
Since it is known that BPS solitons do not exert any static force, 
this result shows that the there is an attractive static force between 
fractional instanton $\lambda_1$ and anti-instanton $\lambda_2$.

\begin{figure}[htbp]
\begin{center}
 \includegraphics[width=0.32\textwidth]{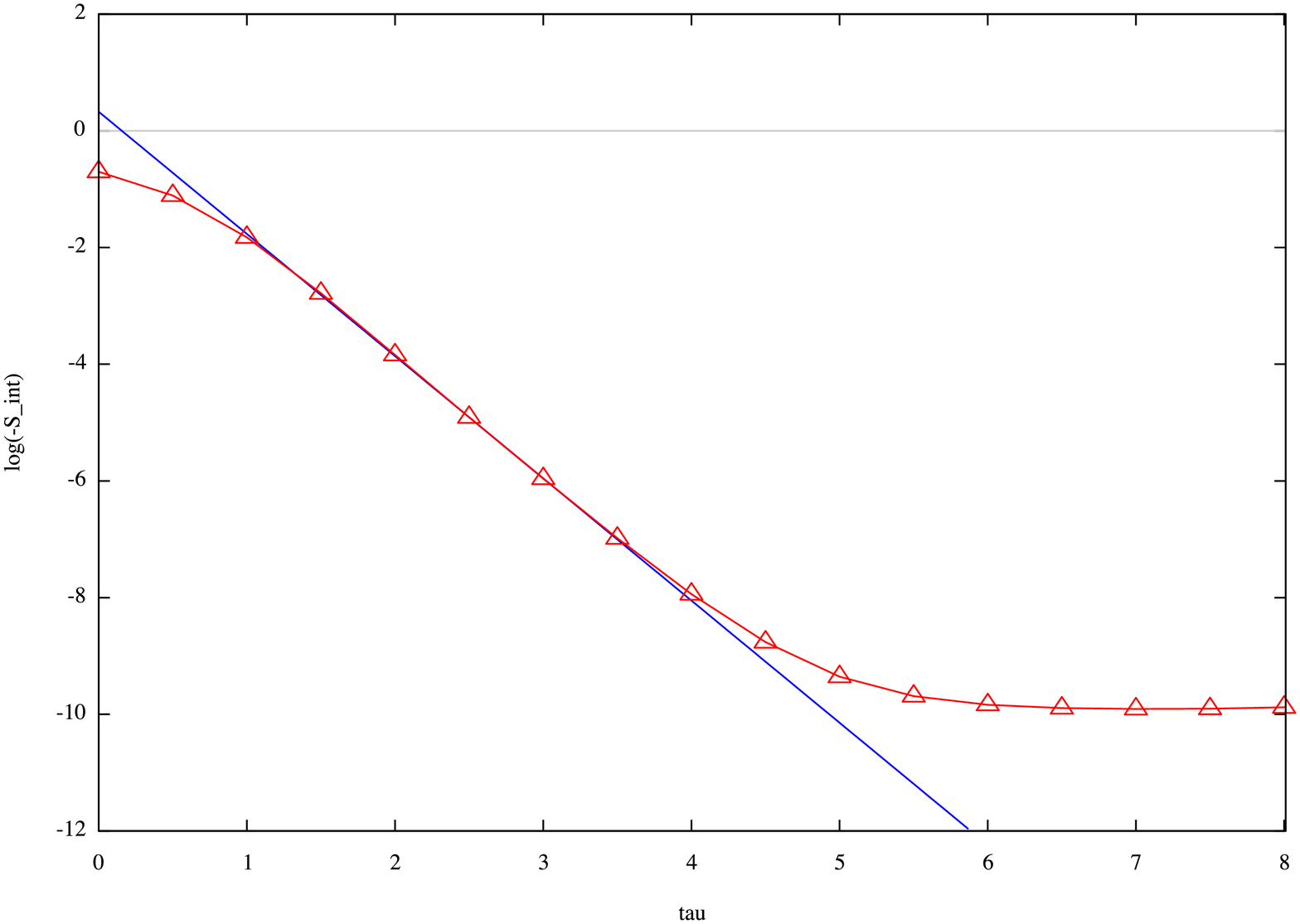}
 \includegraphics[width=0.32\textwidth]{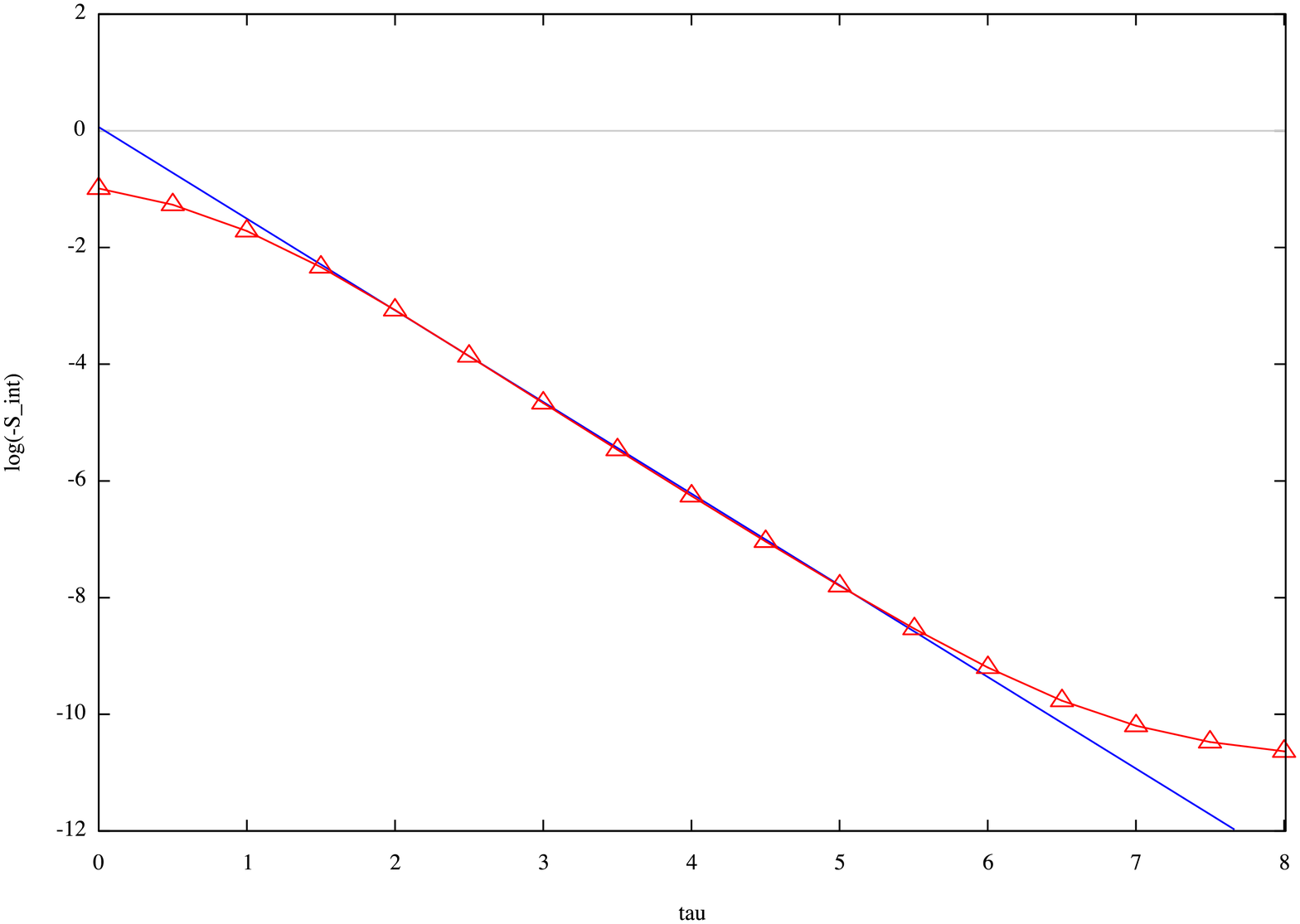}
  \includegraphics[width=0.32\textwidth]{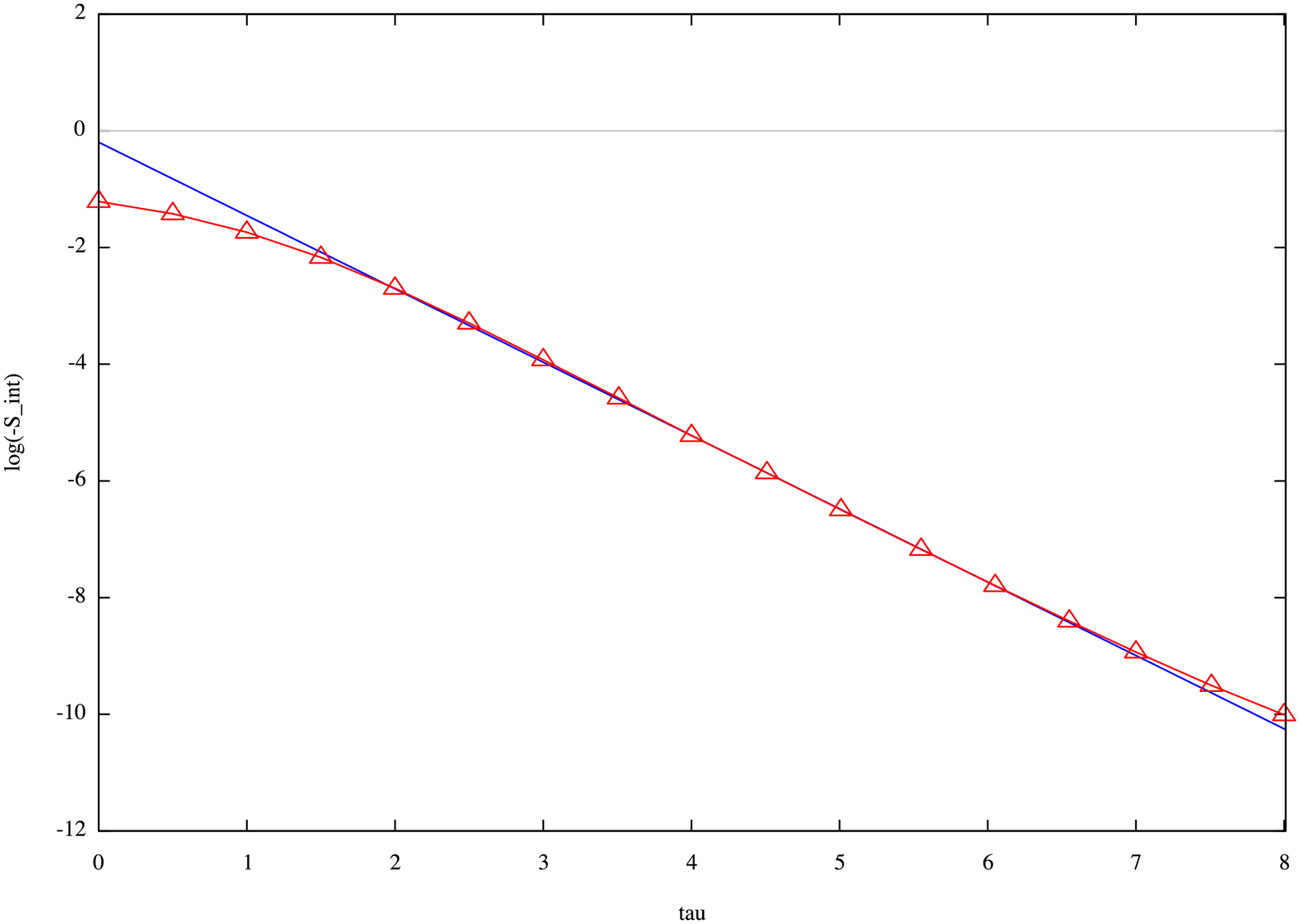}
\end{center}
\caption{Plot of $\log (-S_{\rm int}(N_{\rm F},\tau))$ as a function of $\tau$ 
for $N_{\rm F}=3$(left), $N_{\rm F}=4$(center) and $N_{\rm F}=5$(right) 
for (\ref{eq:11neutral_42model}) (red curves with triangle points).
For the intermediate region of $\tau$, the curve is approximated by 
$-(2\pi/N_{\rm F})\tau+ C(N_{\rm F})$ (blue curves).}
\label{Lsint_554}
\end{figure}

In order to study mutual interactions between constituent 
fractional (anti-)instantons more quantitatively, 
we define the interaction energy density as the energy 
density $s(x^{1})$ minus the two fractional-instanton density 
and two fractional-anti-instanton density 
$2s_{\nu=1/N_{\rm F}}+2s_{\nu=-1/N_{\rm F}}$,
\begin{equation}
s_{\rm int}(x^{1}) 
=s(x^{1})
-(2s_{\nu=1/N_{\rm F}}+2s_{\nu=-1/N_{\rm F}})\,.
\end{equation}
The integrated interaction energy is then given by
\begin{equation}
S_{\rm int}(N_{\rm F}, \lambda_{1},\lambda_{2},\lambda_{3},\lambda_{4})\, 
=\, \int dx^{1}\, s_{\rm int}(x^{1})\,.
\end{equation}
Let us first study interaction between $\lambda_1$ and $\lambda_2$. 
By varying $\tau$ with $\lambda_{3}=\lambda_{4}=10^{-5}$ fixed, 
we show the logarithm of the total interaction energy 
$S_{\rm int}(N_{\rm F}, \tau)$ as a function of $\tau$ 
for $N_{\rm F}=3,4,5$ in Fig.~\ref{Lsint_554}. 
For the intermediate separation of $\tau$, 
$\log (-S_{\rm int}(N_{\rm F},\tau))$ is 
well approximated by the analytic function,
\begin{equation}
\log \left[-S_{\rm int}(N_{\rm F},\tau)\right]\,\,\sim\,\, -\, 
{2\pi\over{N_{\rm F}}}\tau \,\,+\,\, C(N_{\rm F})\,,
\end{equation}
where $C(N_{\rm F})$ is a $y$-intercept. 
In Fig.~\ref{Lsint_554} we simultaneously depict these analytic 
functions. 
The slope $2\pi/N_{\rm F}$ of this line is equivalent to that 
of the elementary neutral bion in the ${\mathbb C}P^{N_{\rm F}-1}$ 
model \cite{Misumi:2014jua}, which contains only one fractional 
instanton and one fractional anti-instanton. 
This means that the interaction energy is dominated by the 
interaction between $\lambda_{1}$ and $\lambda_{2}$, and not 
by the other interactions. 
\begin{figure}[htbp]
\begin{center}
 \includegraphics[width=0.49\textwidth]{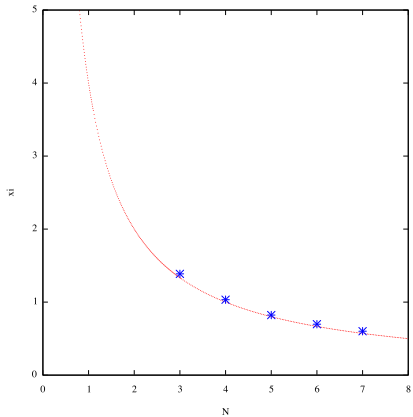}
\end{center}
\caption{The constant $\exp[C(N_{\rm F})]$ as a function of 
$N_{\rm F}$ for $N_{\rm F} = 3,4,5,6,7$ (blue points) .
It is well approximated by $4/N_{\rm F}$ (red curve).}
\label{xi_Nf}
\end{figure}
In Fig.~\ref{xi_Nf} we plot $\exp[{C(N_{\rm F})}]$ 
as a function of $N_{\rm F}$ for $N_{\rm F}=3,4,5,6,7$. 
By fitting the $N_{\rm F}$-dependence of the constant $C(N_{\rm F})$, 
we obtain the interaction energy formula 
identical to the ${\mathbb C}P^{N_{\rm F}-1}$ model 
\cite{Misumi:2014jua}, which means $C_{N_{\rm F}}\sim 4/N_{\rm F}$. 
(We note that we have included the factor $1/(2\pi)$ in the 
definition of Lagrangian in Eq.~(\ref{eq:mdl:reduced-L}) 
in this paper compared to our previous paper \cite{Misumi:2014jua}.). 
Namely we find 
\begin{equation}
S_{\rm int}(N_{\rm F}, \tau)\sim -{4\over{N_{\rm F}}} 
\,\,e^{-(2\pi/N_{\rm F})\tau} \,,
\label{eq:int_energy_cpn}
\end{equation}
in the intermediate region where none of constituent instantons 
are too close.

Next we study the interaction between fractional instanton 
$\lambda_3$ and anti-instanton $\lambda_4$ in outer pair. 
To isolate the interaction between the outer pair, we take 
the annihilation limit of inner pair, fixing the 
parameters $\lambda_1=\lambda_2=10^5$. 
Then practically no remnant is left from the inner pair. 
By varying the separation between $\lambda_3$ and $\lambda_4$, 
we find that the interaction energy is given precisely by the same 
formula as in the case of inner pair $\lambda_1$ and $\lambda_2$ 
in Eq.~(\ref{eq:int_energy_cpn}). 
By using this formula for outer pair, we can see that the 
contribution from outer pair becomes tiny in the geometrical 
configurations in Figs.~\ref{sq22554}, \ref{sq66554}, and 
\ref{sq00554}. 
This justifies to neglect outer pair interaction enrgy in 
analyzing the interaction energy of inner pair 
a posteriori. 

With our level of numerical accuracy, we cannot obtain 
definite results for other possible interactions between 
fractional instanton and anti-fractional instanton residing 
on different color lines, such as $\lambda_1-\lambda_4$, 
and $\lambda_3-\lambda_2$, except that they are at least as 
small as interaction energy between outer pair at the same 
separation. 

\subsection{Charged bions}
\label{sec:CBGS}

We next consider the charged bion in the Grassmann sigma model. 
For $Gr_{4,2}$ model, we found only one irreducible charged bion 
with the fractional instanton number $(-1,1)$ in 
Fig.~\ref{fig:charged-bion}(a), 
which is given by the moduli matrix in Eq.~(\ref{eq:charged_bion}). 
This is an exact non-BPS solution, 
since BPS and anti-BPS sectors reside on color branes 
which do not share any common flavors, and are non-interacting.

\begin{figure}[htbp]
\begin{center}
 \includegraphics[width=0.4\textwidth]{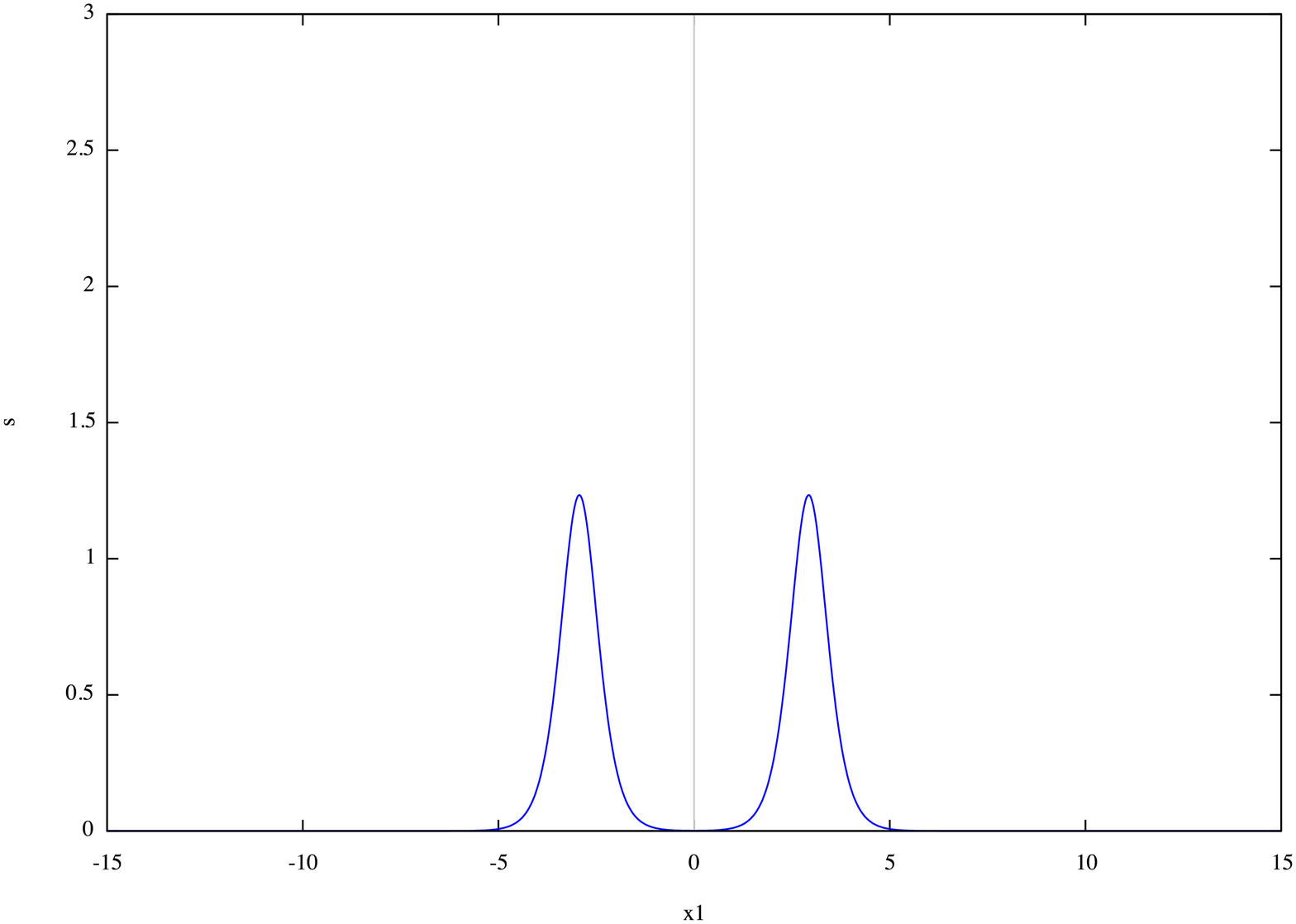}
 \includegraphics[width=0.4\textwidth]{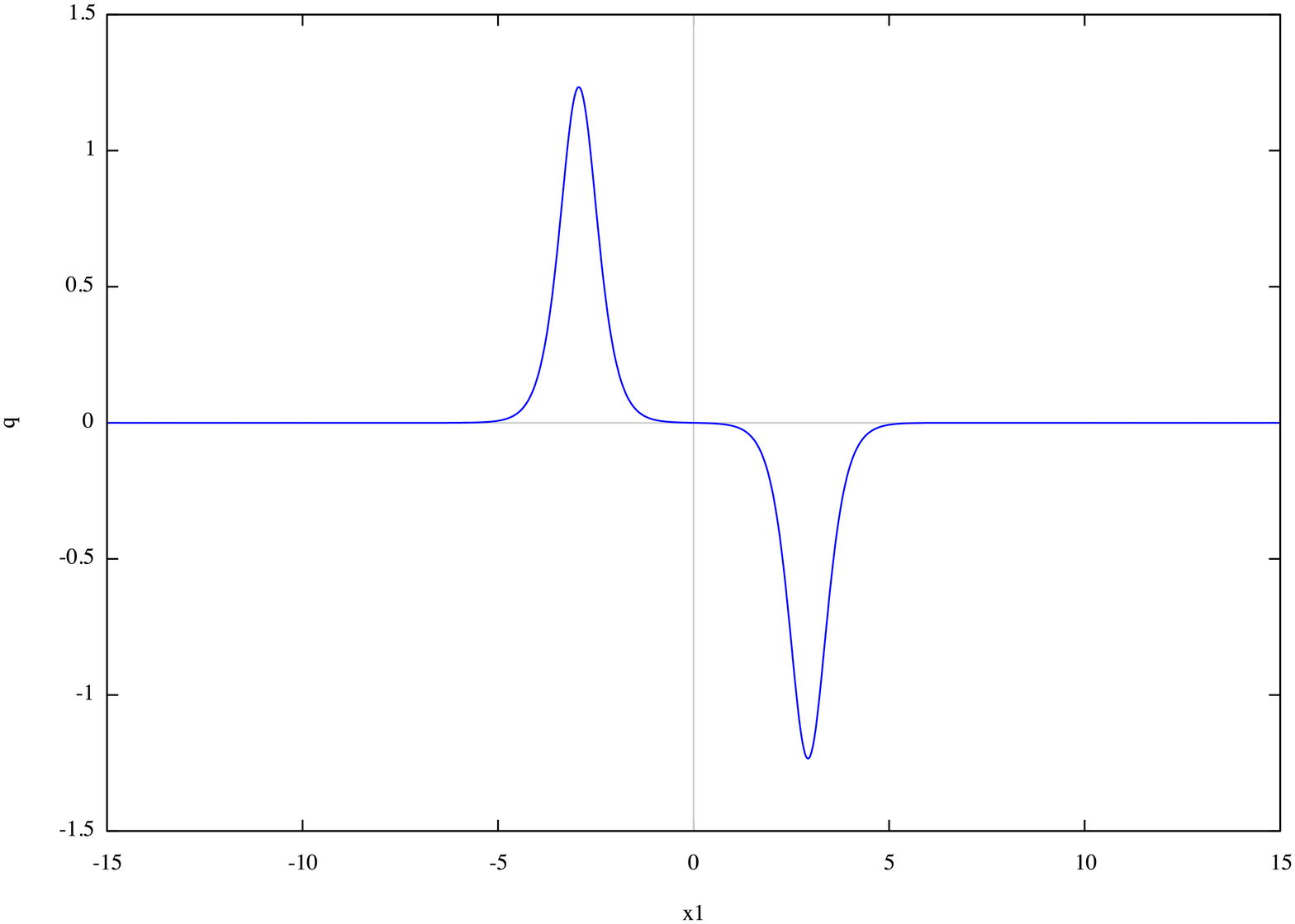}
\end{center}
\caption{Energy density $s(x^{1})$ (left) and topological charge 
density $q(x^{1})$ (right) for the configuration of 
Eq.~(\ref{eq:charged_bion}) for 
$\lambda_{1} =10^{2},\,\lambda_{2}=10^{2},\,N_{\rm F}=4$.
}
\label{cb_sq224}
\end{figure}

\begin{figure}[htbp]
\begin{center}
 \includegraphics[width=0.4\textwidth]{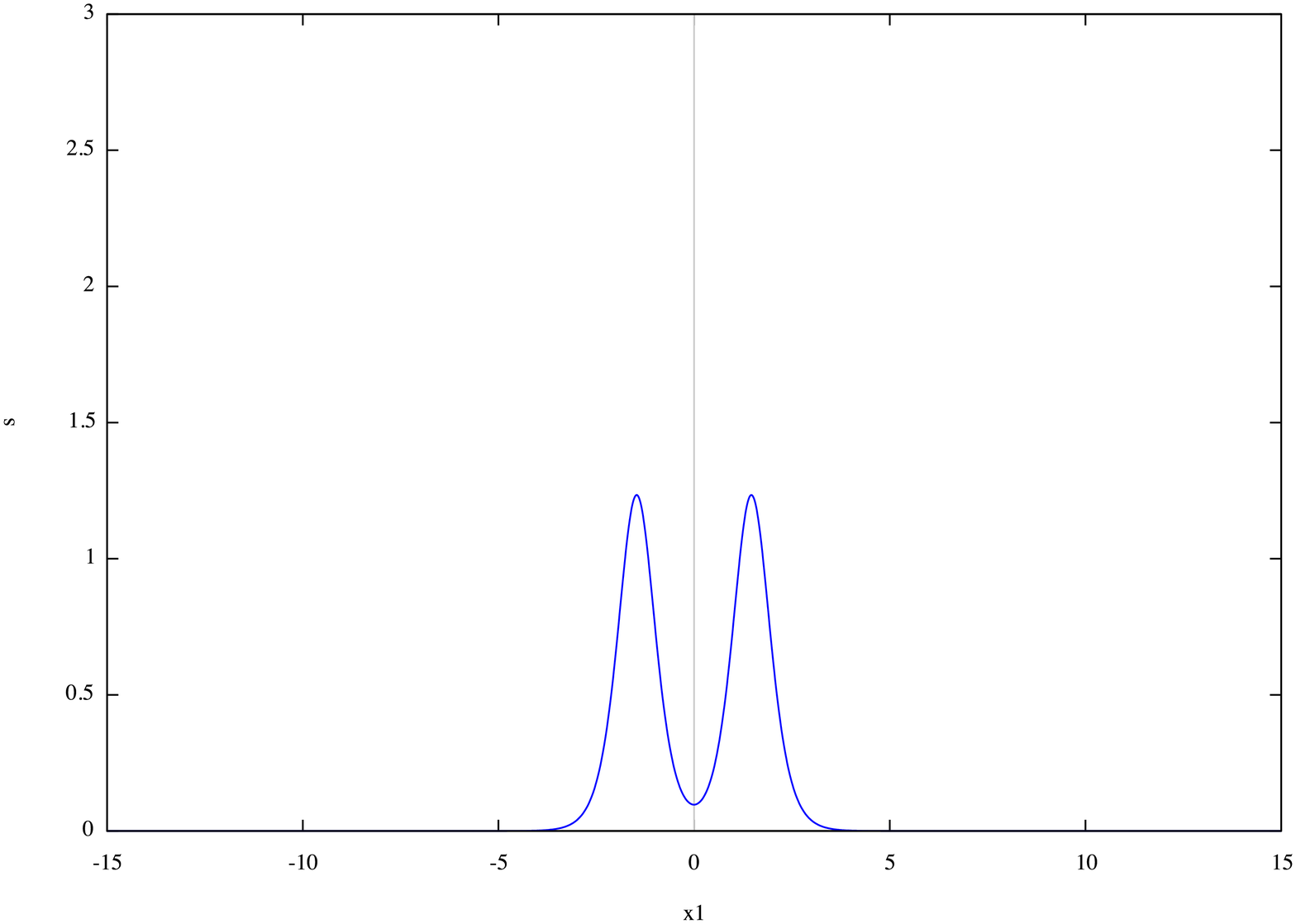}
 \includegraphics[width=0.4\textwidth]{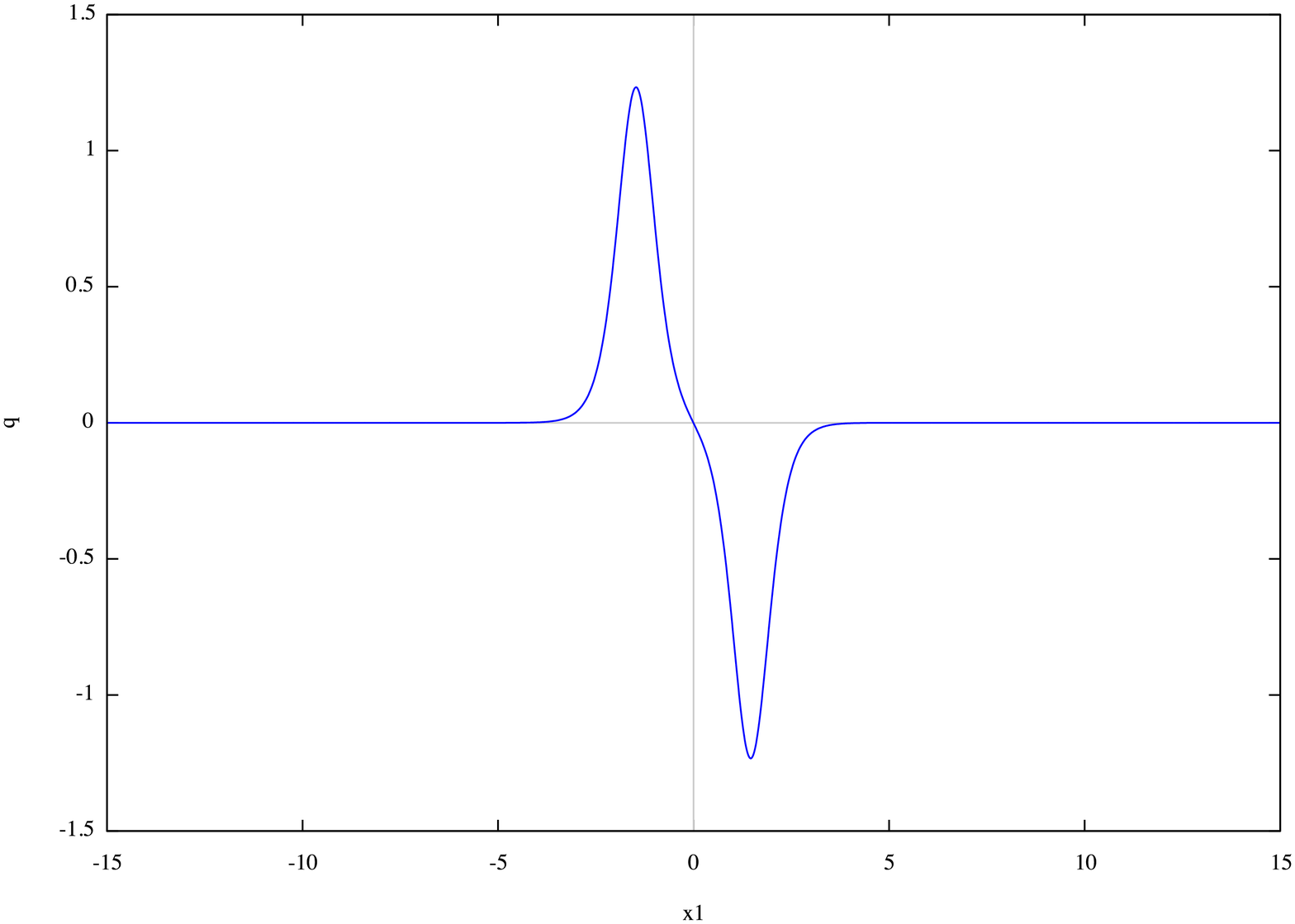}
\end{center}
\caption{Energy density $s(x^{1})$ (left) and topological charge 
density $q(x^{1})$ (right) for the configuration of 
Eq.~(\ref{eq:charged_bion}) for 
$\lambda_{1} =10,\,\lambda_{2}=10,\,N_{\rm F}=4$.}
\label{cb_sq114}
\end{figure}

\begin{figure}[htbp]
\begin{center}
 \includegraphics[width=0.4\textwidth]{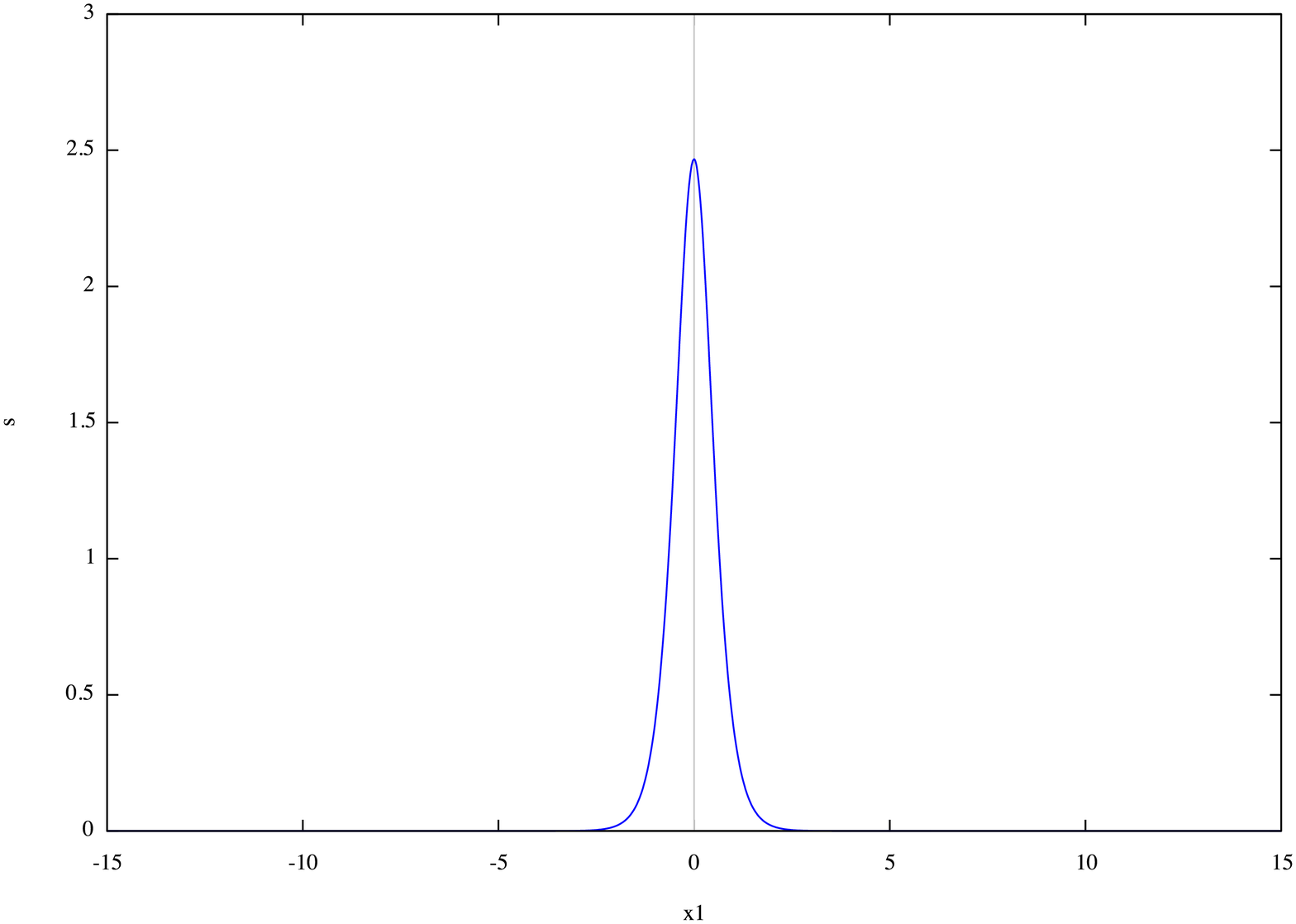}
 \includegraphics[width=0.4\textwidth]{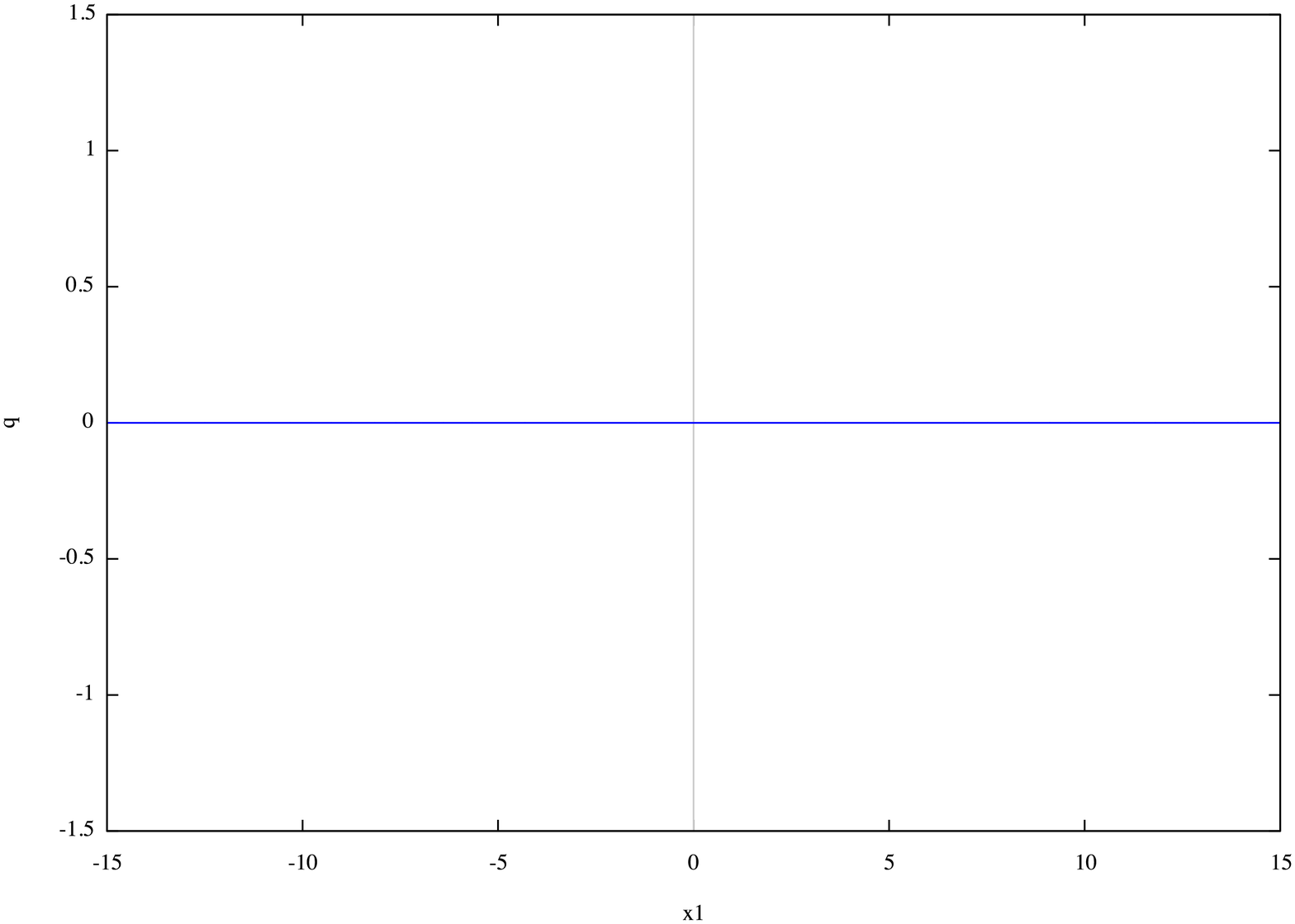}
\end{center}
\caption{Energy density $s(x^{1})$ (left) and topological charge 
density $q(x^{1})$ (right) for the configuration of 
Eq.~(\ref{eq:charged_bion}) for 
$\lambda_{1} =1,\,\lambda_{2}=1,\,N_{\rm F}=4$.}
\label{cb_sq004}
\end{figure}

To find out the properties of the solutin in detail, we study 
it numerically using our formula in Eq.~\ref{eq:H_formula}. 
We observe analytically that the 
energy and charge densities are independent of the moduli 
parameters $\theta_{1}$ and $\theta_{2}$. 
For the symmetric case $\lambda\equiv\lambda_{1}=\lambda_{2}$,
the separation between the fractional instanton (at 
$-(N_{\rm F}/2\pi) \log\lambda_{1}$) and fractional 
anti-instanton (at $(N_{\rm F}/2\pi) \log\lambda_{2}$) is 
given by $\tau=(N_{\rm F}/\pi)\log \lambda$.
We depict energy and charge densities for three sets of parameters,
\begin{align}
&\lambda_{1} =10^{2},\,\lambda_{2}=10^{2},\,N_{\rm F}=4,
 \\
&\lambda_{1} =10,\,\lambda_{2}=10,\,N_{\rm F}=4, 
 \\
&\lambda_{1} =1,\,\lambda_{2}=1,\,N_{\rm F}=4.
\end{align}
Figs.~\ref{cb_sq224}, \ref{cb_sq114} and \ref{cb_sq004} show 
the results. 

\begin{figure}[htbp]
\begin{center}
 \includegraphics[width=0.5\textwidth]{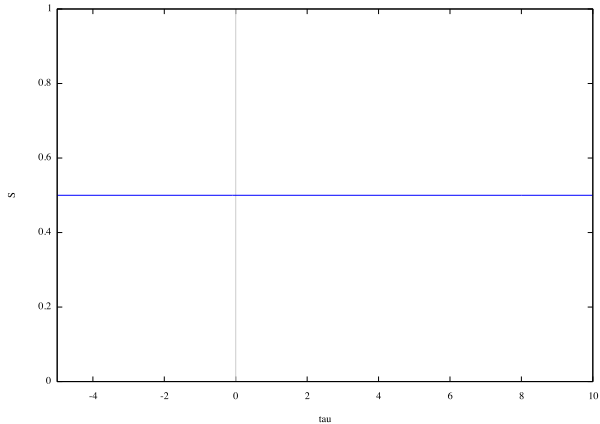}
\end{center}
\caption{The total energy of the charged bions as a function 
of $\tau=(N_{\rm F}/\pi)\log \lambda$ with 
$\lambda=\lambda_{1}=\lambda_{2}$ for $N_{\rm F}=4$ fixed.}
\label{cb_S_tau}
\end{figure}

We find that, unlike the neutral bions, the energy density 
is still nonzero even in the no separation 
limit $\tau=0$ ($\lambda_{1}=\lambda_{2}=1$) of the fraectional 
instanton and anti-instanton. 
As shown in Fig.~\ref{cb_sq004}, lumps of the topological 
charge density annihilates and disappear in this case. 
Although total topological charge happens to vanish, BPS fractional 
instanton and anti-instanton are not of the same species, and 
cannot annihilate each other, as anticipated. 
To show details, we depict the separation ($\tau$) dependence 
of the total energy in Fig.~\ref{cb_S_tau}. 
The total energy is independent of the separation, and keeps 
the constant value $S=2\times1/4=1/2$. 
This result supports the notion that positions as well as 
phases of (anti-)fractional instantons are moduli of the exact 
solution 
of the charged bion in the Grassmann sigma model, even though 
it is non-BPS. 
It is consistent with the argument that the instability and 
ambiguity in the neutral bion amplitude encode the IR renormalon 
in the perturbative expansion series, 
while the charged bions are not directly relevant. 
Rather, it should contribute to the dynamics such as confinement 
of the theory.


\section{Summary and Discussion}
\label{sec:SD}

In this paper, we have considered topologically trivial 
configurations in the Grassmann sigma model on 
${\mathbb R}^{1}\times S^{1}$ with the 
${\mathbb Z}_{N_{\rm F}}$ symmetric twisted boundary conditions, 
to study properties of bions composed of multiple fractional 
instantons. 
By formulating these models as gauge theories, we proposed to 
use the moduli matrix to classify bion configurations. 
By embedding these models to D-brane configurations 
in type-II string theories, we have found that D-brane 
configurations together with the moduli matrix are useful 
to classify all possible bion configurations in the 
${\mathbb C}P^{N_{\rm F}-1}$ models and the Grassmann sigma models. 
We have found that the Grassmann sigma models 
admit neutral bions made of BPS and anti-BPS fractional 
instantons each of which has a topological charge greater 
(less) than one (minus one), nevertheless it cannot be 
decomposed into (anti-)instanton and the rests.  
We have found that Grassmann models admit charged bions, 
while the ${\mathbb C}P^{N_{\rm F}-1}$ models do not admit them. 
We have also constructed exact solutions of charged bions 
in the Grassmann model. 
We have calculated the energy density and topological charge 
density of the bion configurations in these models numerically, 
and have obtained their interactions. 
The dependence of these interactions on the separations 
between fractional instanton constituents is studied explicitly.

We have studied the Grassmann sigma model 
without fermions.
On the other hand, fermions can be coupled to the Grassmann 
sigma model. This is the case of 
the supersymmetric Grassmann sigma model, 
which can be formulated from supersymmetric 
gauge theories. 
In this case, fermions are localized at 
the fractional instantons and contribute to 
the interactions between bions.

In this paper, we have concentrated on the 
${\mathbb Z}_{N_{\rm F}}$ symmetric 
twisted boundary conditions, where the $z$-dependence of
the moduli matrix appears in powers of  $e^{2\pi z/N_{\rm F}}$.
Consequently, the topological charge of fractional instantons
is proportional to $1/N_{\rm F}$, and the flavor branes are equally spaced in D-brane configurations.
Considering more general boundary conditions 
remains as a future problem.
The classification of fractional instantons and bions by 
the moduli matrix and D-brane configurations is essentially the same 
 as far as the boundary conditions of different flavor components are not degenerated. 
More precisely,  
when  
the boundary condition of the scalar fields  
of the flavor component $f$
is $H_f (x_1,x_2+1) = H_f (x_1,x_2) e^{i \alpha_f}$, 
the elementary fractional instantons have the instanton charges 
$\alpha_{f+1} - \alpha_f$.
Consequently, the $z$-dependence
of the moduli matrix of the flavor component $f$
appears in the form of $e^{\alpha_f z}$, 
and  
flavor branes are not equidistant in D-brane configurations.  
On the other hand, 
unboken gauge symmetry occurs 
for gauge theory on ${\mathbb R}^3 \times S^1$ 
with partially degenerated twisted boundary conditions. 
We can prepare corresponding situations 
by putting boundary conditions of some flavors to coincide.
In this case, kinks carry non-Abelian moduli and can be 
called non-Abelian kinks \cite{Shifman:2003uh,Eto:2005cc,Eto:2008dm}.
In the brane picture, the s-rule admits 
at most $n$ color branes  can sit on $n$ coincident flavor branes, 
and there remains a $U(n)$ gauge symmetry on the color branes 
which is a source of non-Abelian moduli for non-Abelian kinks. 
We will consider this situation with ``non-Abelian bions" 
in a future publication.

One of future directions is to extend our method to bion 
configurations in other nonlinear sigma models. 
Since nonlinear sigma models on Hermitian symmetric space 
(including Grassmann and ${\mathbb C}P^{N_{\rm F}-1}$) 
can be formulated as gauge theories \cite{Higashijima:1999ki}, 
the moduli matrix can be used for these cases, 
although embedding to brane configurations 
is not yet available. 
In gauge theory perspective, changing the gauge group from 
$U(\NC)$ studied in this paper to other groups is also one 
possible direction. 
The (hyper-)K\"ahler quotients for $G=SO$ and $USp$ were 
obtained in Ref.~\cite{Eto:2008qw}, in which fractional 
instantons were also studied.
We may study bions in these cases.

As for the relation between two and four dimensional theories, 
the effective theory on a non-Abelian vortex in four 
dimensions \cite{Hanany:2003hp,Auzzi:2003fs,Eto:2005yh} 
is the two-dimensional ${\mathbb C}P^{N_{\rm F}-1}$ model, 
thereby the Yang-Mills instantons and monopoles 
are ${\mathbb C}P^{N_{\rm F}-1}$ instantons and kinks 
inside a vortex, respectively 
\cite{Tong:2003pz,Shifman:2004dr,Hanany:2004ea,
Eto:2004rz,Eto:2006pg,Fujimori:2008ee}.
Therefore, bions in Yang-Mills theory can exist 
inside the vortex as the ${\mathbb C}P^{N_{\rm F}-1}$ bions 
when the vortex world-sheet is wrapped around $S^1$. 
With this regards, a non-Abelian vortex in gauge theories 
with gauge group $G$ admits a $G/H$ nonlinear sigma model 
on its world-sheet \cite{Eto:2008yi}.
In particular, the cases of $G=SO(N)$ and $USp(2N)$ 
were studied extensively \cite{Eto:2008qw,Eto:2009bg}.
With twisted boundary conditions, it admits monopoles as 
kinks on $G/H$ sigma models in the vortex theory 
\cite{Eto:2011cv}, in which kinky brane-like picture 
was obtained.
Quark matter in high density QCD also admits a non-Abelian 
vortex \cite{Balachandran:2005ev} whose effective theory is 
the ${\mathbb C}P^{2}$ model 
\cite{Eto:2009bh} (see Ref.~\cite{Eto:2013hoa} 
as a review).
A bound state of a kink and an anti-kink appears quantum mechanically inside a vortex, 
representing a meson of a monopole and an 
anti-monopole. 
The quark-hadron duality between the confining phase at low density and the Higgs phase at high density may be explained through a non-Abelian vortex \cite{Eto:2011mk}.
Bions inside a non-Abelian vortex 
wrapped around $S^1$
in these cases 
are interesting to study in future.

We may compactify two or more directions in higher dimensional theories. 
For instance, the ${\mathbb C}P^{\NF-1}$ model, 
Grassmann sigma model and corresponding gauge theories 
on ${\mathbb R}^2 \times S^1 \times S^1$ 
admit not only fractional lumps (vortices) 
from one of $S^1$'s which are string-like,  
linearly extended structure of the fractional instantons 
studied in this paper but also their intersections 
with Yang-Mills instanton charge \cite{Fujimori:2008ee}. 
These configurations are called Amoebas in mathematics 
and reduce to domain wall junctions \cite{Eto:2005cp} 
for a small $S^1$ radii limit. 
Bions in this theory will have more varieties because 
fractional solitons have networks in two directions, 
and hopefully are useful for four dimensional 
gauge theories.


\begin{acknowledgments}
We are grateful to Mithat \"{U}nsal and Gerald Dunne for their 
interest and valuable comments and correspondences on their 
related work during the entire course of our study. 
T.\ M.\ and N.\ S.\ thank Philip Argyres, Alexei Cherman, Falk Bruckmann, 
and Tin Sulejmanpasic and other participants of CERN theory 
institute 2014, "Resurgence and Transseries in quantum, 
gauge and string theories" for the fruitful discussion 
and useful correspondence. 
T.\ M.\  is in part supported by the Japan Society for the 
Promotion of Science (JSPS) Grants Number 26800147. 
The work of M.\ N.\ is supported in part by Grant-in-Aid for 
Scientific Research (No. 25400268) and by the ``Topological 
Quantum Phenomena''  Grant-in-Aid for Scientific Research on 
Innovative Areas (No. 25103720) from the Ministry of Education, 
Culture, Sports, Science and Technology  (MEXT) of Japan.
N.\ S.\  is supported by Grant-in Aid for Scientific Research 
No. 25400241 from the Ministry of Education, 
Culture, Sports, Science and Technology  (MEXT) of Japan. 

\end{acknowledgments}

\begin{appendix}
\section{The solution of the constraint 
$H H^\dagger=v^2{\bf 1}_{N_{\rm C}}$
}
\label{app:constraint}

Since the gauge invariant quantity $\Omega$ is a 
nonnegative hermitian $N_{\rm C}\times N_{\rm C}$ matrix, 
we can diagonalize it by a unitary matrix $U$ to obtain a 
nonnegative diagonal matrix $\Omega_{\rm d}$ as given in 
Eq.(\ref{eq:diagonal_omega}). 
Since $\Omega$ depends on both $z$ and $\bar z$, matrices 
$U$ and $\Omega_{\rm d}$ depend also on $z$ and $\bar z$. 
Any (non-integer) $\alpha$ powers of a hermitian matrix 
can be defined as 
\begin{equation}
\Omega^{\alpha} = U \Omega_{\rm d}^{\alpha} U^{\dagger}. 
 \label{eq:power_omega}
\end{equation}
One can choose $\alpha=-1/2$ to define the inverse square root 
$\Omega^{-1/2}$ as a solution for $S^{-1}$. 
Thus we find a possible representative of the physical scalar 
field $H$ as in Eq.(\ref{eq:H_formula}). 
It is easy to see that the solution satisfies the remaining 
constraint (\ref{eq:mdl:D-term-cond}) 
\begin{eqnarray}
H H^\dagger
&=& U \Omega_{\rm d}^{-1/2} U^{-1} H_0 
H_0^\dagger U \Omega_{\rm d}^{-1/2} U^{-1} 
=v^2 U \Omega_{\rm d}^{-1/2} U^{-1} \Omega U \Omega_{\rm d}^{-1/2} U^{-1} 
\nonumber \\
&=&v^2 U \Omega_{\rm d}^{-1/2} U^{-1} H_0 U \Omega_{\rm d} U^\dagger 
 U \Omega_{\rm d}^{-1/2} U^{-1} 
=v^2 {\bf 1}_{N_{\rm C}}. 
 \label{eq:H_constraint}
\end{eqnarray}

The $V$-transformations defined in 
Eq.(\ref{eq:vtx:V-trans}) give different representatives 
for the moduli matrix $H_0'=VH_0, V\in GL(N_{\rm C}, {\mathbb C})$. 
The $V$-transformed $H_0'$ gives a covariant $\Omega'$ 
in Eq.(\ref{eq:vtx:V-transOmega}) 
\begin{eqnarray}
v^2\Omega'=
H_0'H_0^{'\dagger}
=VH_0 H_0' V^\dagger
=v^2V\Omega V^\dagger
=v^2VU\Omega_{\rm d} U^\dagger V^\dagger. 
 \label{eq:Vtransf_omega}
\end{eqnarray}
We can diagonalize $\Omega'$ with a unitary matrix $U'$ 
to give a diagonal matrix $\Omega_{\rm d}'$ 
\begin{eqnarray}
\Omega'=U'\Omega_{\rm d}' U^{'\dagger}. 
 \label{eq:Vtransf_omega_diag}
\end{eqnarray}
Therefore we can obtain another scalar field $H'$ 
\begin{eqnarray}
H'=
U'\Omega_{\rm d}^{'-1/2}U^{'\dagger}H_0', 
 \label{eq:V_transf_H}
\end{eqnarray}
which satisfies the constraint 
\begin{eqnarray}
H' H^{'\dagger}
=v^2 {\bf 1}_{N_{\rm C}}. 
 \label{eq:H_constraint2}
\end{eqnarray}
However, $VU$ is not unitary, and is different from $U'$. 
Even the eigenvalues are different $\Omega_{\rm d}'\not=\Omega_{\rm d}$. 
Therefore the resulting solution $H'$ of the constraint obtained from 
the formula (\ref{eq:H_formula}) is in general different from $H$. 
Although relations between $U', \Omega_{\rm d}'$ and $U, \Omega$ 
are complicated, we find that the following matrix $\tilde U
$ 
\begin{eqnarray}
\tilde U=\frac{1}{v^2}H'H^{\dagger}
 \label{eq:unitary_matrix}
\end{eqnarray}
is a unitary matrix 
\begin{eqnarray}
\tilde U \tilde U^\dagger
&=&\frac{1}{v^4}(U'\Omega_{\rm d}^{'-1/2}U^{'\dagger} V H_0 
H_0^\dagger U\Omega_{\rm d}^{-1/2}U^{\dagger}) 
(U\Omega_{\rm d}^{-1/2}U^{\dagger} H_0H_0^\dagger V^\dagger 
U'\Omega_{\rm d}^{'-1/2}U^{'\dagger}) 
\nonumber \\
&=&U'\Omega_{\rm d}^{'-1/2}U^{'\dagger} V 
\Omega V^\dagger 
U'\Omega_{\rm d}^{'-1/2}U^{'\dagger} 
=
 {\bf 1}_{N_{\rm C}}. 
 \label{eq:unitarity_Vtransf}
\end{eqnarray}
This relation together with the relations in 
Eqs.(\ref{eq:H_constraint}) and (\ref{eq:H_constraint2}) implies 
\begin{eqnarray}
H'=\tilde U H. 
 \label{eq:unitarity_Vtransf}
\end{eqnarray}
Namely $H'$ and $H$ are $U(N_{\rm C})$ gauge transformation of each other. 

\end{appendix}



\end{document}